\documentclass[epj,nopacs]{svjour}

\usepackage{graphics}
\usepackage{atlasphysics} 
\usepackage{cite}
\usepackage[parsep]{collref}
\usepackage[figuresleft]{rotating}
\usepackage{epstopdf}
\usepackage{hyperref}
\usepackage{xspace}
\usepackage{mydefs}
\usepackage{subfigure}
\usepackage[switch]{lineno}

\begin{document}
\title{Measurement of inclusive jet and dijet cross sections in proton-proton collisions 
at 7~TeV centre-of-mass energy with the ATLAS detector}
\author{The ATLAS Collaboration}
\institute{}
\date{\today}
\abstract{Jet cross sections have been measured for the first time 
in proton-proton collisions at a centre-of-mass energy of 7~TeV using the ATLAS detector. 
The measurement uses an integrated luminosity of 17~nb$^{-1}$ recorded at the Large Hadron Collider.
The \AKT algorithm is used to identify jets, with two jet resolution parameters, $R=0.4$ 
and $0.6$. The dominant uncertainty comes from the jet energy scale, which is determined to within 
7\% for central jets above 60~GeV 
transverse momentum. 
Inclusive single-jet differential cross sections are presented as functions of jet transverse 
momentum and rapidity. 
Dijet cross sections are presented as functions of dijet mass and the angular variable $\chi$. 
The results are compared to expectations based on next-to-leading-order QCD, which agree with the data,
providing a validation of the theory in a new kinematic regime.
}
\titlerunning{Inclusive jet and dijet cross sections}
\maketitle
\section{Introduction}
\label{sec:intro}

At the Large Hadron Collider (LHC), jet production is the dominant high transverse-momentum (\pt) 
process and as such gives the first glimpse of physics at the TeV scale. 

Jet cross sections and properties are key observables in high-energy particle physics. They have been measured at 
\ee, \ep, \ppb, and \pp colliders, as well as in \gp and \gg collisions.  They have provided precise measurements 
of the strong coupling constant, have been used to obtain information about the structure of the proton and photon, 
and have become important tools for understanding the strong interaction and searching for physics beyond the 
Standard Model (see, for example, \cite{Arnison:1983dk,Alitti:1990aa,Adeva:1990nu,Chekanov:2001bw,Heister:2002tq,Chekanov:2002be,Abdallah:2004uq,Abbiendi:2005vd,Chekanov:2005nn,Abulencia:2007ez,:2007jx,:2008hua,Aaltonen:2008eq,Abazov:2009nc,:2009he,Aaron:2009vs,:2009mh,Abramowicz:2010ke,Abazov:2010fr}). Searches for new physics
using jets in 7 TeV collisions were recently published~\cite{Collaboration:2010bc,angles}.
In this paper, we present the first measurements of inclusive single-jet and dijet cross sections using the ATLAS 
detector.  The measurements are performed using a data set taken early in LHC running, from 
30 March to 5 June 2010, corresponding to an integrated luminosity of $17$~nb$^{-1}$. 
The measurement involves a determination of the
trigger and reconstruction efficiencies of ATLAS for jets, as well as a first determination of
the calorimeter response to jet energy.

The paper is organised as follows. The detector is described in the next section, followed by the definition
of the cross sections to be measured (Section~\ref{sec:xsec}), a discussion of the simulations 
used in the measurement (Section~\ref{sec:mc}) and the theoretical
predictions to which the data are compared (Section~\ref{sec:theory}). 
The evaluation of the trigger efficiency is given in 
Section~\ref{sec:trigger}. The following
two sections (Sections~\ref{sec:JES} and \ref{sec:JESunc}) 
describe the evaluation of the main uncertainty in the measurement, coming from the jet energy scale. 
The event selection and data correction are then described (Sections~\ref{sec:offline} and \ref{sec:correction}), 
followed by the results and conclusions.

\section{The ATLAS Detector}

The ATLAS detector covers almost the entire solid angle around
the collision point with layers of tracking detectors, calorimeters, and muon
chambers. For the measurements presented in this paper, 
the inner detector, the calorimeters, and the trigger are of particular importance.
These components, and the rest of the detector, are described in detail elsewhere~\cite{Aad:2008zzm}. 

The inner detector has full coverage in $\phi$ 
and covers the pseudorapidity\footnote{
Pseudorapidity is defined as $\eta = - \ln({\rm tan}(\theta/2))$. 
The ATLAS reference system is a Cartesian 
right-handed coordinate system, with the nominal collision point at the origin. The anti-clockwise beam 
direction defines the positive $z$-axis, while the positive $x$-axis is defined as pointing from the collision 
point to the centre of the LHC ring and the positive $y$-axis points upwards.  The azimuthal angle $\phi$ is 
measured around the beam axis, and the polar angle $\theta$ is measured with respect to the $z$-axis.} 
range $|\eta|<2.5$. It consists of a silicon pixel detector, a silicon microstrip detector, and a transition 
radiation tracker, all  
immersed in a 2 T magnetic field. These tracking detectors are used to reconstruct tracks and vertices, 
including the primary vertex.

High granularity liquid-argon (LAr) electromagnetic sampling calorimeters, with excellent energy and 
position resolution, cover the pseudorapidity
range $|\eta|<$~3.2 (the barrel covers $|\eta|<1.475$ and the two end-caps cover
\\   
$1.375<|\eta|<3.2$). 
The hadronic calorimetry in the range $|\eta|<$~1.7 is provided by a scintillating-tile 
calorimeter, which is separated into a large barrel ($|\eta|<1.0$) and two smaller extended barrel cylinders, 
one on either side of
the central barrel ($0.8<|\eta|<1.7$). In the end-caps ($|\eta|>$~1.5), LAr hadronic
calorimeters match the outer $|\eta|$ limits of the end-cap electromagnetic calorimeters. The LAr
forward calorimeters provide both electromagnetic and hadronic energy measurements, and extend
the coverage to $|\eta| < 4.9$.

The trigger system uses three consecutive trigger levels to select signal and reject background 
events. The Level-1 (L1) trigger is based on custom-built hardware to process the incoming data with 
a fixed latency of 2.5~$\mu s$. This is the only trigger level used in this analysis. In order to 
commission the trigger software, the higher level triggers
also recorded decisions on events, but these decisions were not applied to reject any data. 
The events in this analysis were accepted either by the system of minimum-bias trigger
scintillators (MBTS) or by the calorimeter trigger. 

The MBTS detector~\cite{Aad:2010rd} consists of 32 scintillator counters that are each 2~cm thick, 
which are organised into two disks with one on each side of the ATLAS detector. The scintillators are installed on the 
inner face of the end-cap calorimeter cryostats at $z = \pm 356$~cm such that the disk surface is perpendicular to 
the beam direction. This leads to a coverage of $2.09 < |\eta| < 3.84$.
The MBTS multiplicity is calculated for each side independently, and allows events containing jets 
to be triggered with high efficiency and negligible bias. 

The L1 calorimeter trigger uses coarse detector information to 
identify the position of interesting physics objects above a given energy threshold.
The ATLAS jet trigger is based on the selection of jets according to their transverse energy, \ET. 
The L1 jet reconstruction uses so-called 
jet elements, which are formed 
from the electromagnetic and hadronic calorimeters with a granularity of $\Delta\phi \times  \Delta\eta = 
0.2 \times 0.2$ for $|\eta|<3.2$. The jet finding is based on a sliding window algorithm with steps of one jet element, 
and the jet \ET is computed in a window of configurable size around the jet.

Recorded events are fully reconstructed offline, using object-oriented analysis software running on a distributed computing grid.

\section{Cross Section Definition}
\label{sec:xsec}

Jets are identified using the \AKT jet algorithm~\cite{Cacciari:2008gp} implemented in the 
\fastjet~\cite{Cacciari:2005hq,fastjet} package. 
This algorithm constructs, for each input object (e.g. a parton, particle or energy cluster) $i$, 
the quantities $d_{ij}$ and $d_{iB}$ as follows:
\begin{eqnarray}
d_{ij} &=& \min(k_{t i}^{-2},k_{t j}^{-2})\frac{(\Delta R)^2_{ij}}{R^2}, \\
d_{iB} &=& k_{t i}^{-2},
\label{eq:dist}
\end{eqnarray} 
where
\begin{equation}
(\Delta R)^2_{ij} = (y_i - y_j)^2 + (\phi_i - \phi_j)^2, 
\end{equation}
$k_{t i}$ is the transverse momentum of object $i$ with respect to the beam direction, $\phi_i$ is its 
azimuthal angle,
and $y_i$ is its rapidity, defined as $y = \frac{1}{2}\ln[(E+p_z)/(E-p_z)]$, 
where $E$ denotes the energy and $p_z$ is the component of the momentum along the beam direction.
A list containing all the $d_{ij}$ and $d_{iB}$ values is compiled. If the smallest entry is a $d_{ij}$, 
objects $i$ and $j$ are 
combined (their four-vectors are added) and the list is updated. If the smallest entry is a $d_{iB}$, 
this object is considered a complete 
``jet'' and is removed from the list. As defined above, $d_{ij}$ is a distance measure between
two objects, and $d_{iB}$ is a similar distance between the object and the beam. Thus the variable $R$ is a resolution
parameter which sets the relative distance at which jets are resolved from each other as compared to the beam. 
In this analysis, two different values for the $R$ parameter are chosen: $R=0.4$ and $R=0.6$; using two values allows
comparison to QCD calculations subject to rather different soft (non-perturbative) QCD corrections. 
The \AKT algorithm is well-motivated since it can be implemented in next-to-leading-order (NLO) perturbative QCD (pQCD)
calculations, is infrared-safe to all orders, and produces geometrically well-defined (``cone-like'') jets. 

The jet cross section measurements are corrected for all experimental effects, 
and so refer to the ideal ``truth'' final-state 
of a proton-proton collision (see, for example~\cite{Buttar:2008jx}), 
where jets are built from stable particles, i.e. those with a proper lifetime longer than 10~ps. This definition
includes muons and neutrinos from decaying hadrons.

Inclusive single-jet double-differential cross sections are measured as a function of 
jet $\pt$ and $y$ for all jets in the kinematic region $\pt > 60\gev, |y| < 2.8$.
This ensures that jets lie well within the high efficiency plateau region for the triggers used, as described in 
Section~\ref{sec:trigger}, and that the jets are in a region where the jet energy scale is well understood, 
as described in Section~\ref{sec:JES}.

The dijet double-differential cross section is measured as a function of the invariant mass of the dijet system, 
$\twomass{1}{2}$, 
binned in the maximum rapidity of the two leading (i.e. highest $\pt$) jets, $|y|_{\mathrm{max}} = \max(|y_{1}|, |y_{2}|)$. It is also measured as a function of 
the angular variable 
\begin{equation}
\chi = \exp( |y_1 - y_2| ) \approx \frac{1+\cos \theta^*}{1-\cos \theta^*}
\label{eq:chi}
\end{equation}
binned in the dijet mass $\twomass{1}{2}$.
Here the subscripts 1,2 label the highest and second highest $\pt$ jet in the event within $|y|<2.8$, 
respectively,
and $\theta^*$ is the polar scattering angle of the outgoing jets in the dijet centre-of-mass frame.
The approximation in the expression is exact for massless jets perfectly balanced in $\pt$.
The leading jet is required to lie in the $\pt, |y|$ kinematic region defined above.
The subleading jet is required to lie in the same rapidity region and to have $\pt > 30\gev$, which
ensures that both the jet reconstruction efficiency and purity\footnote{The efficiency and purity were determined 
using Monte Carlo with a requirement that truth and reconstructed jets match to within $\Delta R = 0.3$.} 
are above 99\%. 
This cut is also important to limit misidentification of the subleading jet due to less precise jet energy resolution 
for $\pt < 30\gev$ (see Section ~\ref{sec:correction}).
Allowing for some imbalance in the \pt of the two jets improves the stability of the NLO 
calculation~\cite{Frixione:1997ks}.    

The dijet mass is plotted in the allowed rapidity region only above the minimum mass where it is no longer 
biased by the $\pt$ and rapidity cuts on the two leading jets.  The minimum unbiased mass depends on the 
$|y|_{\mathrm{max}}$ bin, which determines the maximum opening angle in rapidity allowed.  The biased spectrum 
below this mass is not measured due to its particular sensitivity to the jet energy scale uncertainty through 
the jet $\pt$ cut.

The variable $\chi$ is plotted up to a maximum of 30, restricting the angular 
separation in rapidity to $|y_1 - y_2| < \ln(30)$.  In the rotated coordinate system ($y^{*}$, $y_{\mathrm{boost}}$), 
where $y^{*}=0.5 \cdot (y_{1}-y_{2})$, 
and $y_{\mathrm{boost}}=0.5 \cdot (y_{1}+y_{2})$ is the boost of the dijet system with respect to the laboratory frame, 
this restricts the acceptance to $|y^{*}|<0.5 \ln(30)$.  An orthogonal acceptance cut 
$|y_{\mathrm{boost}}|<1.1$ is then made on the $\chi$ distribution in order to reject events in which 
both jets are boosted into the forward or backward direction.  This reduces the sensitivity to parton density 
function (PDF) uncertainties at low $x$, where $x$ is the fraction of the momentum of the proton 
carried by the parton participating 
in the hard scattering, and in turn enhances sensitivity to differences that could arise from 
deviations from the matrix element predictions of pQCD.  The $\chi$ spectrum is plotted only in 
mass bins above the minimum unbiased mass.

The kinematic constraints mean that the region of $x$ probed by these measurements varies in the approximate 
range $5\times 10^{-4} < x < 0.4$ for the inclusive jet measurements, 
and $1.4 \times 10^{-2} < x < 0.3$ for the dijet measurements.

\section{Monte Carlo Samples}
\label{sec:mc}

Samples of simulated jet events in proton-proton collisions at $\sqrt{s} = 7$~TeV were produced using several
Monte Carlo (MC)
generators. The \pythiasix.421~\cite{pythia} event generator is used for the baseline comparisons and corrections. 
It implements leading-order (LO) pQCD matrix elements for $2 \rightarrow 2$ processes,
$\pt$-ordered parton showers calculated in a leading-logarithmic approximation,
an underlying event\footnote{The term underlying event is used to mean particles produced in the same proton-proton 
collision, but not originating from the primary hard partonic scatter or its products.}  simulation using 
multiple-parton interactions, and uses the Lund string model for hadronisation. 
For studies of systematic uncertainties, jet samples were produced using the \herwigsix~\cite{herwig2} generator, 
which also employs LO pQCD matrix elements, but uses an 
angle-ordered parton shower model and a cluster hadronisation model. The underlying event for the \herwigsix samples 
is generated using the
\jimmy~\cite{Butterworth:1996zw} package using multiple-parton interactions. 
The \herwigpp~\cite{Bahr:2008pv}, \alpgen~\cite{Mangano:2002ea}, and 
\sherpa~\cite{Gleisberg:2008ta} programmes were also used for various cross-checks.
The samples are QCD $2\rightarrow 2$ scattering samples created using a tuned set of parameters denoted as 
ATLAS MC09~\cite{ATL-PHYS-PUB-2010-002} with the MRST2007LO${}^{*}$~\cite{Martin:2009iq,Sherstnev:2007nd} modified leading-order PDFs, unless stated otherwise.

The generated samples are passed through a full simulation~\cite{Collaboration:2010wq} of the ATLAS detector 
and trigger based on \geant~\cite{Agostinelli:2002hh}.
The Quark Gluon String model \cite{QGS} was used for the fragmentation of the nucleus, and 
the Bertini cascade model \cite{Bertini} 
for the description 
of the interactions of the hadrons in the medium of the nucleus. 
The parameters used in \geant are described in more 
detail elsewhere~\cite{geanthadronic}.
Test-beam measurements for single pions have shown that these simulation settings best describe 
the response and resolution in the barrel~\cite{Abat:1263861,Adragna:1214935,TB1,NIMA621134} and 
end-cap~\cite{TB2,1748-0221-2-05-P05005} calorimeters.  

Finally, the events are reconstructed and selected using the same analysis chain as 
for the data with the same trigger, event selection, and jet selection criteria.

\section{Theoretical Predictions}
\label{sec:theory}

Several NLO pQCD calculations are available for jet production in proton-proton collisions.
NLOJET++~4.1.2~\cite{Nagy:2003tz} was used to calculate the QCD $2\rightarrow2$ scattering process at NLO 
for comparison with data. JETRAD~\cite{Giele:1993dj} 
was used for cross-checks. The CTEQ 6.6~\cite{Nadolsky:2008zw} NLO parton densities were used for the central 
value and uncertainties, and the MSTW 2008~\cite{Martin:2009iq}, NNPDF 2.0~\cite{Ball:2008by} and 
HERAPDF 1.0~\cite{H1:2009wt} parton density sets were used as cross-checks. 
The default renormalisation and factorisation scales ($\mu_{R}$ and $\mu_{F}$ respectively) 
were defined to be equal to the $\pt$ of the leading jet in the event.
To estimate the potential impact of higher order terms not included in the calculation, 
$\mu_{R}$
was varied from half to twice the default scale.  
To estimate the impact of the choice of the scale at which the PDF evolution is separated 
from the matrix element, 
$\mu_{F}$
was similarly varied. These two scales were varied independently apart 
from a constraint that the ratio of the two scales be between
1/2
and 2, applied to avoid introducing 
large logarithms of the ratio of the scales. In addition, the effect of the uncertainty in the
strong coupling constant, $\alpha_s(M_Z)$, was estimated by calculating the cross section using $\alpha_s(M_Z)$ 
values within the uncertainty range, and using PDFs fitted using these values.
To efficiently calculate all these uncertainties, the \applgrid~\cite{Carli:2010rw} program was used.

The NLO calculations predict partonic cross sections, which are unmeasurable.
For comparison with data at the particle level, soft (non-perturbative) corrections must be applied. 
This was done using leading-logarithmic parton shower Monte Carlo programs,
by evaluating the ratio of the cross section before and after hadronisation and underlying event simulation
and dividing the NLO theory distributions by this factor. The \pythiasix and \herwigsix models 
described above were used, as well as
a variety of alternative tunes of 
\pythiasix~\cite{Skands:2009zm,Buckley:2009bj}
as a cross-check. The central value used is that from the \pythiasix MC09 sample, and the uncertainty is estimated 
as the maximum spread of the other models investigated. To calculate the particle and parton-level 
theory distributions, the 
\rivet~\cite{Buckley:2010ar} package was used. 
The soft QCD corrections depend significantly on the value of $R$ (0.4 or 0.6), since
wider jets are affected more by the underlying event, whereas narrower jets are more likely to lose particles
due to hadronisation. The size of these effects, and their dependence on jet size, increases with decreasing \pt.
The corrections are within 5\% of unity over most of the kinematic
region, but drop to -10\% for the lowest $\pt$ jets with $R=0.4$, and rise to about 15\% for the lowest $\pt$ 
jets with $R=0.6$.

\section{Trigger Efficiency}
\label{sec:trigger}

The MBTS\_1 trigger, which requires a single MBTS counter 
over threshold, was operational in the early data-taking period.  
It was used to trigger approximately 2\% of the integrated luminosity of the data sample analysed.  
It has negligible inefficiency (as measured in randomly triggered events~\cite{Aad:2010rd}) 
for the events considered in this analysis, which all contain several charged tracks.
As the instantaneous luminosity increased, this trigger had a large prescale factor applied. Consequently
subsequent events -- 
comprising approximately 98\% of the data sample studied -- were triggered by the jet trigger.

\begin{figure}
\begin{center}
\includegraphics[width=0.5\textwidth]{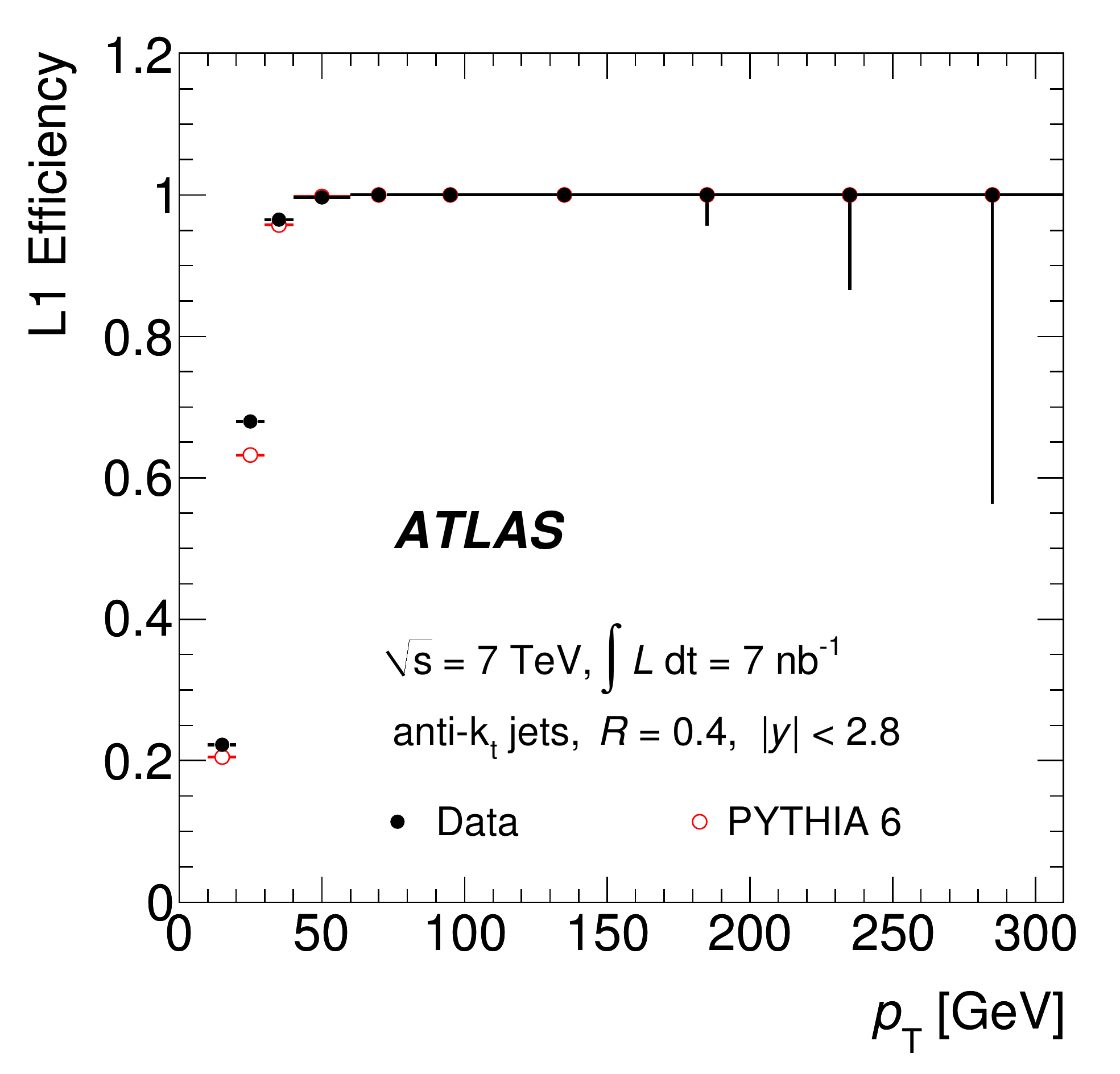}

\includegraphics[width=0.5\textwidth]{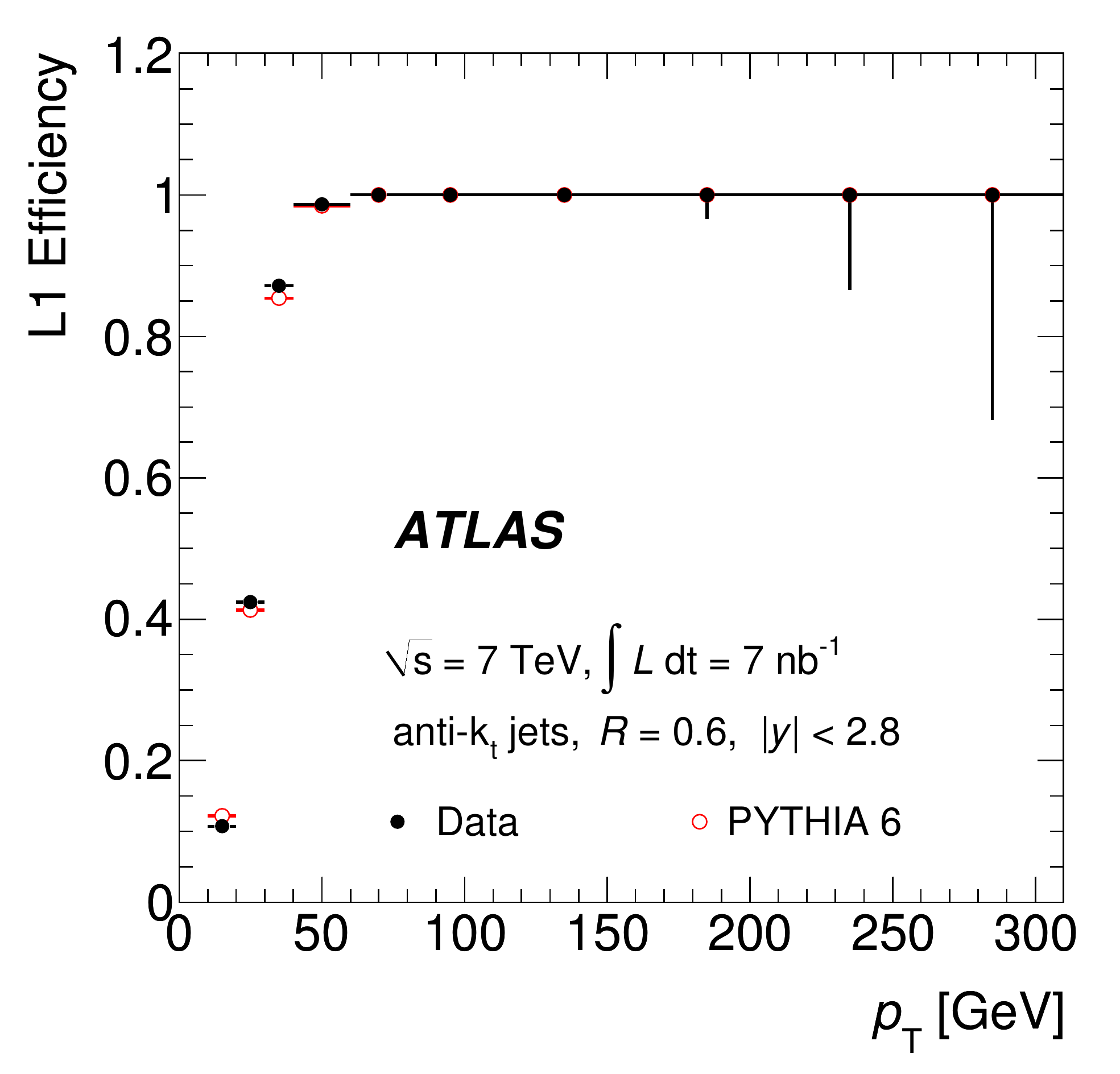}
\caption{
Inclusive-jet L1 trigger efficiency as a function of reconstructed jet \pt for jets identified using the 
$\AKT$ algorithm with (upper) $R=0.4$ and (lower) $R=0.6$.}
\label{fig:trigger}
\end{center}
\end{figure}

The lowest threshold L1 jet trigger, which is used in this analysis, employs 
a $0.4 \times 0.4$ window size in $\eta-\phi$ and 
requires a jet with $\pt > 5\gev$ at the electromagnetic scale  (see Section~\ref{sec:JES}).
The inclusive jet trigger efficiency was measured with respect to the MBTS\_1 trigger, 
which provides an unbiased reference as described above. Its efficiency is shown as a function of the final 
reconstructed \pt 
for single jets ($R=0.4$ and $0.6$) in Fig.~\ref{fig:trigger}. The efficiency is 
compared to 
that predicted from MC simulation, demonstrating that the modelling of the trigger efficiency curve is good. 
The trigger efficiency for jets with $\pt > 60\gev$ and $|y|<2.8$ is above 99\%. All events considered here contain
at least one jet in this region.

\section{Jet Energy Scale Calibration}
\label{sec:JES}

The input objects to the jet algorithm in the data and in the detector-level simulation are 
topological energy clusters in the calorimeter~\cite{ATLAS-LARG-PUB-2008-002}. These clusters are seeded by 
calorimeter cells with energy $|E_{\rm cell}|>4\sigma$ above the noise, where $\sigma$ is
the RMS of the noise.  
All directly neighbouring cells are added, then neighbours of neighbours are iteratively added 
for all cells with signals above a secondary threshold $|E_{\rm cell}|>2\sigma$. Finally the energy in 
all further immediate neighbours is added. Clusters are split or merged based on the position
of local minima and maxima.
The cell energies are summed to give the cluster energy, and the clusters are treated as massless.
The baseline calibration for these clusters corrects their energy to the electromagnetic (EM) scale. 
The EM scale is 
established using test-beam measurements for electrons and muons in the electromagnetic 
and hadronic calorimeters \cite{Ctbelectron,LArReadiness,ATLAS-TCAL-2010-01-003}. 
It provides a good estimate of the energy deposited in the calorimeter by photons and electrons, 
but does not correct for detector effects on the calorimeter measurement, including:
\begin{itemize}
 \item calorimeter non-compensation (the ATLAS calorimeters' response to hadrons is
lower than their response to electrons of the same energy),
 \item energy losses in inactive regions of the detector (``dead material''),
 \item particles for which the shower is not totally contained in the calorimeter.
\end{itemize}
In addition, the baseline calibration does not correct for:
\begin{itemize}
 \item particles that are clustered into the truth jet but for which the corresponding cluster is not in the reconstructed jet,
 \item inefficiencies in energy clustering and jet reconstruction.
\end{itemize}
After a jet is identified, its energy is calibrated to account for these effects, as follows.

The jet energy calibration is carried out in 45 bins of $\eta$ as a function of $\pt$ and is based upon 
MC simulation. The simulation has been validated using test-beam and collision data. 
Jets with pseudorapidity up to 1.2 are 
considered central, while jets with $1.2 < |\eta| < 2.8$ belong to the end-cap region\footnote{The end-cap region 
includes the transition in the ATLAS detector between the barrel and the end-cap, 
which needs special treatment because of its geometry and material composition.}. 

The jet energy scale (JES) is obtained using reconstructed calorimeter jets matched to MC particle jets
\\   
(truth jets, but excluding muons and neutrinos) within a cone of $\Delta R=0.3$. Each jet is
required to be isolated, such that there are no other jets with $\pt > 7$~GeV within a cone of $2.5 \times R$ 
around the jet axis.
The distribution of the response of the 
calorimeter jets
matched with MC particle-level jets, in bins of particle-level jet \pt and $\eta$, 
is used to determine the average jet energy 
response as the mean value of a Gaussian fit.

\begin{figure}
  \centering
  \includegraphics[width=0.5\textwidth]{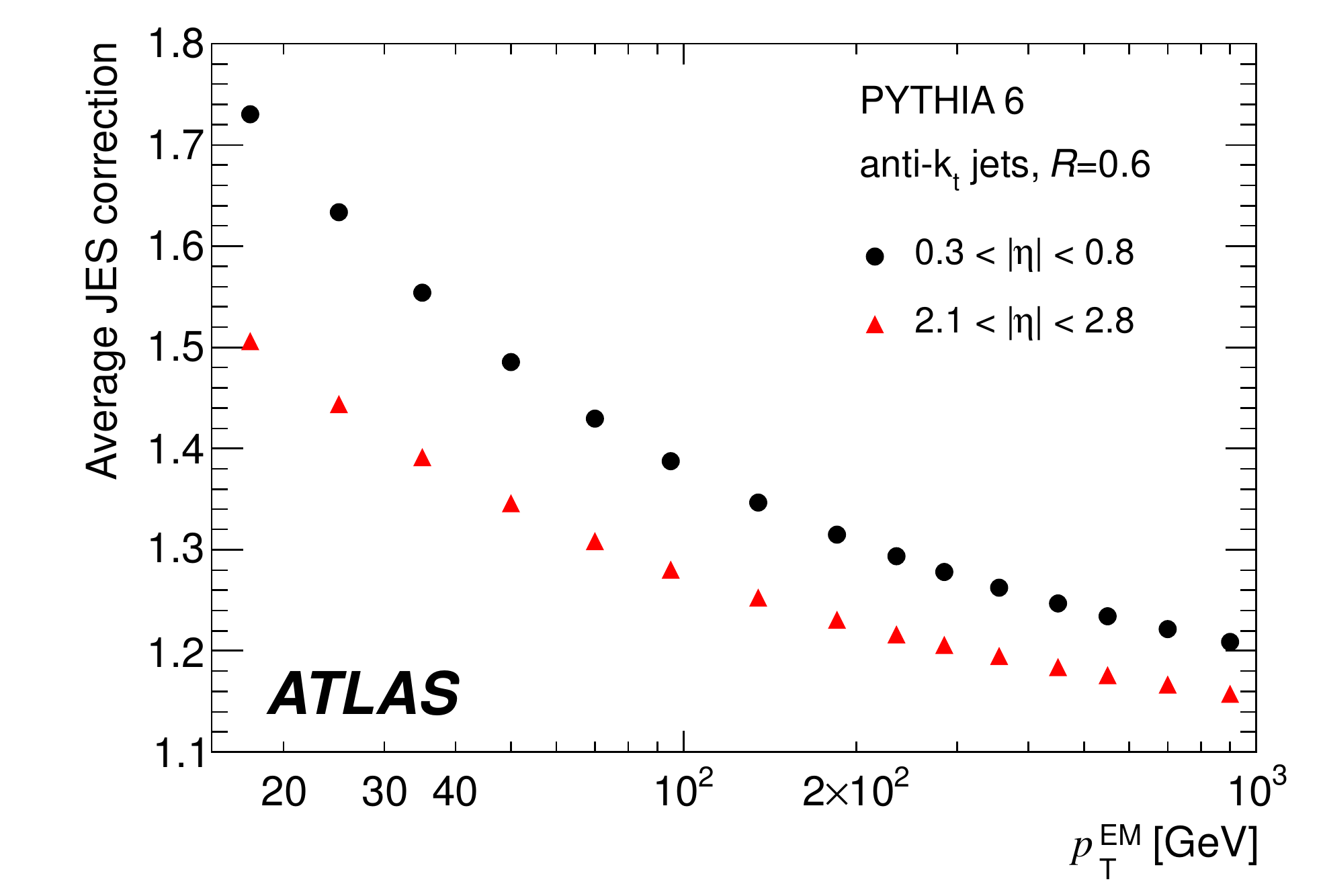}
  \caption{Average jet energy scale correction, evaluated using \pythiasix, 
as a function of jet transverse momentum 
at the EM scale for jets in the central barrel
(black circles) and end-cap (red triangles) regions, 
shown in EM scale \pt bins and $\eta$ regions.}
  \label{fig:NI}
\end{figure}

The correction is obtained by evaluating the transfer function between the energy of the particle-level 
and EM scale jets,
inverting it, and refitting the resulting distribution in bins of reconstructed $\pt$ to obtain a correction
which can be applied in such bins.
The JES correction is shown in Fig.~\ref{fig:NI} as a function of the jet \pt at the EM scale, for \AKT jets with $R=0.6$, 
for two of the rapidity bins.
The size of the overall correction to the $\pt$ of the jets is below 75\%, and for central jets 
with $\pt > 60\gev$ it is below 50\%. 

\section{Uncertainty on the Jet Energy Scale}
\label{sec:JESunc}

The JES systematic uncertainty
is derived combining information from test-beam data, LHC collision
data and MC simulations.

The pseudorapidity bins used for the estimate of the jet energy scale
uncertainty divide the detector in five $|\eta|$ regions with boundaries 
at $0.0, 0.3, 0.8, 1.2, 2.1$ and 
$2.8$. This binning closely matches the binning in $y$ used in the final 
cross section measurement, which follows the calorimeter 
geometry\footnote{For massless objects, rapidity and pseudorapidity
are identical.}.

Only jets with a particle-level jet $\pt>20$~GeV, and a measured $\pt>10$~GeV after calibration, are 
considered. 
No isolation requirement
is imposed in the evaluation of the uncertainty in the JES.



\subsection{Experimental Conditions and Calibration Method}
\label{sub:Detector}

Several sources of uncertainty related to the detector and experimental conditions have been considered: 

\begin{itemize}

\item {\bf{Material and Geometry}} 
The effect of additional dead material on the jet energy 
scale has been evaluated with a dedicated geometry model in the simulation, which includes 
the presence of additional material in front of the barrel calorimeters. 
Test-beam measurements~\cite{Aharrouche:2010zz} and comparisons of 900 GeV data to 
simulations~\cite{atlas:2010knc} have been used to conservatively estimate 
the largest possible change in the amount of material. The contribution to the JES uncertainty
from this source is around 2\% of the jet energy.

 \item {\bf{Noise Thresholds}} The uncertainty on the JES due to possible discrepancies between data and the
description of the calorimeter electronic noise in the Monte Carlo was
evaluated using
MC simulation samples reconstructed with signal-to-noise thresholds for topological cluster 
seeds and cell neighbours modified to be 10\% higher and 10\% lower than their nominal values. 
The stability observed in the noise in special monitoring runs where calorimeter signals were studied in
the absence of genuine signals, and the comparison of the noise distribution between data and MC simulation,
indicate that this 10\% variation provides a conservative estimate of the uncertainty on the noise description. 
The maximum contribution to the JES from this source occurs at low jet $\pt$ values, where it is around 3\% 
of the jet energy.

\item {\bf{Beamspot}} The jet reconstruction for the JES calibration uses $(x,y,z)=(0,0,0)$ as a reference to calculate 
the direction and $\pt$ of the input jet constituents. 
If the beamspot is shifted with respect to this position, and if this shift is not correctly modelled, 
the jet $\pt$ could be biased. 
The variation of the JES from differences in 
the beamspot position between data and MC simulation is evaluated using a 
sample generated with a shifted beamspot of $(x,y,z)=(1.5,2.5,-9)$ mm. This shift covers the shift 
in the current average coordinates observed from data collected by ATLAS from 
LHC collisions: $(x,y,z)=(-0.4,0.62,-1.3)$ mm. The contribution to the JES uncertainty is below 1\% 
of the jet energy.

\item {\bf{EM scale}} For the LAr calorimeters, the EM
scale has been measured in test-beam studies, translating into a $3\%$ uncertainty 
in the scale for in situ operation of the 
calorimeter~\cite{Aharrouche:2010zz,Aharrouche:2008zz,LArReadiness}.

For the tile calorimeter, the EM scale uncertainty
of 4\% is obtained by comparing test-beam muons, 
cosmic-ray muons and simulation~\cite{ATLAS-TCAL-2010-01-003}.


These uncertainties are scaled according to the average fraction of jet energy 
deposited, respectively, in the electromagnetic 
and hadronic calorimeter as a function of \pt, and
combined to form the uncertainty on the EM scale.

\item {\bf{Closure test of the JES calibration}} Any deviation from unity (non-closure) in \pt and 
energy response with respect to the particle jet after the application of the JES corrections 
to the nominal MC sample 
implies that the kinematics of the calibrated calorimeter
jet are not restored to that of the corresponding particle jets.
This can be caused by, for example, the fact that the JES calibration is 
derived using isolated jets, while the systematic 
uncertainty is estimated for inclusive jets.

The systematic uncertainty due to the non-closure of the calibration procedure in any given 
bin is taken as the largest deviation of the response from unity seen either in energy or \pt in that bin.
The contribution to the uncertainty from this source is below 2\% of the jet energy.

\item {\bf{JES uncertainty from dijet balance studies}}
The JES uncertainty for the higher rapidity regions of the barrel and for the end-cap region
is determined using the JES uncertainty for 
the central barrel region (0.3~$ < |\eta| < $~0.8) as a baseline, and adding a contribution from the 
calibration of the jets with respect to it.
This contribution is evaluated by measuring the relative \pt balance 
of forward jets in dijet events against reference central jets~\cite{ATLAS-CONF-2010-055}. 
The $\eta$ intercalibration uncertainty is determined for jets where the average \pt ($\ptavg$) 
of the two leading jets is between 50~GeV and 110~GeV and the resulting uncertainty is applied to all \pt. 
Since the main sources of uncertainty have been shown to decrease for higher $\pt$ and energy values than those 
considered in the jet $\eta$-intercalibration study, this leads to a conservative estimate of the uncertainty 
in the end-cap region for most of the jets considered.


The ratio of the calorimeter response between the reference jet (lying in the region
$0 < |\eta| < 0.8$   
) and the 
probe jet, as a function of the probe jet $\eta$, is shown in Fig.~\ref{fig:etaIntercalibCoarse}
for both data and simulation. Two contributions to the uncertainty are derived -- that due to the difference
between data and simulation, and that due to the deviation from unity in the data.
The combined contribution to the JES uncertainty from these sources is below 3\% 
of the jet energy.


\begin{figure}
  \includegraphics[width=0.5\textwidth]{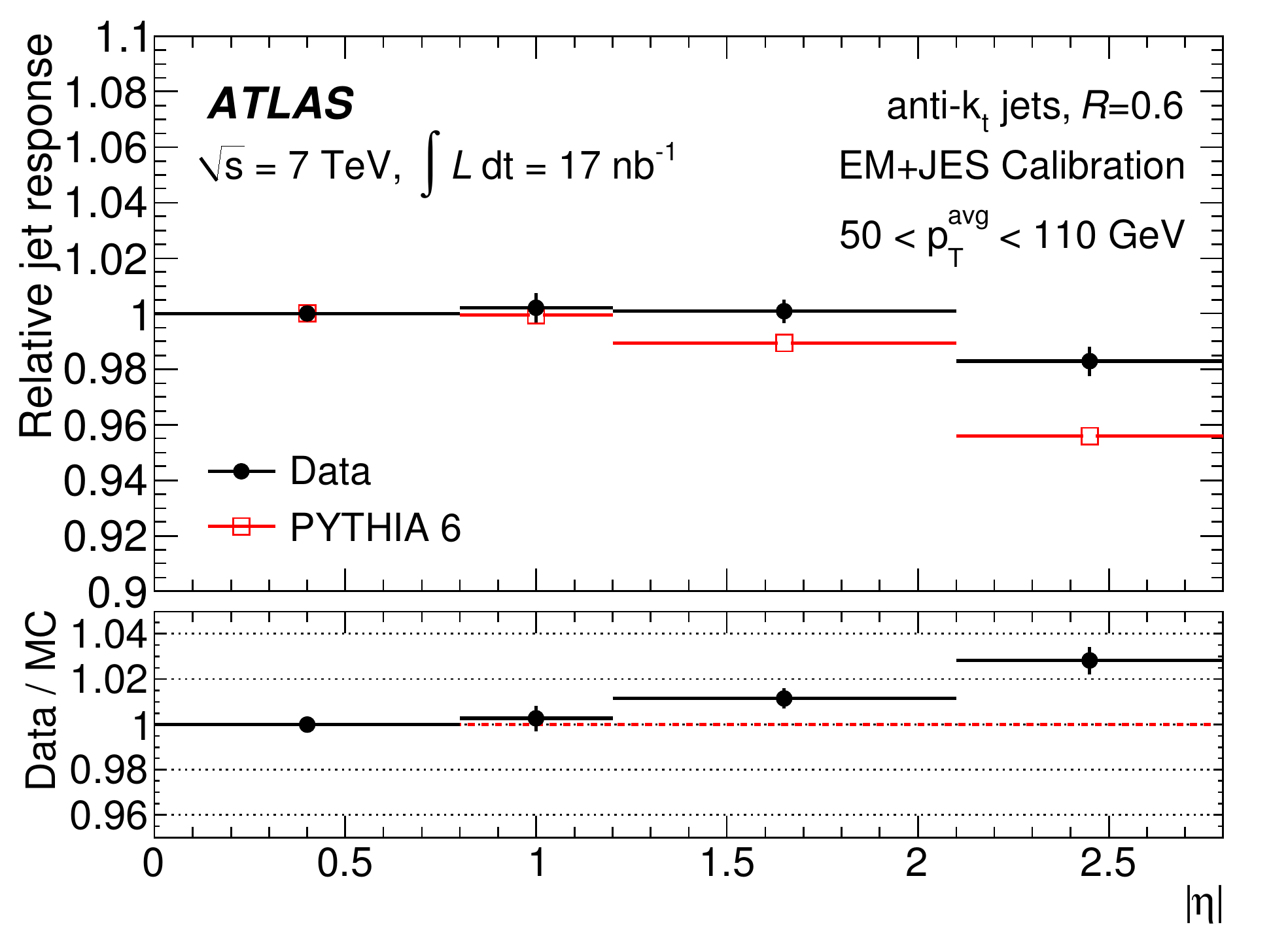}
\caption{
Jet $\pt$ response ($p_T^{\mathrm{jet, probe}}/p_T^{\mathrm{jet, reference}}$) after the EM scale plus JES (EM+JES) 
calibration against jets in the central reference region (0 $ < |\eta| < $ 0.8), obtained by exploiting
the $\pt$ balance as a function of $\eta$ in data and simulation. The lower plot indicates the ratio of the data to the simulation result.
  \label{fig:etaIntercalibCoarse}}
\end{figure}

\end{itemize}

\subsection{Hadronic Shower Model}
\label{sub:PhysicsLists}

The contributions to the JES uncertainty from the
hadro\-nic   
shower model are
evaluated using two MC samples, one in which the Bertini nucleon cascade is not used,
and one in which the Fritiof model~\cite{Fritiof} is used instead of the Quark Gluon String fragmentation 
model.

ATLAS test-beam data for single pions with energies ranging from 2 to 180~GeV have been 
compared to simulations using these two sets of parameters~\cite{ATL-CAL-PUB-2010-001}.
The measured mean single pion response was shown to lie between these two descriptions
over the whole $\pt$ range of particles in jets.
These models lead to variations of within $\pm 4\%$ in the calorimeter response to hadrons.
This is confirmed by studies comparing single isolated hadrons in collision data 
to MC simulation~\cite{ATLAS-CONF-2010-052}.

\subsection{Event Generator Models}
\label{sub:MCUnc}

The contributions to the JES uncertainty from the fragmentation and underlying event models and 
parameters of the MC event generator are obtained using samples generated with
\alpgen + \herwig + \jimmy (which has a different matrix element,  parton shower, hadronisation 
model and underlying event compared to the nominal sample), the \pythiasix MC09 tune modified to use 
Perugia0 fragmentation 
(which has a different underlying-event model with respect to the nominal sample~\cite{Skands:2009zm}) and the
\pythiasix MC09 tune modified to use parameters tuned to LEP data using the \professor~\cite{Buckley:2009bj} software.

The observed deviations of the response from unity are smaller than 4\%.

\subsection{Pile-up}
\label{sec:pileup}

In data-taking periods with higher instantaneous luminosity, 
the effect of pile-up (multiple proton-proton interactions in the same bunch crossing) 
was small, but not negligible. The size of the effect was estimated by studying the dependence of the average
energy density deposited in the calorimeters as a function of the number of reconstructed vertices per event.
No correction is applied for this effect, but it is accounted for in the JES uncertainty. 
For jets with $20 < \pt < 50$~GeV, the pile-up fractional systematic uncertainty is about 1\% in the barrel 
and 1-2\% in the end-caps. For $\pt >50$~GeV, the pile-up uncertainty is only significant for 
$|\eta| > 2.1$, and is smaller than 1\%.

\subsection{Effect of decorrelated JES uncertainty on dijet observables}
\label{sec:uncorr}

Dijet observables, which in a single event can span the entire range in 
rapidity, with one jet in the central region and one in the end-cap region, are sensitive to decorrelations 
in the JES uncertainty as well as to its value at any given bin in rapidity.
Based on results from dijet balance, a 3\% positive 
shift in the jet energy scale was taken at $|y|=2.8$ 
compared to that at $|y|=0$, varying linearly in between and assumed to be symmetric in rapidity. This
shift is interpreted as a 3\% relative JES uncertainty and
is added in quadrature to the other (absolute) sources of uncertainty for 
the dijet cross sections.

\subsection{Combination of JES Uncertainties}

Given that the JES uncertainty is applied to all components of the jet four-momentum, 
the largest deviation from unity in each bin derived from energy or 
\pt response is considered as the contribution to the final JES systematic uncertainty
for each specific systematic effect.

All individual uncertainties are added in quadrature
except that from the closure test, which is conservatively treated
as fully correlated and added linearly.

\begin{figure}
  \centering
  \includegraphics[width=0.5\textwidth]{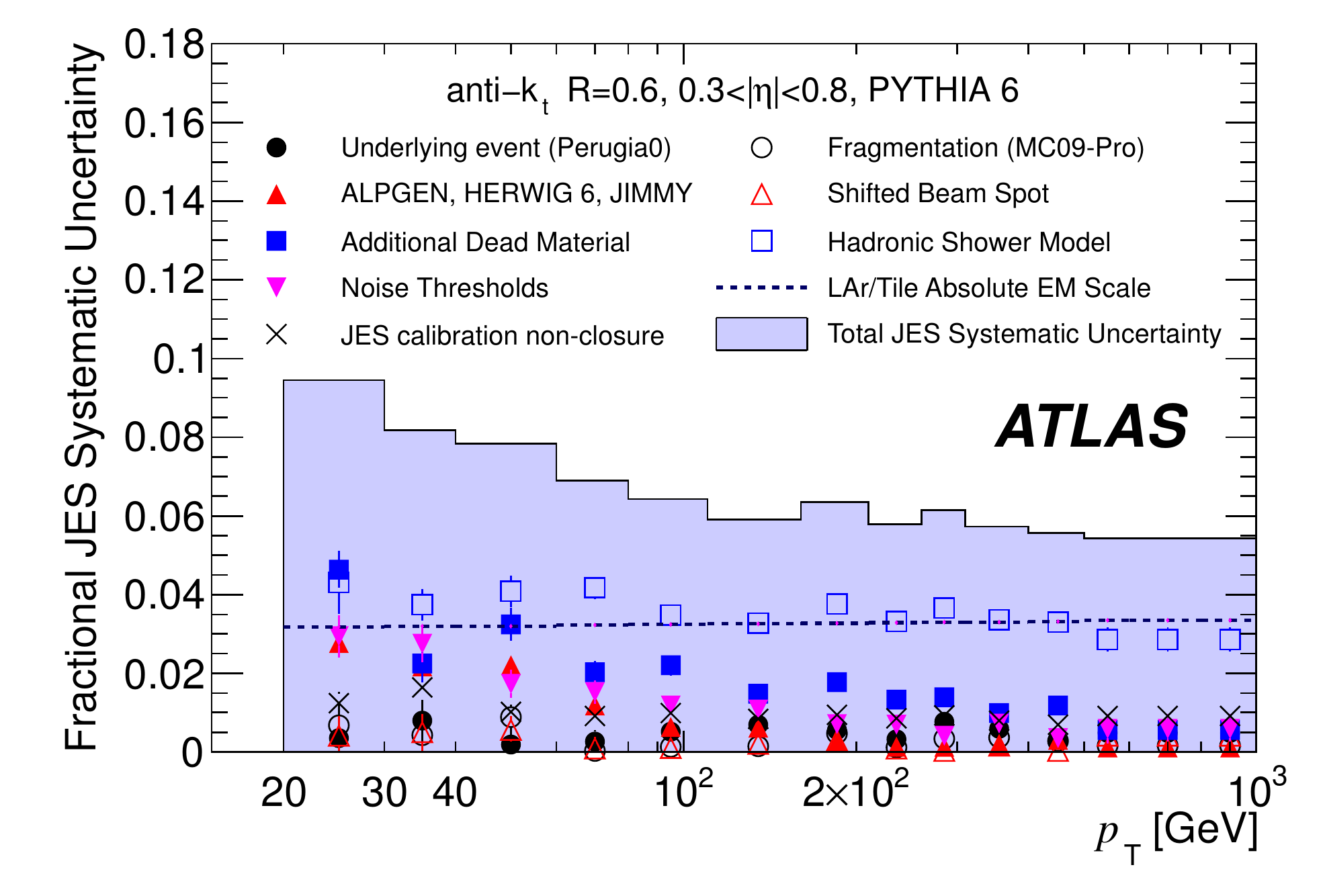}
  \caption{Fractional jet energy scale systematic uncertainty as a function of \pt for jets in the 
pseudorapidity region 0.3~$<|\eta|<$~0.8 in the barrel calorimeter. The total systematic uncertainty is shown as 
the solid light blue area. 
The individual sources are also shown, with statistical errors if applicable.}
  \label{fig:FinalJES1}
\end{figure}

\begin{figure}
  \centering
{
  \includegraphics[width=0.5\textwidth]{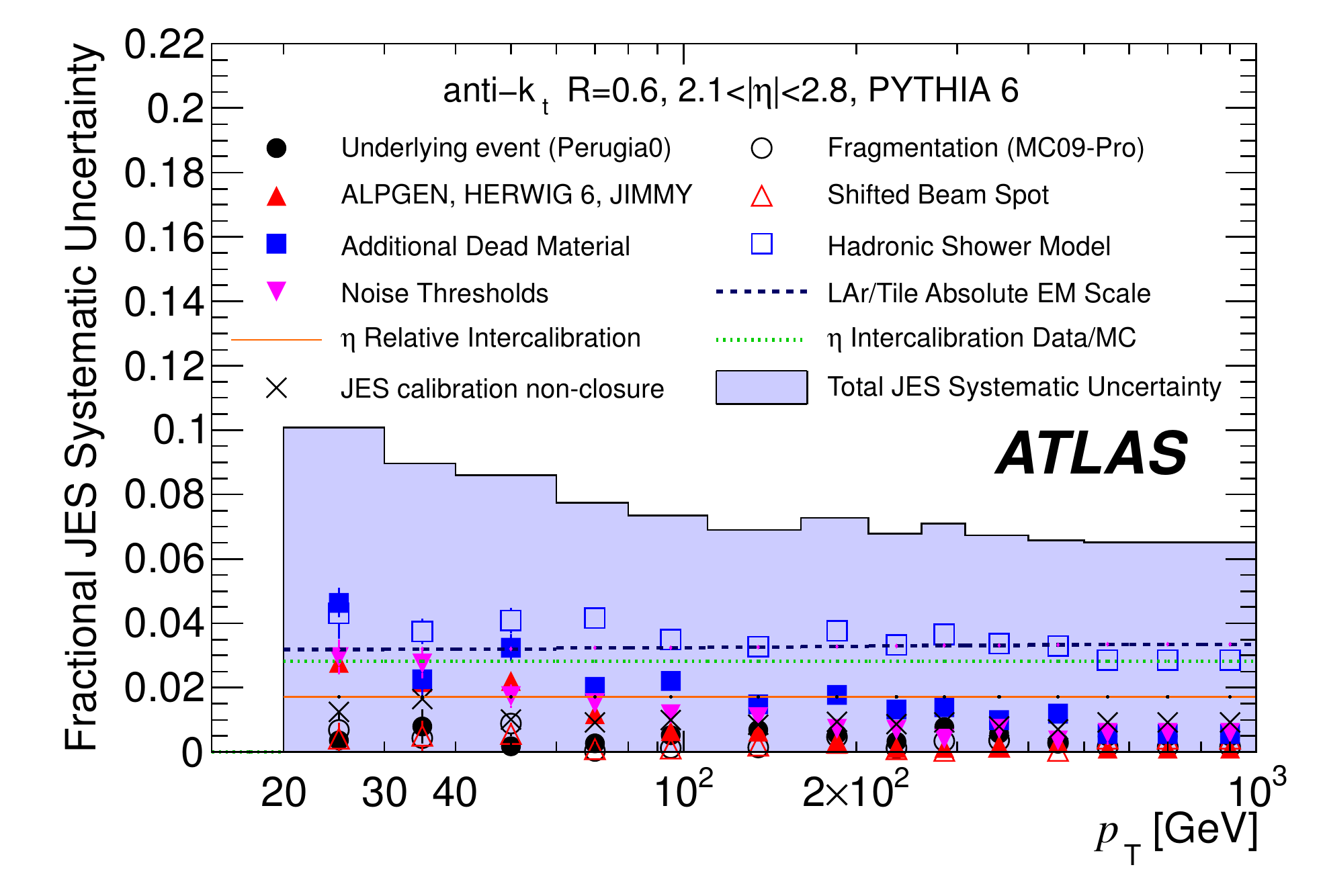}
  \caption{Fractional jet energy scale systematic uncertainty as a function of \pt for jets in the 
pseudorapidity region 2.1~$< |\eta| <$~2.8. 
The total uncertainty is shown as the solid light blue area. 
The JES uncertainty for the end-cap is extrapolated 
from the barrel uncertainty using dijet balance, with the contributions from the deviation 
from unity in the data ($\eta$ relative intercalibration) and the
deviation between data and simulation ($\eta$ intercalibration Data/MC) shown separately. 
The other individual sources
are also shown, with statistical errors if applicable.}
  \label{fig:FinalJES2}

}
\end{figure}

Figure~\ref{fig:FinalJES1} 
shows the final fractional JES systematic uncertainty as 
a function of jet \pt for an example central $\eta$ region. 
Figure~\ref{fig:FinalJES2} shows the forward region, where the contribution from intercalibration with
the central region is also included.
Both the total systematic uncertainty (light blue area) and the
individual contributions are shown, with statistical errors from the fitting procedure if applicable.
The pile-up uncertainty (see Section~\ref{sec:pileup}) and the dijet-specific relative uncertainty (see Section~\ref{sec:uncorr}) are not included in these plots.

The maximum JES uncertainty in the central region amounts to approximately 9\% for 
jets with 30 GeV $<$ \pt $< $ 60 GeV, and 7\% for \pt~$> $~60 GeV. 
The uncertainty is increased to up to 10\% and 8\% respectively for 30 GeV $<$~\pt~$< $~60 GeV and 
\pt~$> $~60 GeV in the end-cap region, where
the central uncertainty is taken as a baseline and the uncertainty due to the intercalibration 
is added.

The dominant contributions to the uncertainty come from the hadronic shower model, the EM scale uncertainty, 
the detector material description, and the noise description.

The same study has been repeated for \AKT jets with resolution
parameter $R=0.4$, and the estimate of the JES uncertainty is comparable
to \AKT jets with $R=0.6$, albeit slightly smaller because of the reduced effect of
the dead material variation and the change in the noise contribution
due to the smaller jet radius.
The JES uncertainty for \AKT jets with $R=0.4$ is between $\approx$ $8\%$ ($9\%$) at low 
jet \pt and $\approx$ $6\%$ ($7\%$) for jets with $\pt>$ 60 GeV in the central (end-cap) region.

The overall JES uncertainty is consistent with the results of detailed comparisons between
collision data and simulation~\cite{ATLAS-CONF-2010-053}.



\section{Event Selection}
\label{sec:offline}

The jet algorithm is run on energy clusters assuming that the event vertex is at the origin. The jet momenta 
are then corrected for the beamspot position.
After calibration, all events are required to have at least one jet within the kinematic region
$\pt > 60\gev, |y| < 2.8$. 
Additional quality criteria are also applied to ensure that jets are not produced by 
noisy calorimeter cells or poorly-calibrated detector regions~\cite{ATLAS-CONF-2010-038}.
Events are required to have at least one vertex with at least five reconstructed tracks connected, 
within 10~cm in $z$ of the beamspot. 
Simulated events are reweighted so that the $z$ vertex distribution agrees with the data.
Of the events passing the kinematic selection, 2.6\% have more than one vertex.
The overall efficiency of these selection cuts, evaluated in
simulation using triggered events with truth jets in the kinematic region of the measurement, 
is above 99\%, and has a small dependence on the kinematic variables.
Background contributions from non-$pp$-collision 
sources were evaluated using unpaired and empty bunches and found to be negligible.

After this selection, 56535 (77716) events remain, for $R=0.4~ (0.6)$,
with at least one jet passing the inclusive jet selection. 
Of these, 45621 (65739) events also pass the dijet selection.

\section{Data Correction}
\label{sec:correction}

The correction for trigger and detector efficiencies and resolutions, other than the energy scale correction 
already applied, is performed in a single step using a bin-by-bin unfolding method evaluated using the MC 
samples. For each measured distribution, the corresponding MC cross section using truth jets (including muons 
and neutrinos) is evaluated in the relevant bins, 
along with the equivalent distributions after the detector simulation and analysis cuts. The ratio of the 
true to the simulated distributions provides a 
correction factor which is then applied to the data. \pythiasix is used for the central
correction. The uncertainty is estimated from the spread of the correction for the different generators, and also
from artificially changing the shape of the simulated distributions by reweighting the MC samples to account
for possible biases caused by the input distribution. 

\begin{figure}
\begin{center}
\includegraphics[width=0.4\textwidth]{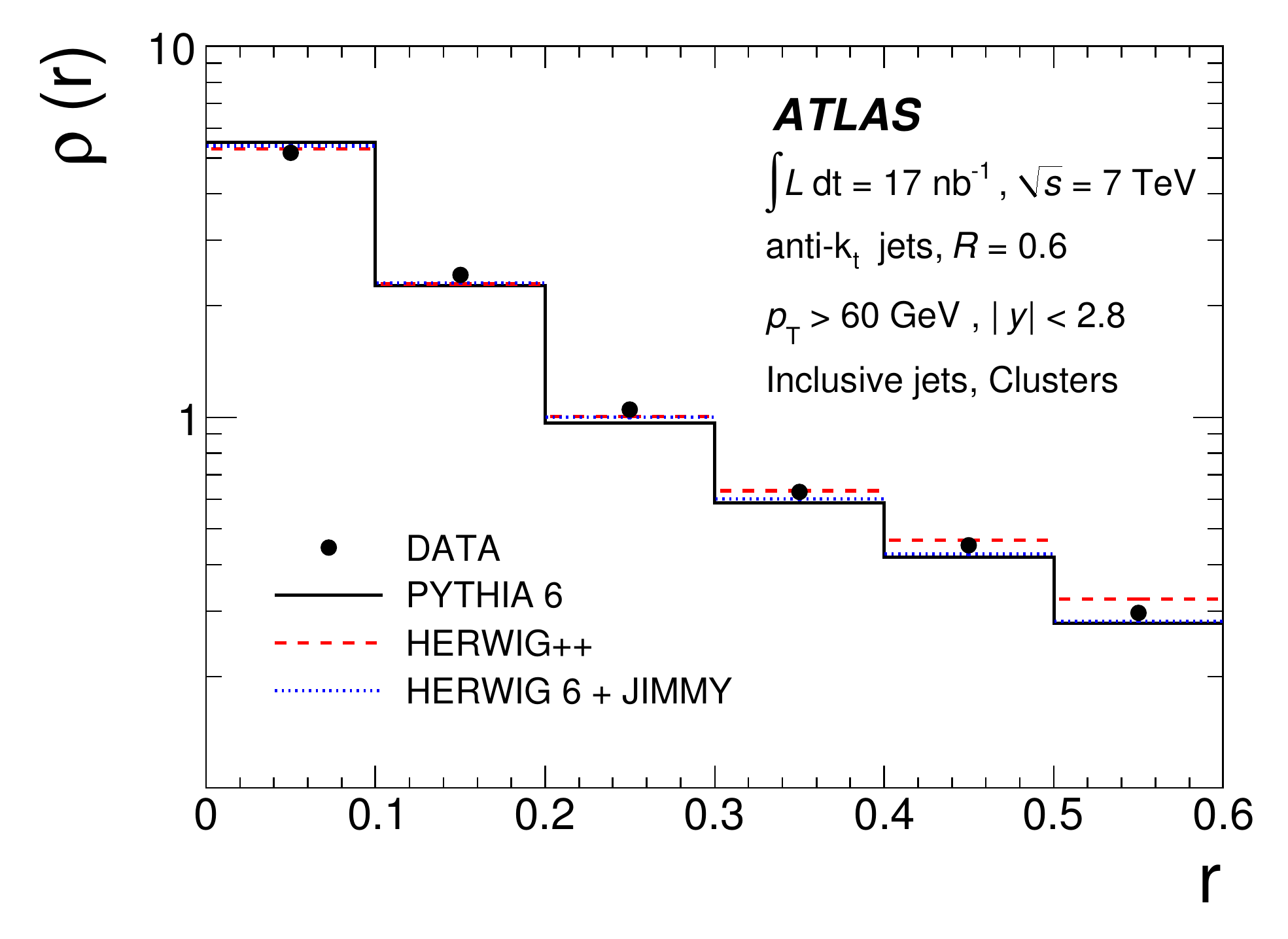}
\includegraphics[width=0.4\textwidth]{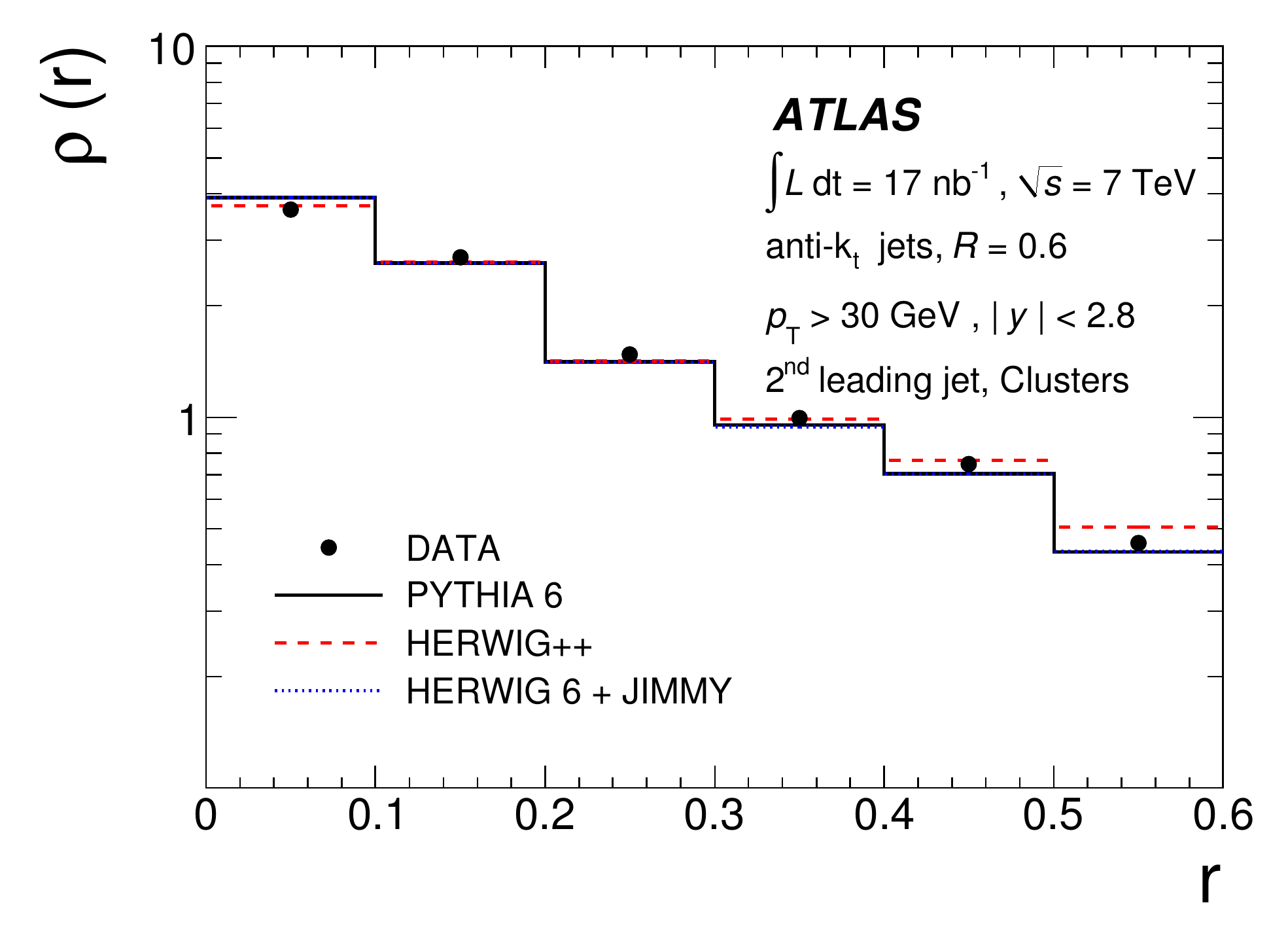}
\includegraphics[width=0.4\textwidth]{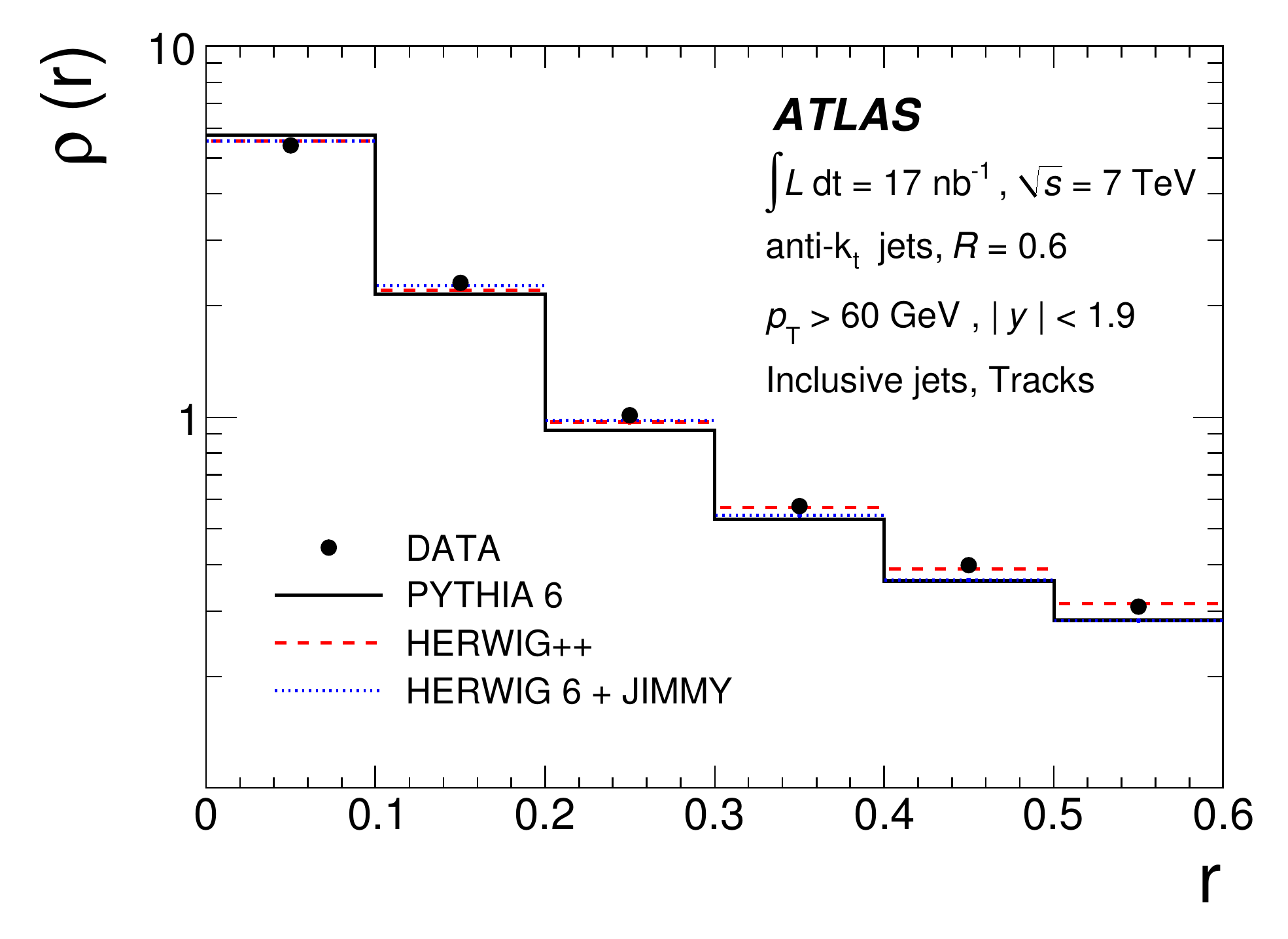}
\includegraphics[width=0.4\textwidth]{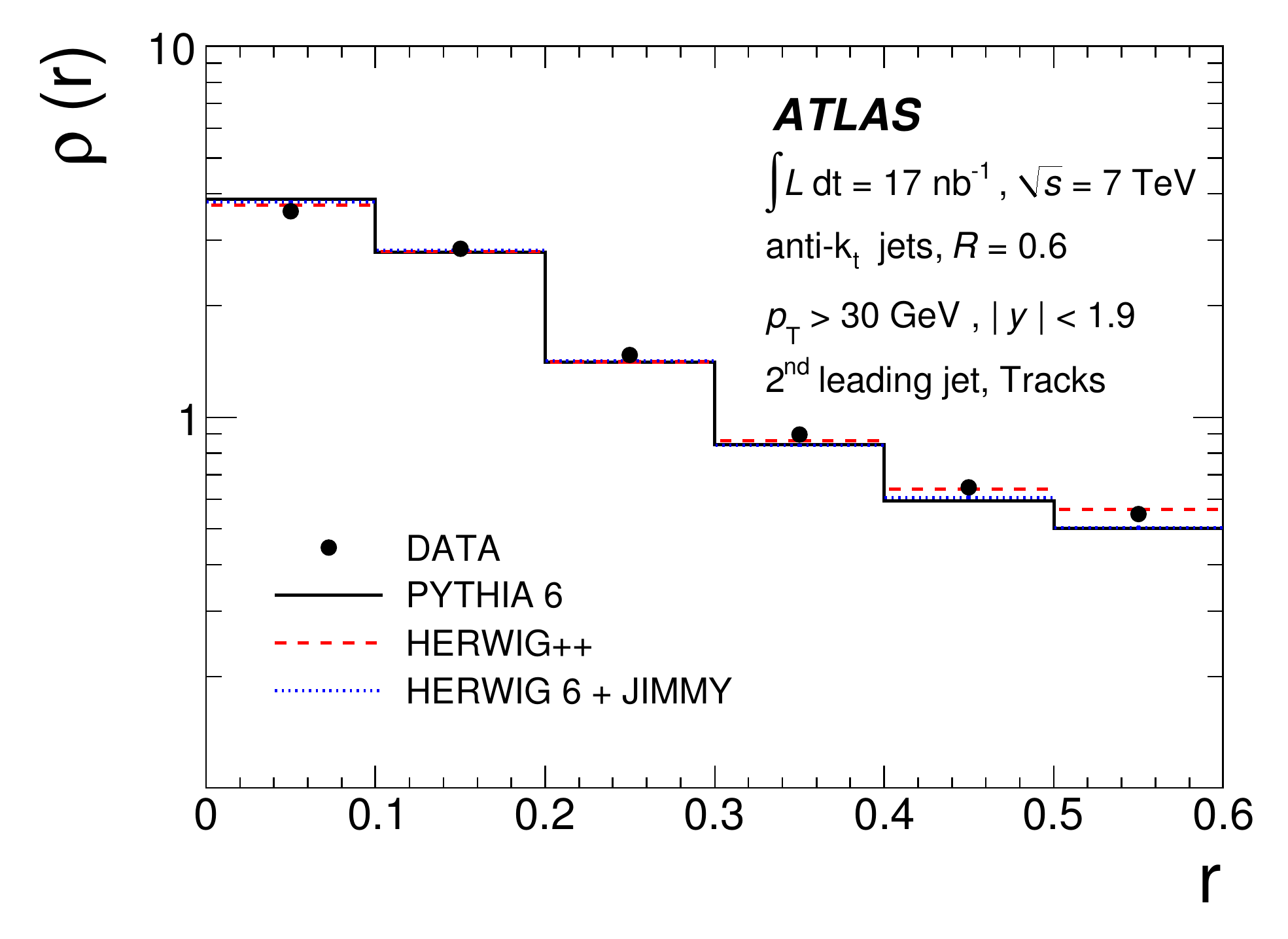}
\caption{
The uncorrected jet shape measured using energy clusters (first two plots) and tracks (third and fourth plots) 
for \AKT jets with $R=0.6$, 
compared to simulation, as a function of the radial distance to the jet axis, $r$. 
The first and third figures show the jet shapes for all jets with $\pt > 60\gev$, and
the second and fourth show the shape for the second highest $\pt$ jet in dijet events.
}
\label{fig:shapes}
\end{center}
\end{figure}

This procedure is justified by the good modelling of the trigger efficiencies (Fig.~\ref{fig:trigger})
and the fact that the \pT and $y$ distributions of the jets are reasonably well described by the 
simulation~\cite{ATLAS-CONF-2010-053}. It is also important that the energy flow around the jet core 
is well understood, both as a validation of the QCD description contained in the event generators and 
as a cross check of the calibration studies previously discussed, most of which are sensitive to the 
distribution of energy amongst particles, and within different angular regions, within the jet.
The energy and momentum flow within jets can be expressed in terms of the differential jet shape, 
defined as the fraction, $\rho(r) = \pt^r/\pt^R$, where $\pt^R$ is the 
transverse momentum within a radius $R$ of the jet centre, and $\pt^r$ is the transverse momentum contained 
within a ring of thickness $\Delta r = 0.1$ at a radius 
$r = \sqrt{(\Delta y)^2 + (\Delta\phi)^2}$ 
around the jet centre, divided by $\Delta r$. The jet shapes evaluated (without any correction for detector effects)
using energy clusters and tracks are shown separately in Fig.~\ref{fig:shapes} for \AKT jets with 
$R=0.6$. The jets simulated by \pythiasix are slightly narrower than the jets in the data, while the \herwigsix + \jimmy and \herwigpp simulations 
provide a somewhat better description. Overall the distribution of energy within the jets is 
reasonably well simulated. A similar level of
agreement has been demonstrated for $R=0.4$ jets. This gives further 
confidence in the calibrations and corrections applied.

\begin{figure*}
\begin{center}
\includegraphics[width=0.48\textwidth]{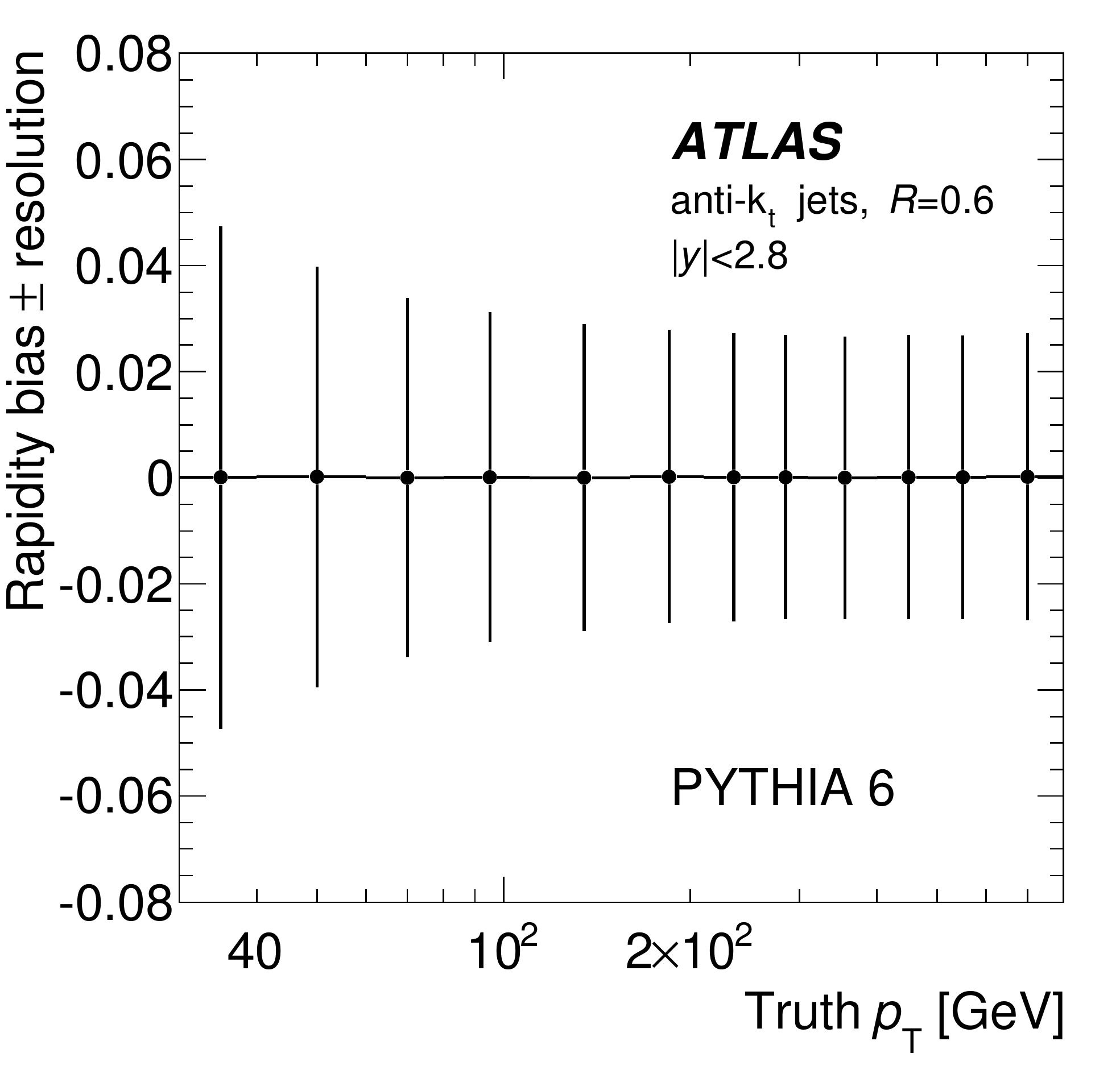}
\includegraphics[width=0.48\textwidth]{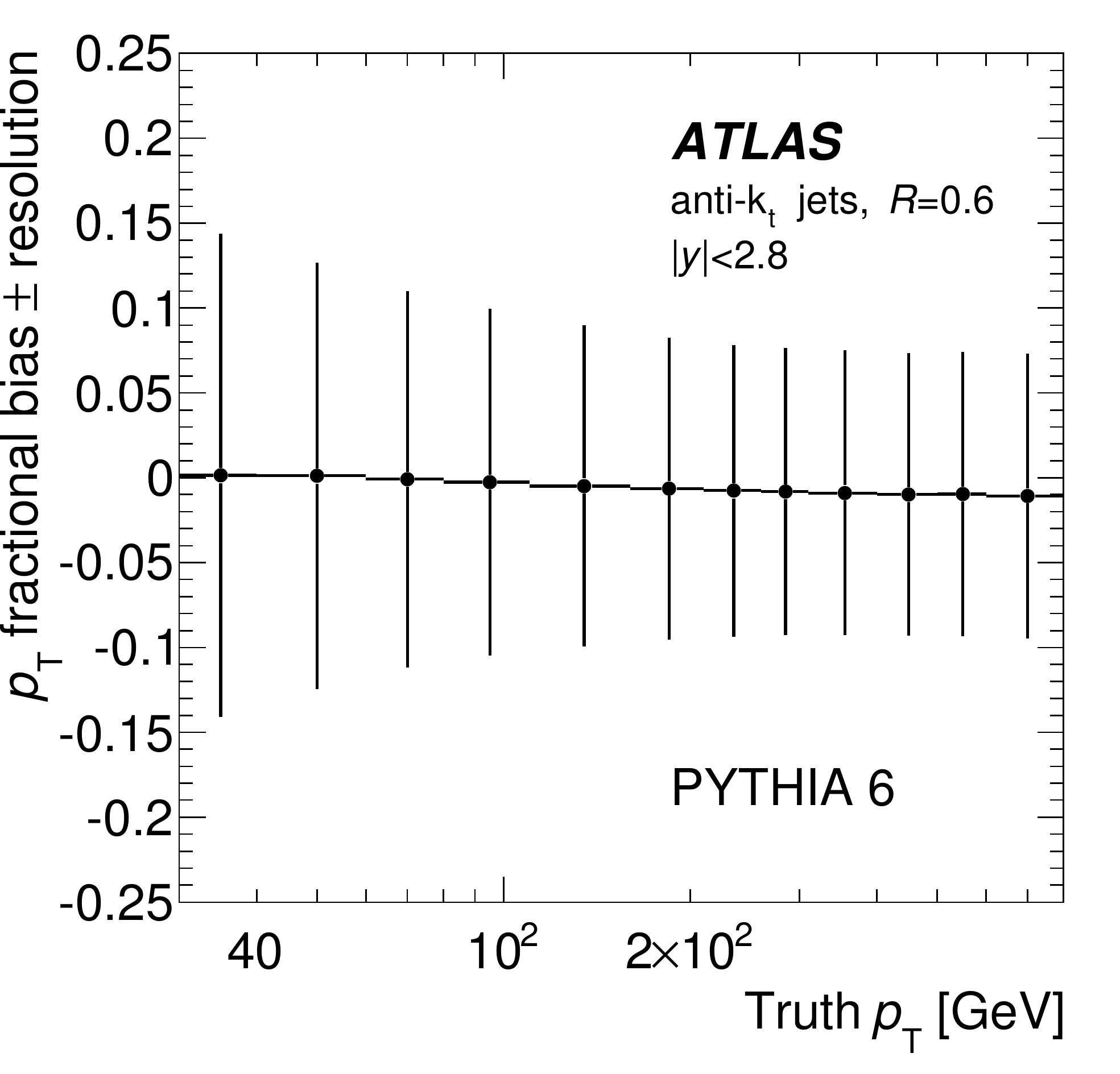}
\includegraphics[width=0.48\textwidth]{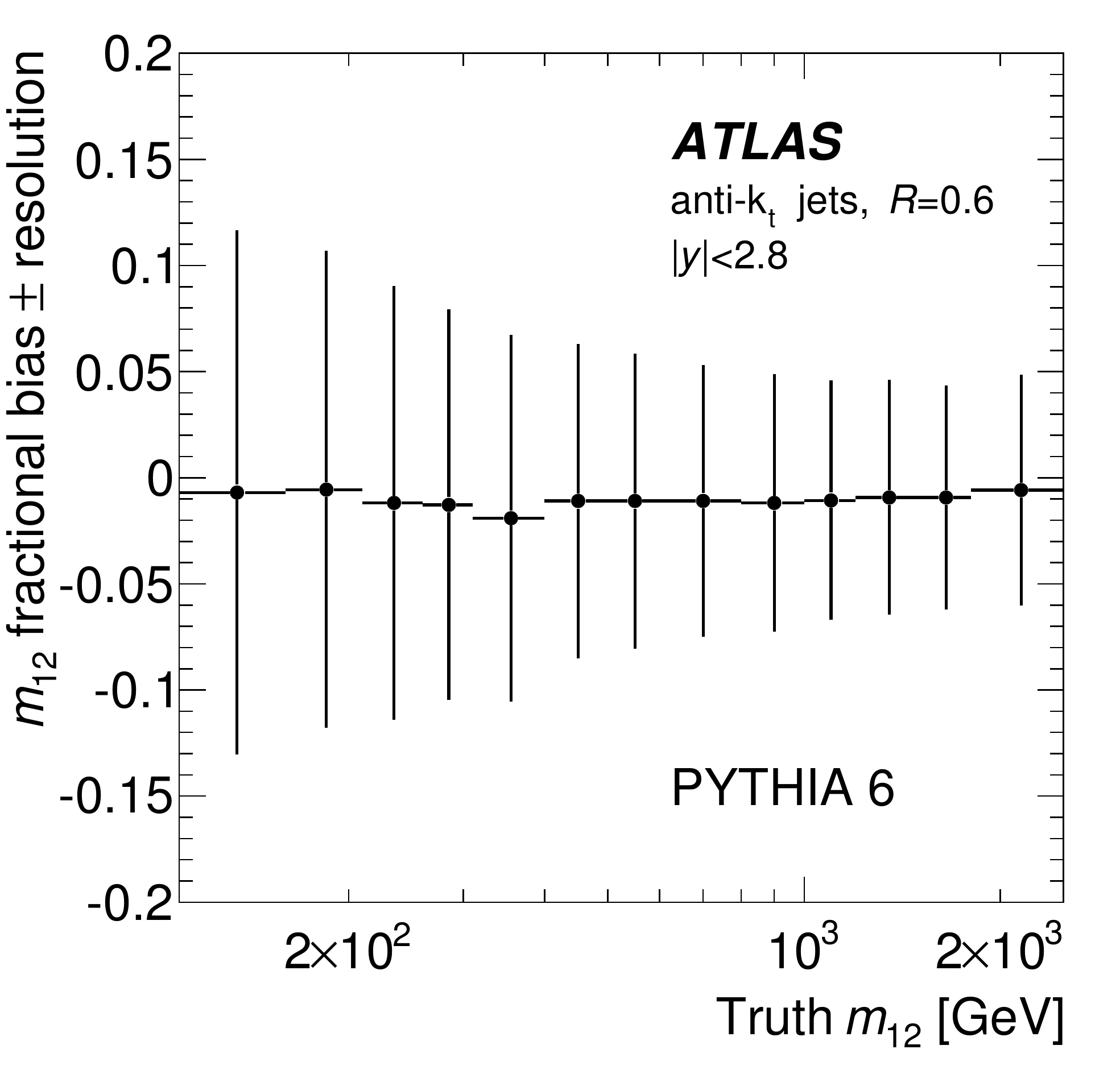}
\includegraphics[width=0.48\textwidth]{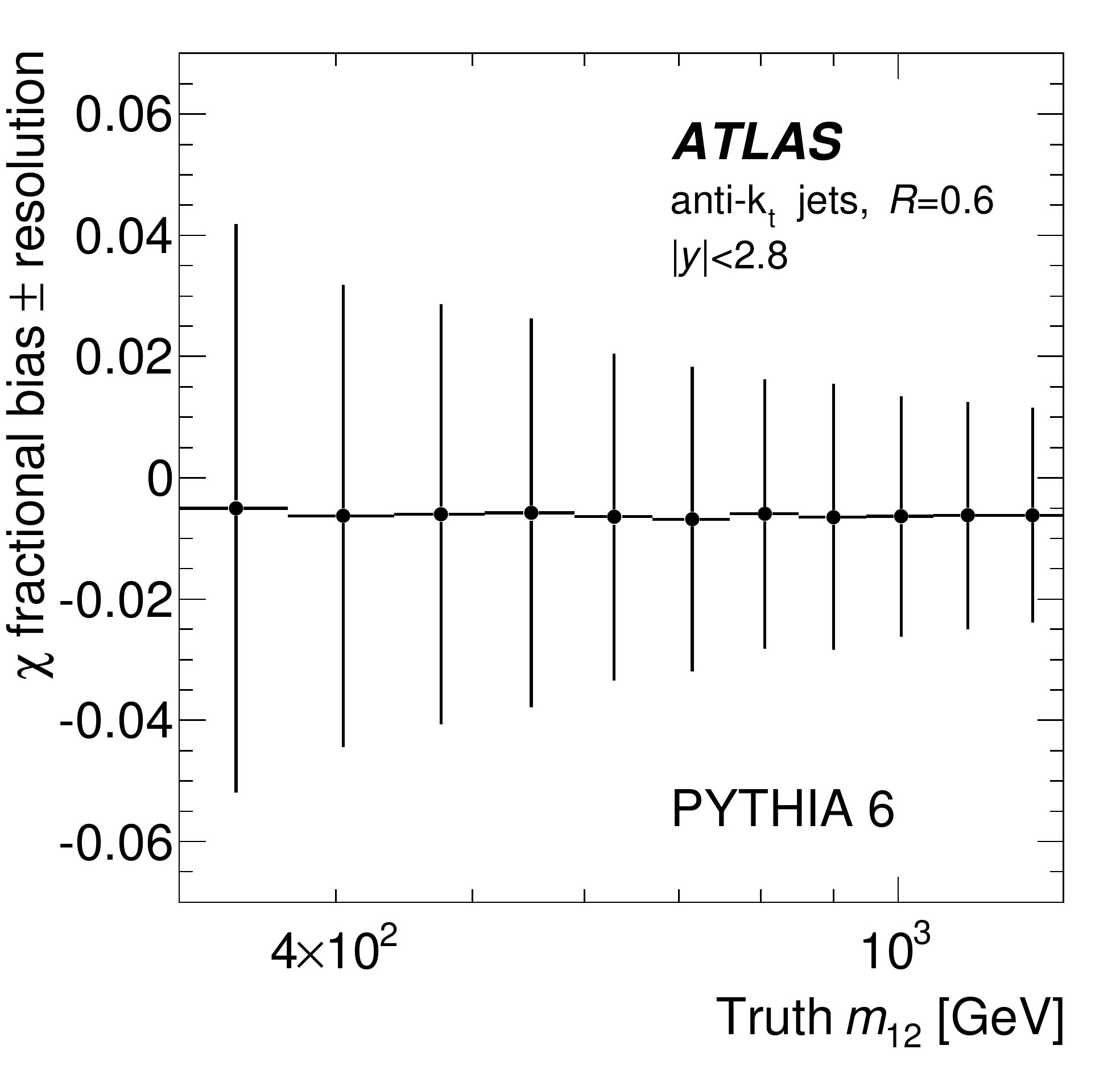}
\caption{The upper two plots show the absolute (fractional) resolution and bias in jet $y$ ($\pt$) as a function of true $\pt$.  The bottom two plots show the fractional resolution and bias in dijet mass $\twomass{1}{2}$ and angular variable $\chi$ as a function of truth $\twomass{1}{2}$ computed from the two leading truth jets.  These are shown for all jets identified using the $\AKT$ algorithm with $R=0.6$ in events passing the final kinematic selection, as predicted by \pythiasix.  The error bar indicates the resolution and the central value indicates the bias.
}
\label{fig:resolutions}
\end{center}
\end{figure*}

The resolutions in $y$, $\pt$, dijet mass $\twomass{1}{2}$, and dijet $\chi$ for \AKT jets with 
$R=0.6$ within $|y|<2.8$, as obtained using \pythiasix, are shown in Fig.~\ref{fig:resolutions}.
The present JES calibration procedure applies an average correction to restore the jet response and 
does not attempt to optimise the jet energy resolution, which can be improved with more sophisticated 
calibration techniques.
From dijet balance and $E/p$ studies of single hadrons, the \pt resolution has been verified 
to within a fractional uncertainty of $\approx$ 14\% \cite{ATLAS-CONF-2010-054}, though 
at a lower $\pt$ than most of the jets considered here.
The effect of varying the nominal \pt resolution by up to 15\% of its nominal value is included in 
the systematic uncertainty on the unfolding correction factors. The uncertainties due to the jet energy 
scale are also propagated to the 
final cross section through this unfolding procedure, by applying variations to simulated samples.
A fit is used to reduce statistical fluctuations in the systematic uncertainties from the jet energy scale for the dijet mass spectrum.

\begin{figure*}
\begin{center}
\includegraphics[width=0.48\textwidth]{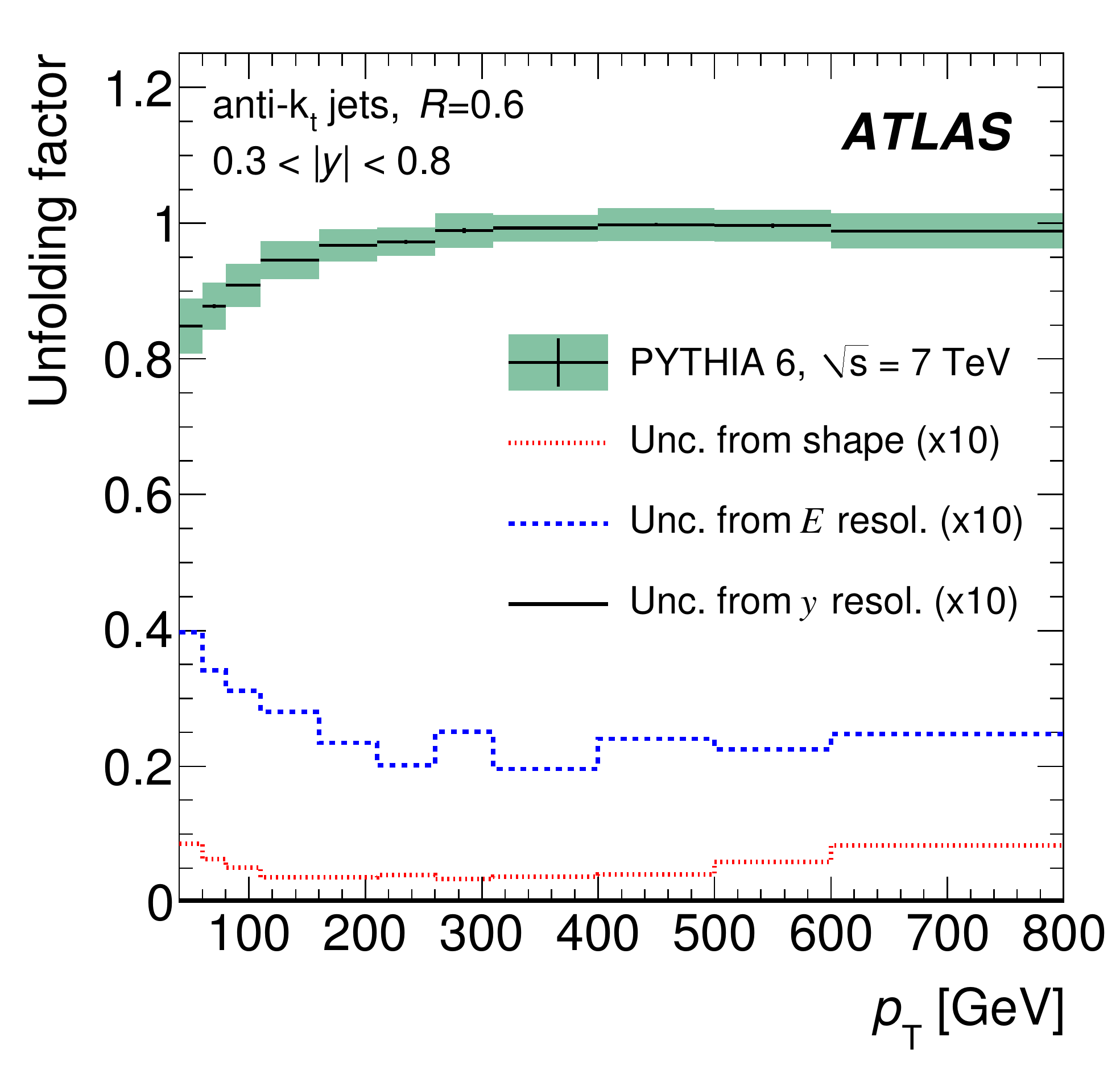}
\includegraphics[width=0.48\textwidth]{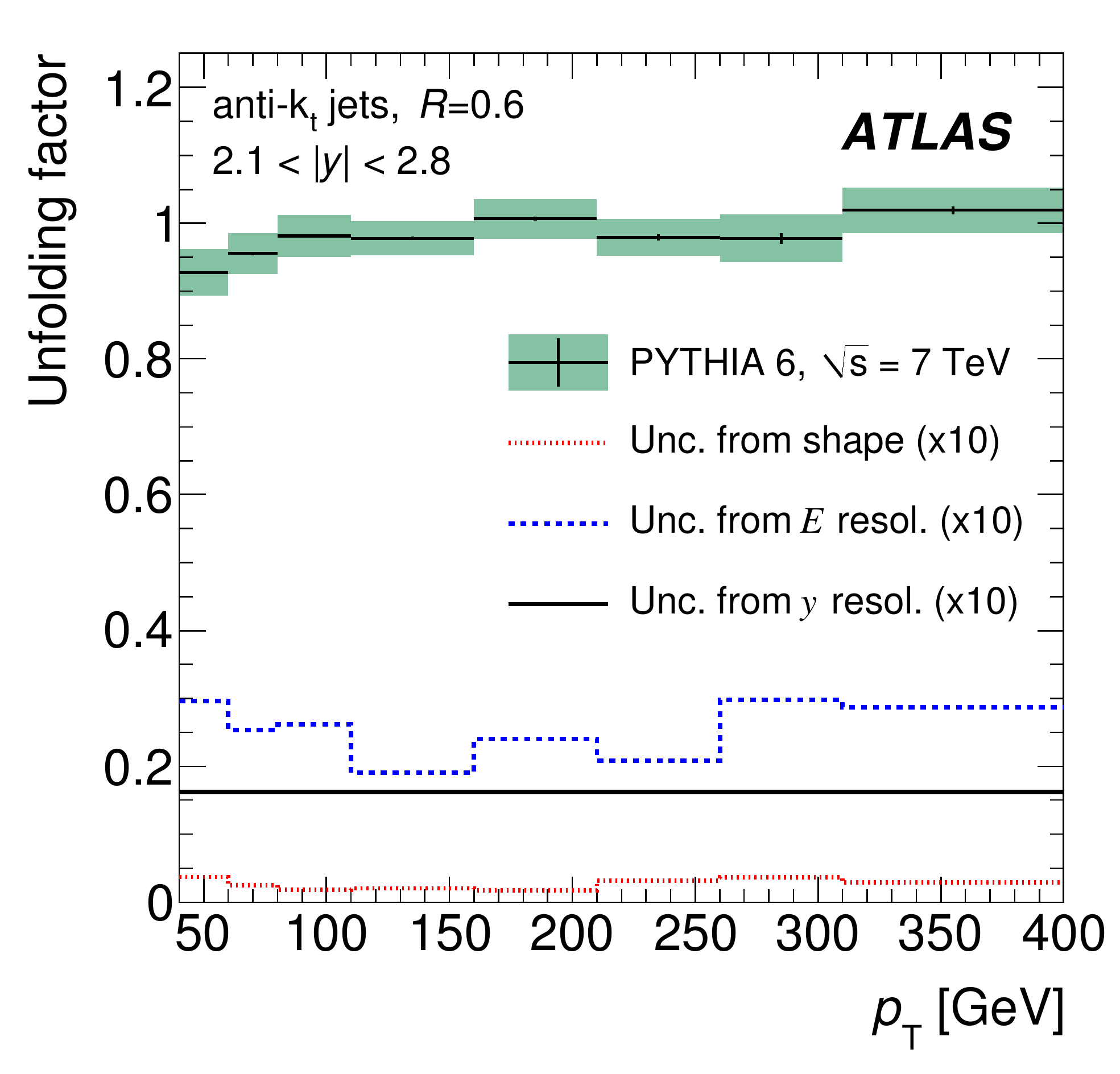}
\caption{Correction factors for the inclusive jet $\pt$ spectrum in the rapidity bins 0.3~$ < |y| < $~0.8 (left) and 2.1~$ < |y| < $~2.8 (right), along with systematic uncertainties due to uncertainties in the jet $\pt$ spectrum shape, jet energy resolution, and jet angular resolution. The band on the correction factor indicates the total systematic uncertainty on the correction.}
\label{fig:unfold}
\end{center}
\end{figure*}

The overall correction factor for the $\pt$ spectrum is below 20\% throughout the kinematic region, and below 10\% for central jets with $\pt > 60\gev$.  As an example, the correction factors for the $\pt$ spectrum with $R=0.6$ are shown along with their systematic uncertainties in Fig.~\ref{fig:unfold} for two rapidity regions.  For the dijet mass spectrum, the correction factors are generally within 15\% while for $\chi$ they are less than 5\%.

The integrated luminosities are calculated during runs\footnote{An ATLAS run is a period of continuous data-taking 
during an LHC proton fill.} by measuring interaction rates using several ATLAS 
devices at small angles to the beam direction, with the absolute calibration obtained from
van der Meer scans. The uncertainty in the luminosity is estimated to
be 11\%~\cite{ATLAS-CONF-2010-060}.

The final systematic uncertainty in the cross section measurements is dominated by the jet energy scale uncertainty.

\section{Results and Discussion}

\begin{figure*}
\begin{center}
\includegraphics[width=0.99\textwidth]{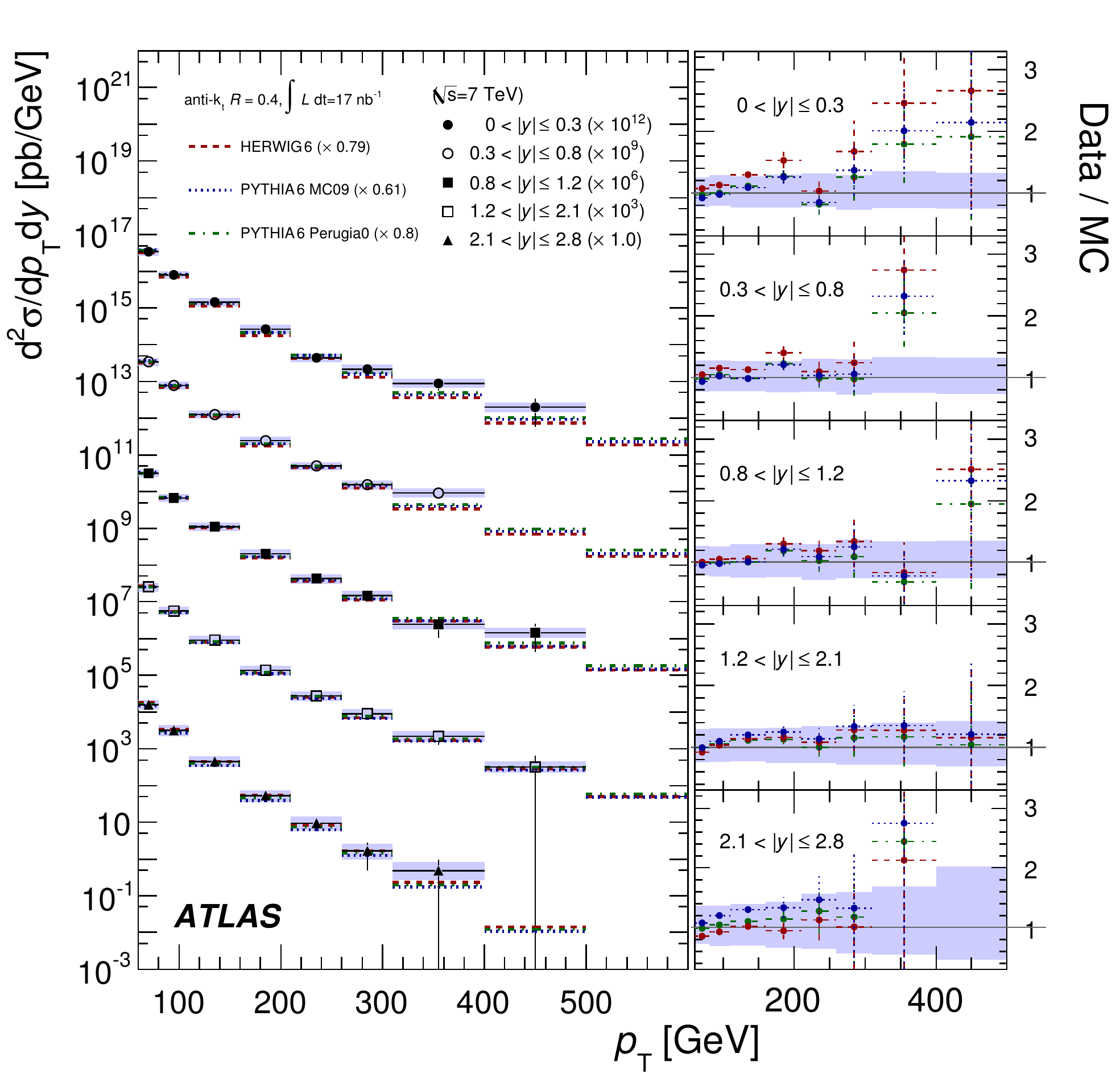}
\caption{Inclusive jet double-differential cross section as a function of $\pt$, for different bins of 
rapidity $y$.
The results are shown for jets identified using the \protect\AKT algorithm with $R=0.4$.
The data are compared to leading-logarithmic parton-shower MC simulations, normalised to the measured
cross section by the factors shown in the legend, fixed to give the best 
normalisation to the inclusive jet measurements.
The bands indicate the total systematic uncertainty on the data.
The error bars indicate the statistical uncertainty, which is calculated as $1/\sqrt{N}$, where $N$ is the number of entries in a given bin.
The insets along the right-hand side show the ratio of the data to the various MC simulations.}
\label{fig:incjetmc04}
\end{center}
\end{figure*}

\begin{figure*}
\begin{center}
\includegraphics[width=0.99\textwidth]{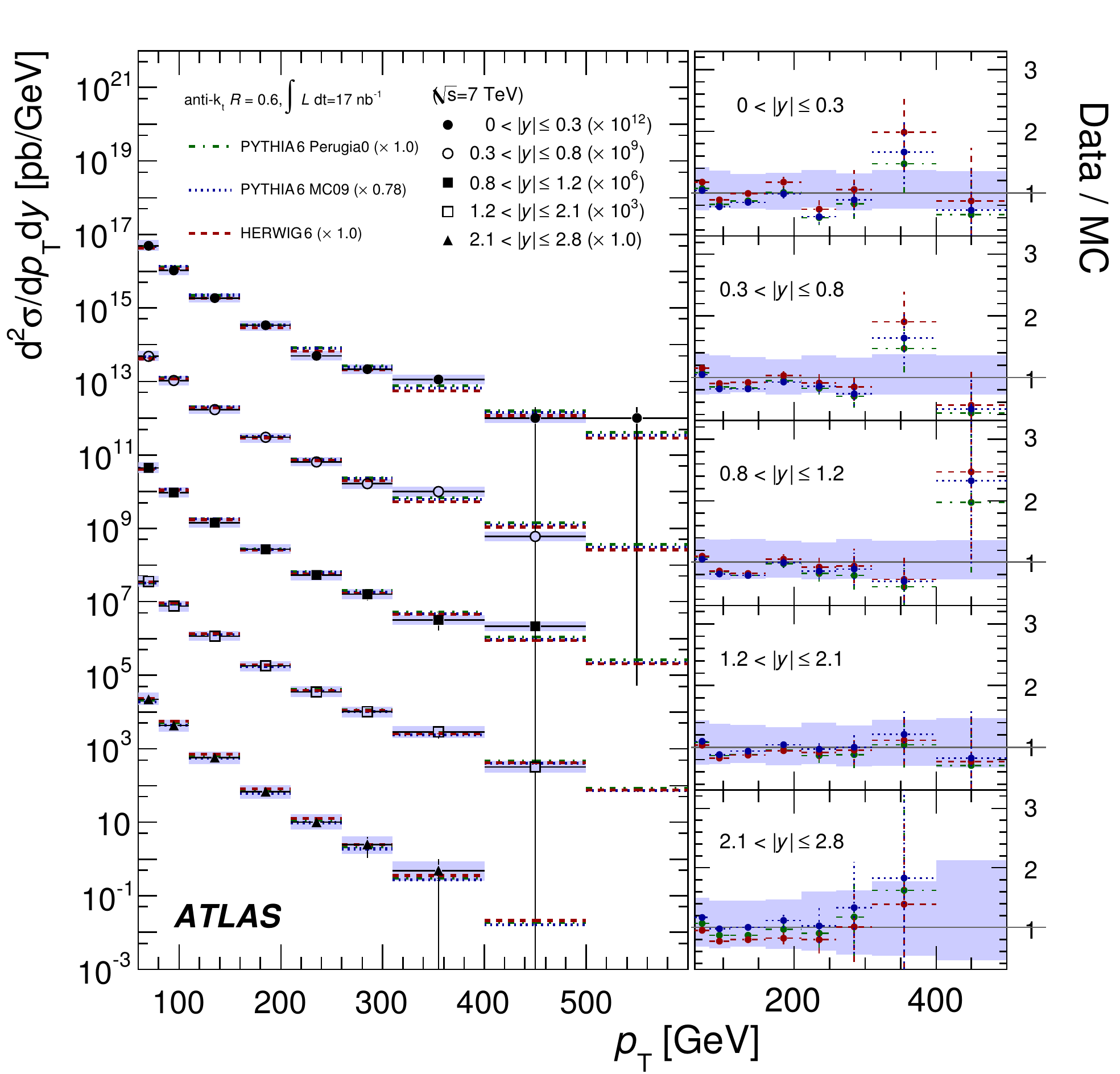}
\caption{Inclusive jet double-differential cross section as a function of $\pt$, for different bins of 
rapidity $y$.
The results are shown for jets identified using the \AKT algorithm with $R=0.6$.
The data are compared to leading-logarithmic parton-shower MC simulations, normalised to the measured 
cross section by the factors shown in the legend, fixed to give the best 
normalisation to the inclusive jet measurements.
The bands indicate the total systematic uncertainty on the data.
The error bars indicate the statistical uncertainty, which is calculated as $1/\sqrt{N}$, where $N$ is the number of entries in a given bin.
The insets along the right-hand side show the ratio of the data to the various MC simulations.
\label{fig:incjetmc06}
}
\end{center}
\end{figure*}

\begin{figure*}
\begin{center}
\includegraphics[width=0.53\textwidth]{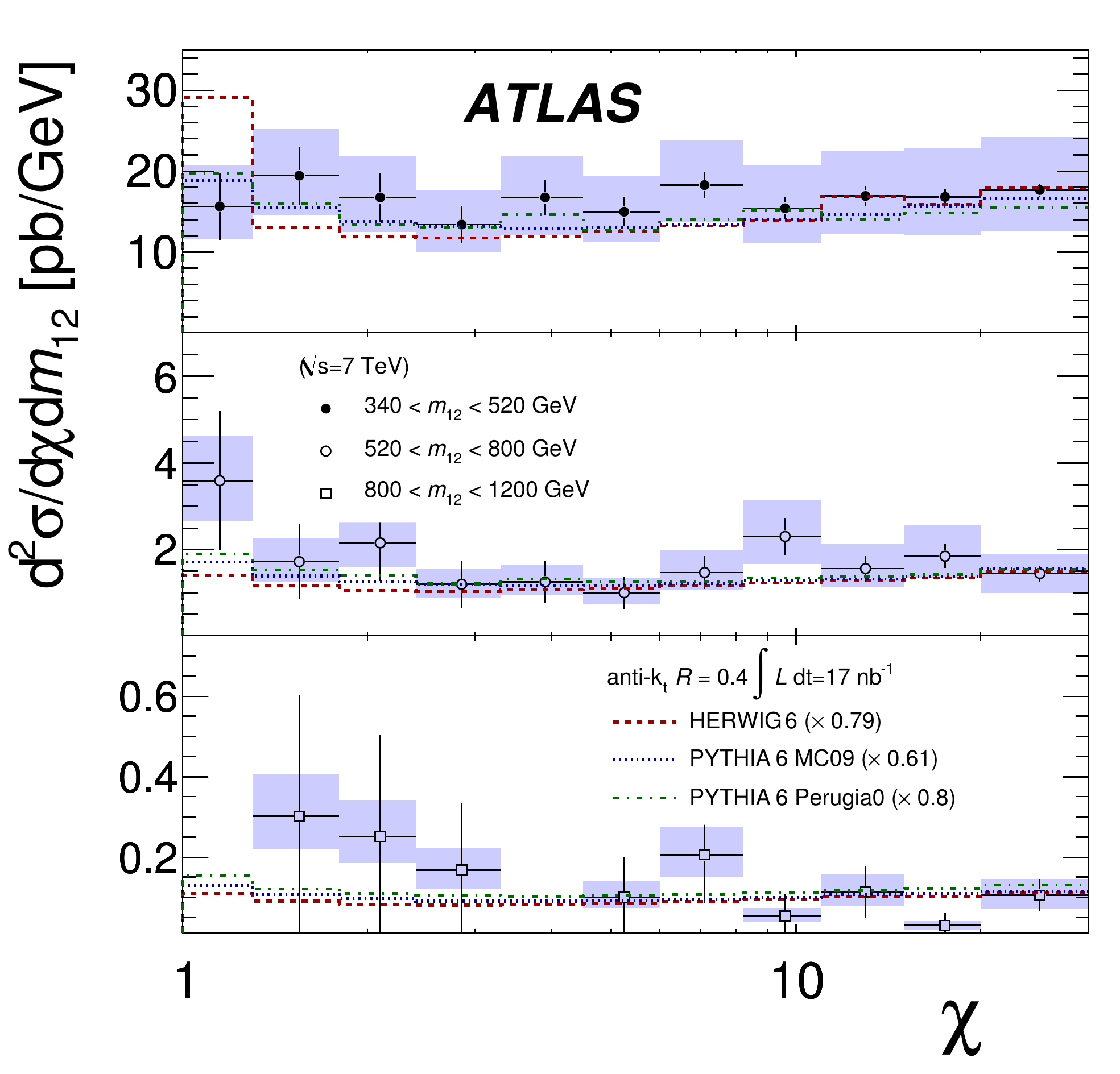}
\caption{Dijet double-differential cross section as a function of angular variable $\chi$ in different regions of dijet mass $\twomass{1}{2}$, 
for jets identified using the \AKT algorithm with $R=0.4$.
The data are compared to leading-logarithmic parton-shower MC simulations, normalised to the measured cross section by the factors shown in the legend, fixed to give the best 
normalisation to the inclusive jet measurements.
The bands indicate the total systematic uncertainty on the data.
The error bars indicate the statistical uncertainty, which is calculated as $1/\sqrt{N}$, where $N$ is the number of entries in a given bin.
\label{fig:dijetchimc04}
}
\end{center}
\end{figure*}

\begin{figure*}
\begin{center}
\includegraphics[width=0.53\textwidth]{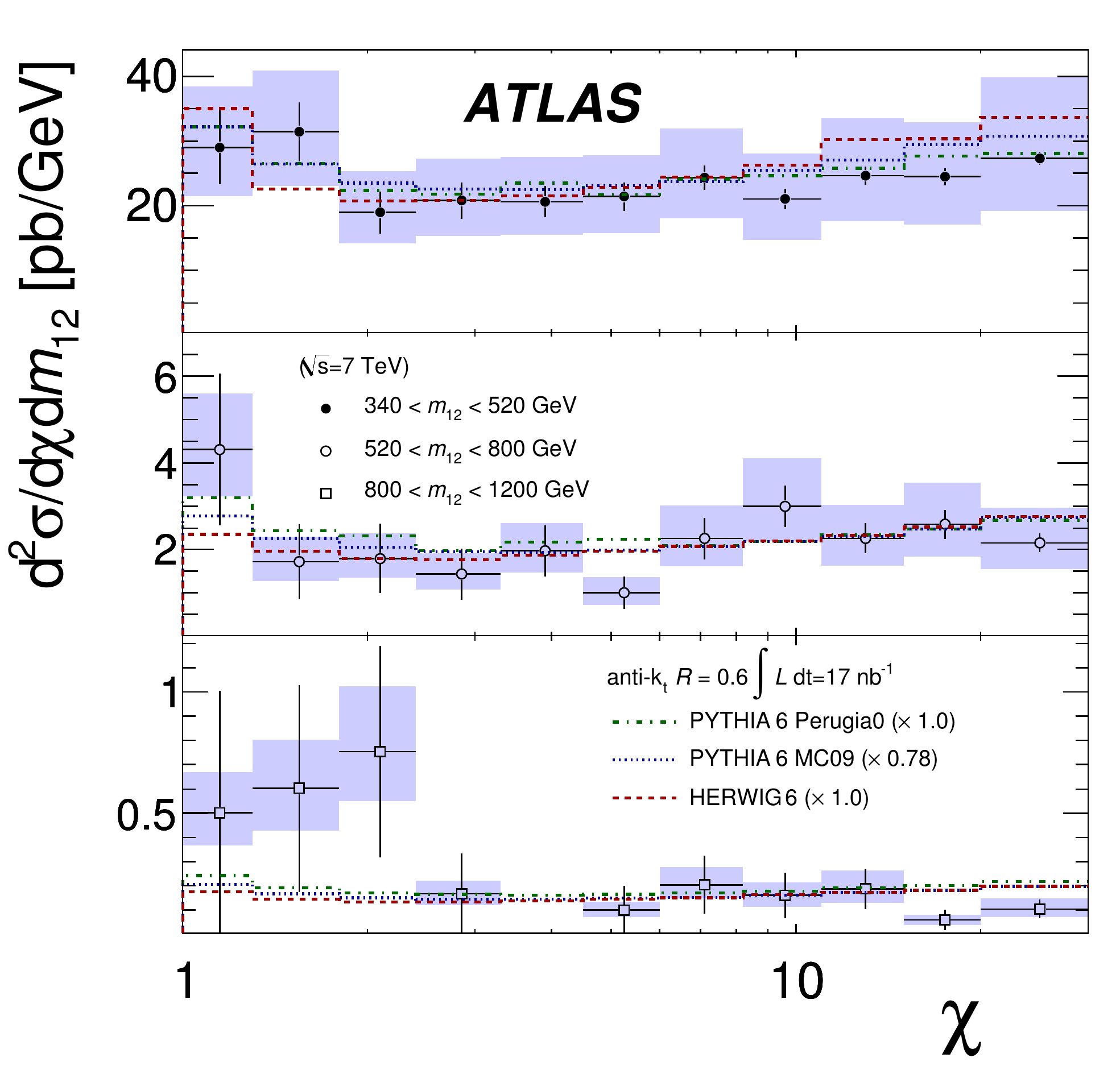}
\caption{Dijet double-differential cross section as a function of angular variable $\chi$ in different regions of dijet mass $\twomass{1}{2}$, 
for jets identified using the \AKT algorithm with $R=0.6$.
The data are compared to leading-logarithmic parton-shower MC simulations, 
normalised to the measured cross section by the factors shown in the legend, fixed to give the best 
normalisation to the inclusive jet measurements.
The bands indicate the total systematic uncertainty on the data.
The error bars indicate the statistical uncertainty, which is calculated as $1/\sqrt{N}$, where $N$ is the number of entries in a given bin.
}
\label{fig:dijetchimc06}
\end{center}
\end{figure*}

The cross sections from the parton-shower 
MC generators considered here are not reliable, since 
these calculations are performed at leading-order. 
However, many important kinematic terms are included in these calculations, 
and, unlike the NLO pQCD calculations, the predictions are made at the particle level.
A comparison of the shapes of the distributions is therefore valuable. 
The expectations for the
corrected $\pt$ and $\chi$ distributions from two different \pythiasix parameter tunes, as well as 
for \herwigsix + \jimmy
programs are compared to the data in Figs.~\ref{fig:incjetmc04}-\ref{fig:dijetchimc06}. 
The normalisation of the simulation is to the inclusive jet 
cross section separately for each $R$ value, and requires the 
factors shown in the legend\footnote{If the $R=0.4$ and $R=0.6$ measurements are fitted simultaneously, the factors
are 0.91, 0.92 and 0.69 for \herwig + \jimmy, \pythiasix Perugia0 tune and \pythiasix MC09 respectively.}.
In general the simulations agree with the shapes of the data distributions.

\begin{figure*}
\begin{center}
\includegraphics[width=0.95\textwidth]{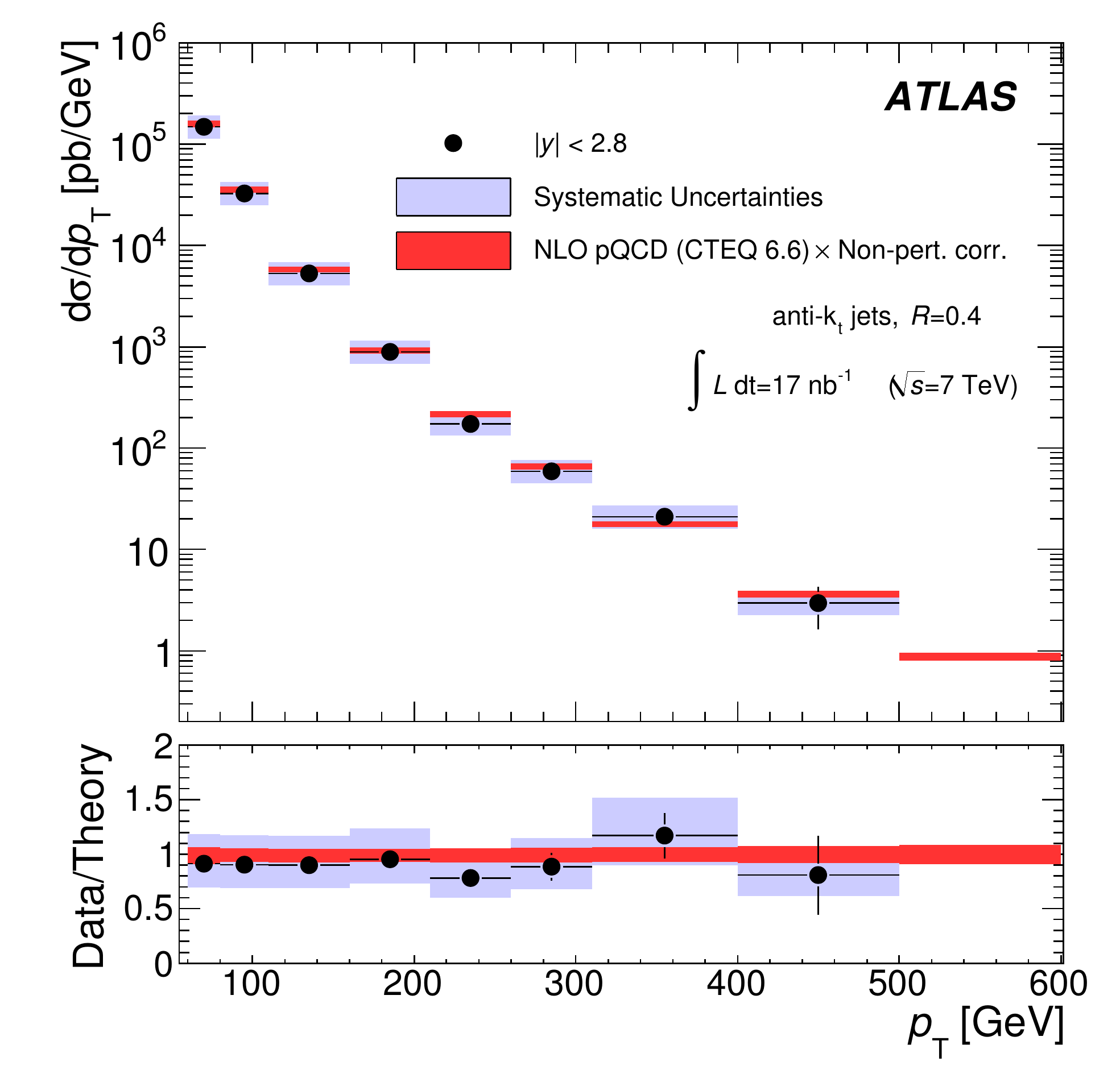}
\caption{Inclusive jet differential cross section as a function of jet $\pt$ integrated over the full
region $|y|<2.8$ for jets identified using the \AKT algorithm with $R=0.4$.
The data are compared to NLO pQCD calculations to which soft QCD corrections have been applied. 
The error bars indicate the statistical 
uncertainty on the measurement, and the grey shaded bands indicate the quadratic sum of the systematic uncertainties, 
dominated by 
the jet energy scale uncertainty.  The statistical uncertainty is calculated as $1/\sqrt{N}$, where $N$ is the number of entries in a given bin.  There is an additional overall uncertainty of 11\% due to the luminosity 
measurement that is not shown. The theory uncertainty shown in red is the quadratic sum of uncertainties from 
the choice of renormalisation and factorisation scales, parton distribution functions, $\alpha_s(M_Z)$, 
and the modelling of soft QCD effects, as described in the text.}
\label{fig:incjetptfullsummary04}
\end{center}
\end{figure*}

\begin{figure*}
\begin{center}
\includegraphics[width=0.95\textwidth]{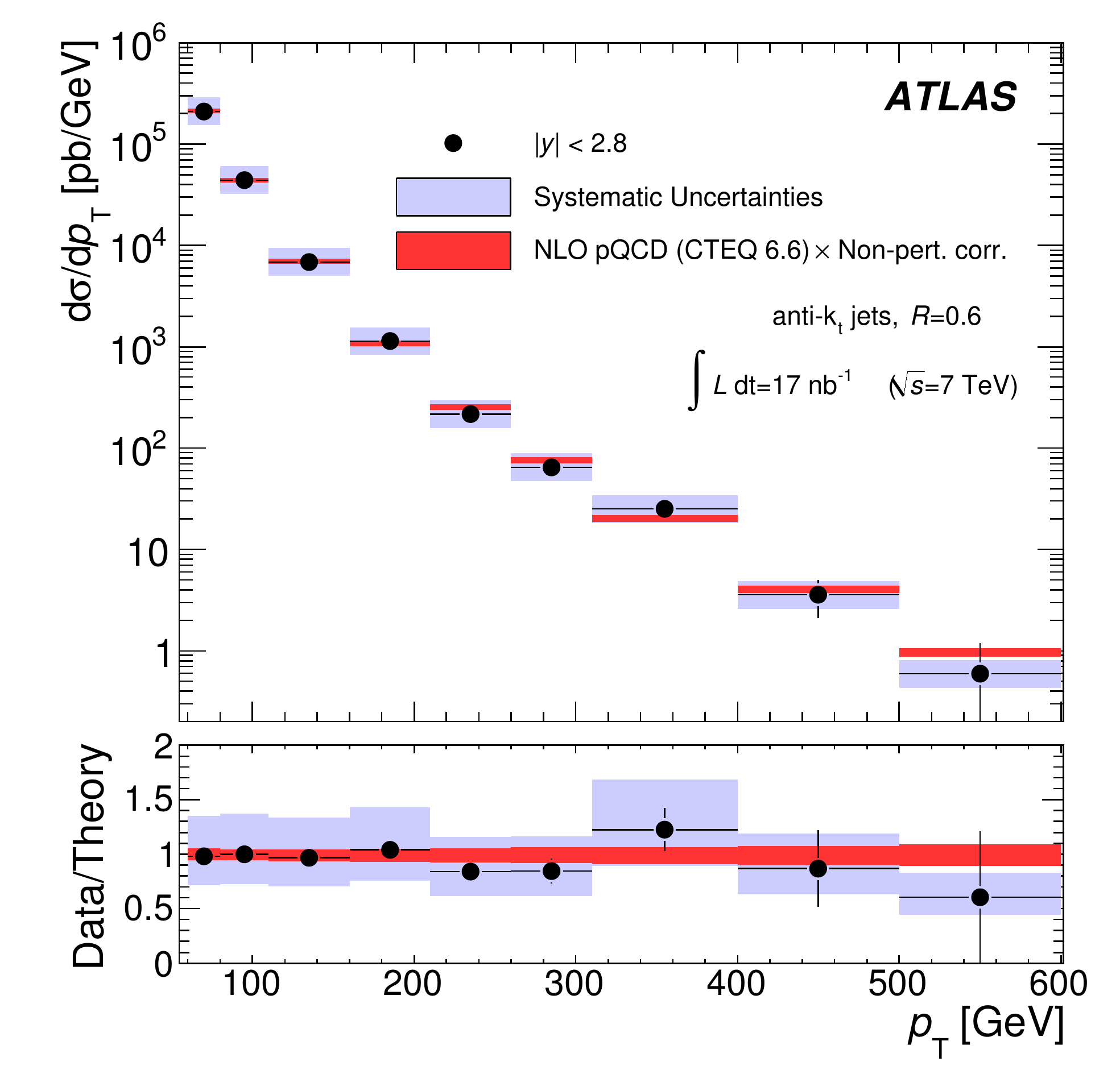}
\caption{Inclusive jet differential cross section as a function of jet $\pt$ integrated over the full
region $|y|<2.8$ for jets identified using the \AKT algorithm with $R=0.6$.
The data are compared to NLO pQCD calculations to which soft QCD corrections have been applied. 
The uncertainties on the data and theory are shown as described in Fig.~\protect\ref{fig:incjetptfullsummary04}.
}
\label{fig:incjetptfullsummary06}
\end{center}
\end{figure*}

The differential inclusive jet cross section in 7~TeV proton-proton collisions is shown in 
Fig.~\ref{fig:incjetptfullsummary04} and Fig.~\ref{fig:incjetptfullsummary06}, 
as a function of jet $\pt$,
for \AKT jets with $R=0.4$ and $R=0.6$ respectively.
The cross section extends from $\pt = 60\gev$ up to around $\pt = 600$~\gev, and falls by more than four 
orders of magnitude over this range. The data are compared to NLO pQCD calculations 
corrected for non-perturbative effects. For both $R=0.4$ and $R=0.6$, data and theory are consistent.

\begin{figure*}
\begin{center}
\includegraphics[width=0.9\textwidth]{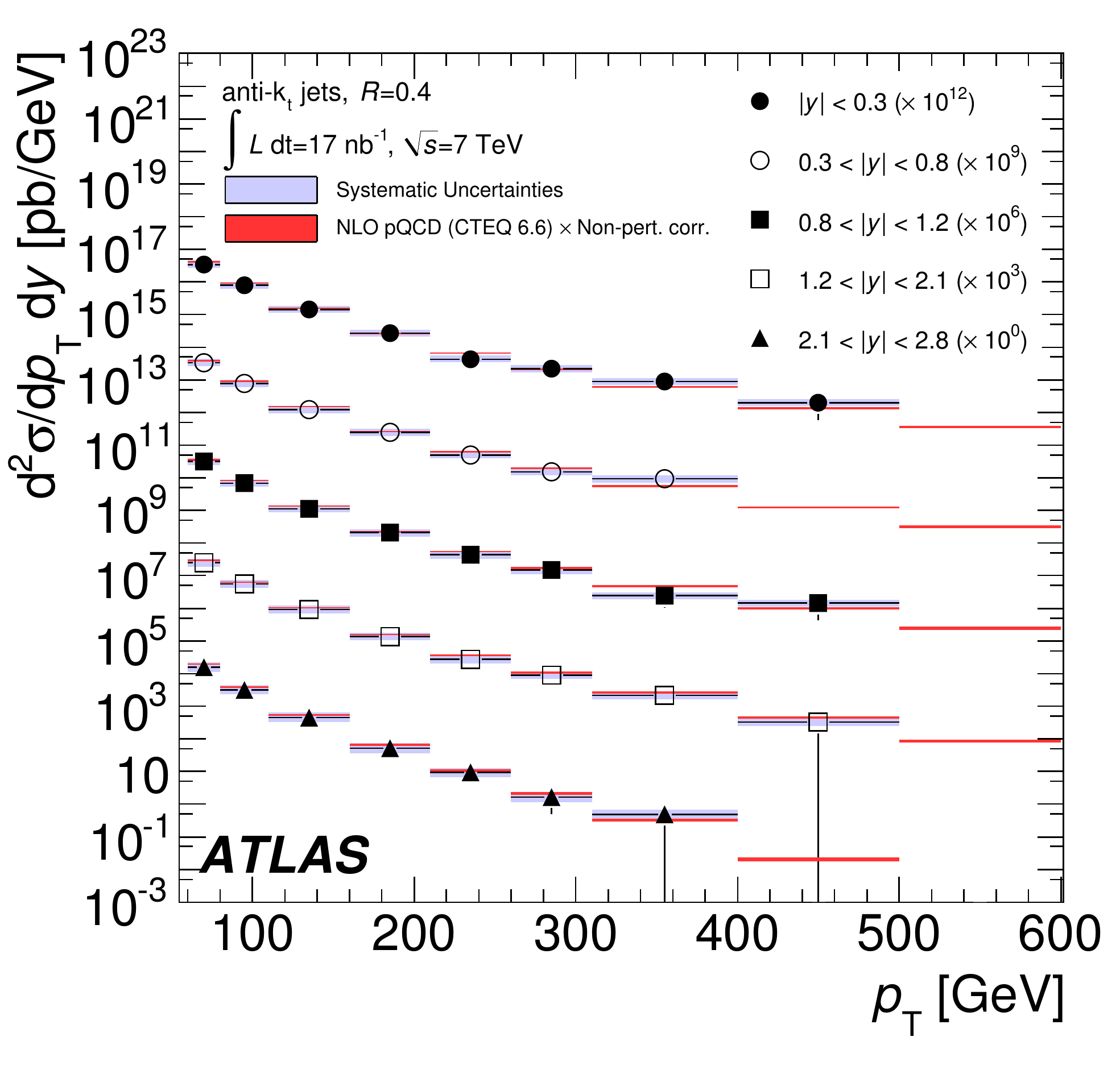}
\caption{Inclusive jet double-differential cross section as a function of jet $\pt$ in different regions of $|y|$ 
for jets identified using the \AKT algorithm with $R=0.4$.
The data are compared to NLO pQCD calculations to which soft QCD corrections have been applied. 
The uncertainties on the data and theory are shown as described in Fig.~\protect\ref{fig:incjetptfullsummary04}.
}
\label{fig:incjetptsummary04}
\end{center}
\end{figure*}

\begin{figure*}
\begin{center}
\includegraphics[width=0.9\textwidth]{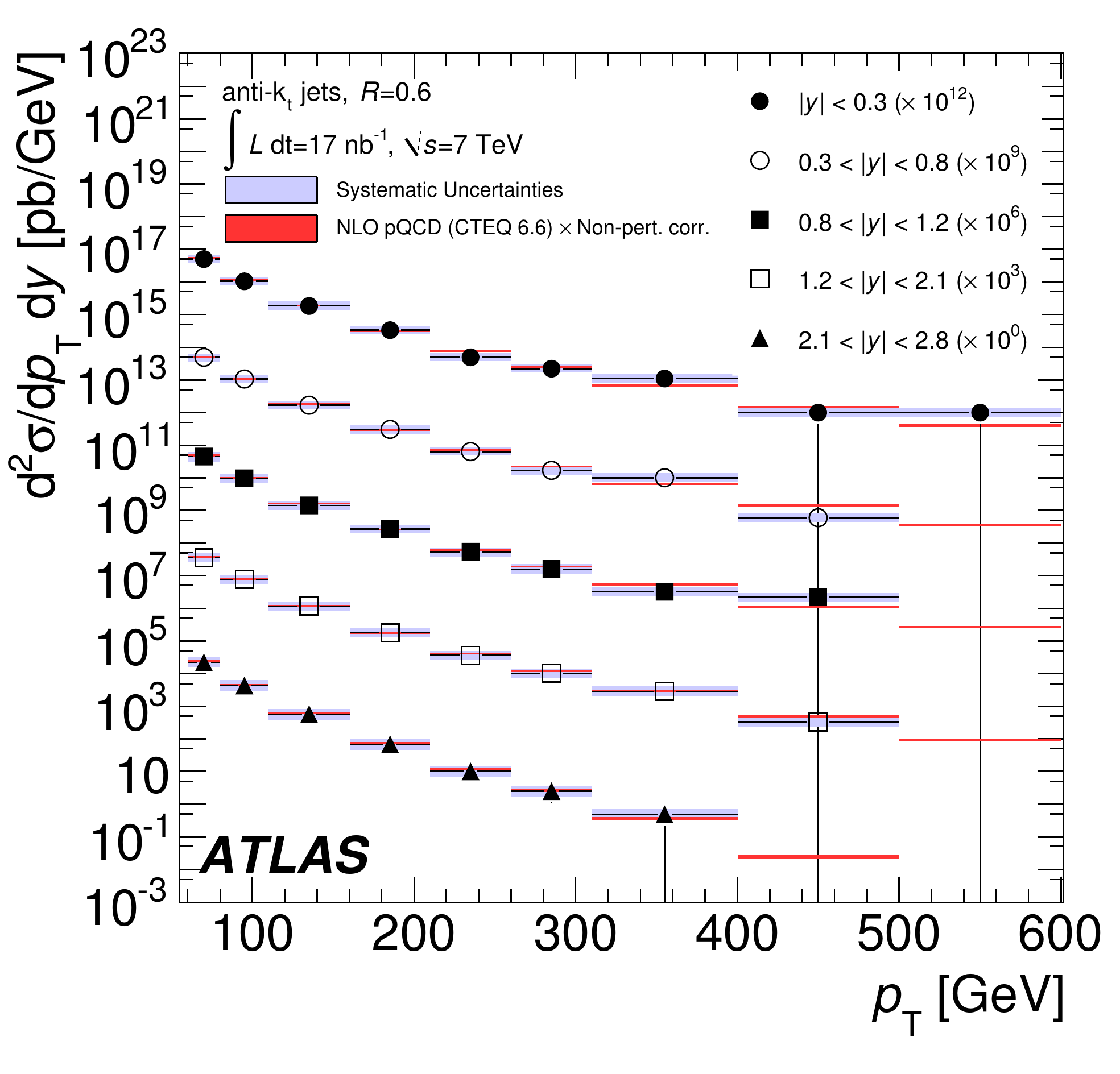}
\caption{Inclusive jet double-differential cross section as a function of jet $\pt$ in different regions of $|y|$ 
for jets identified using the \AKT algorithm with $R=0.6$.
The data are compared to NLO pQCD calculations to which soft QCD corrections have been applied. 
The uncertainties on the data and theory are shown as described in Fig.~\protect\ref{fig:incjetptfullsummary04}.}
\label{fig:incjetptsummary06}
\end{center}
\end{figure*}

\begin{figure*}
\begin{center}
\includegraphics[width=0.9\textwidth]{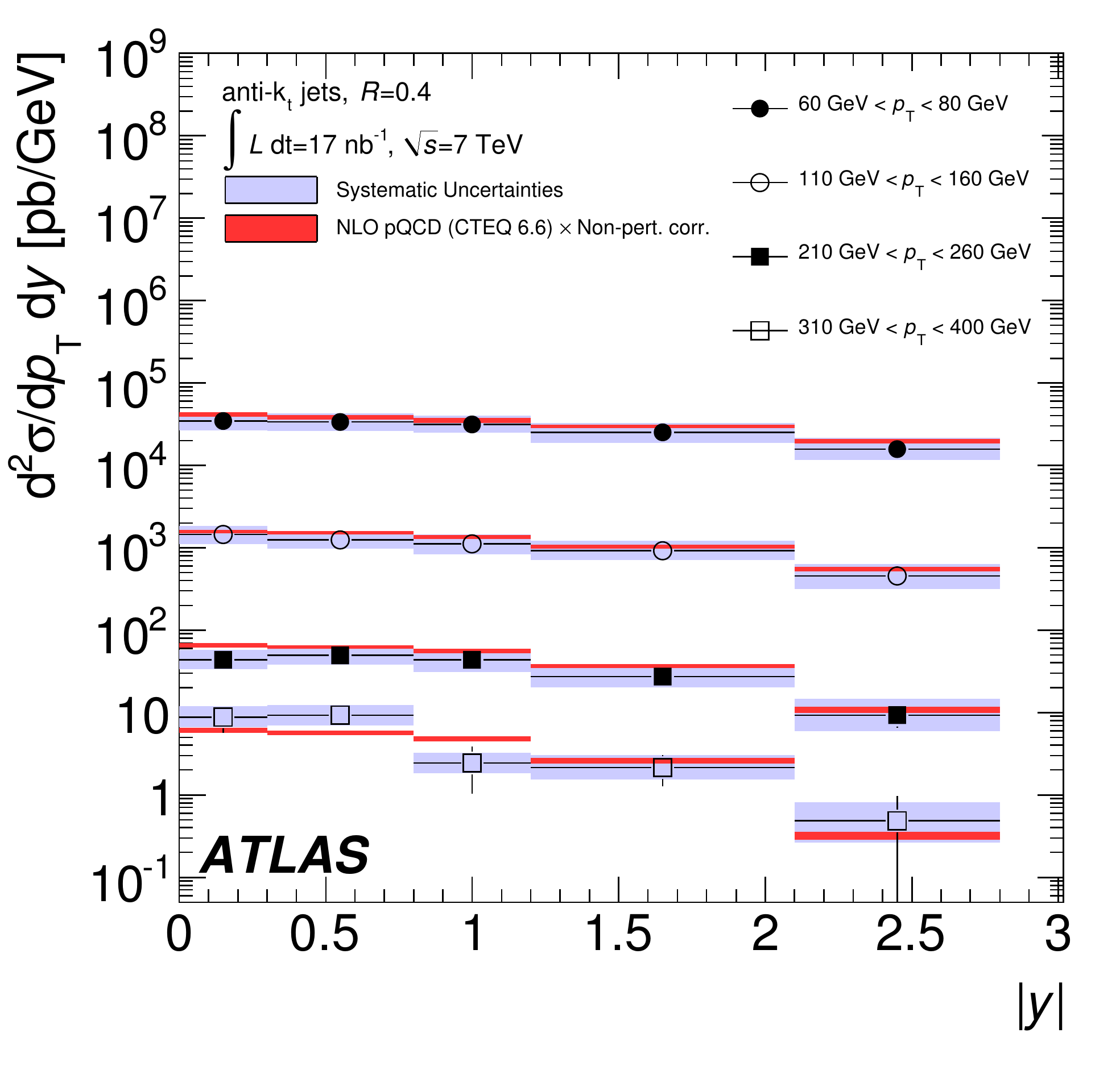}
\caption{Inclusive jet double-differential cross section as a function of jet $|y|$ in different regions of $\pt$ 
for jets identified using the \AKT algorithm with $R=0.4$.
The data are compared to NLO pQCD calculations to which soft QCD corrections have been applied.
The uncertainties on the data and theory are shown as described in Fig.~\protect\ref{fig:incjetptfullsummary04}.}
\label{fig:incjetysummary04}
\end{center}
\end{figure*}

\begin{figure*}
\begin{center}
\includegraphics[width=0.9\textwidth]{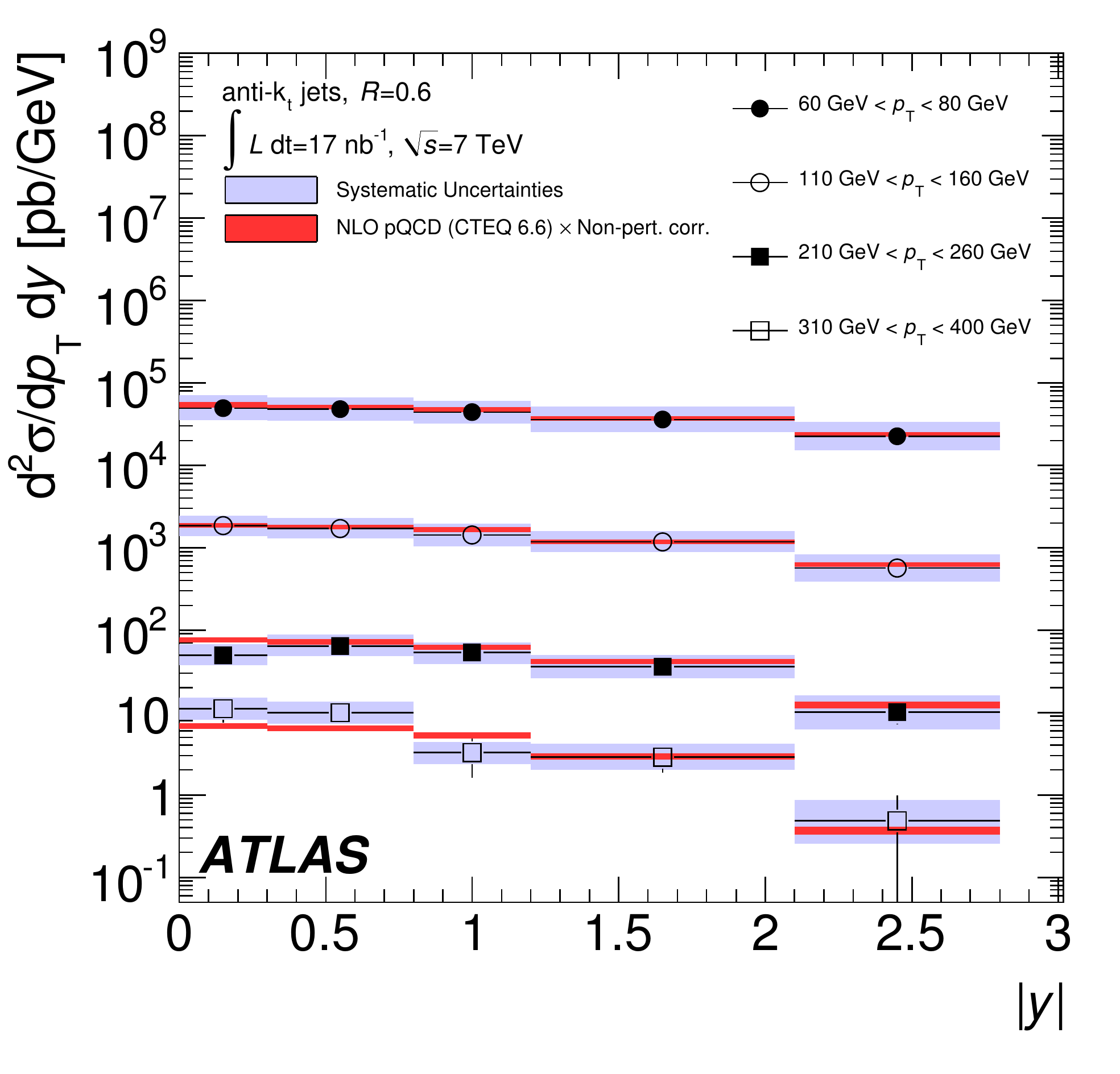}
\caption{Inclusive jet double-differential cross section as a function of jet $|y|$ in different regions of $\pt$ 
for jets identified using the \AKT algorithm with $R=0.6$.
The data are compared to NLO pQCD calculations to which soft QCD corrections have been applied.
The uncertainties on the data and theory are shown as described in Fig.~\protect\ref{fig:incjetptfullsummary04}.}
\label{fig:incjetysummary06}
\end{center}
\end{figure*}

\begin{figure*}
\begin{center}
\includegraphics[width=0.9\textwidth]{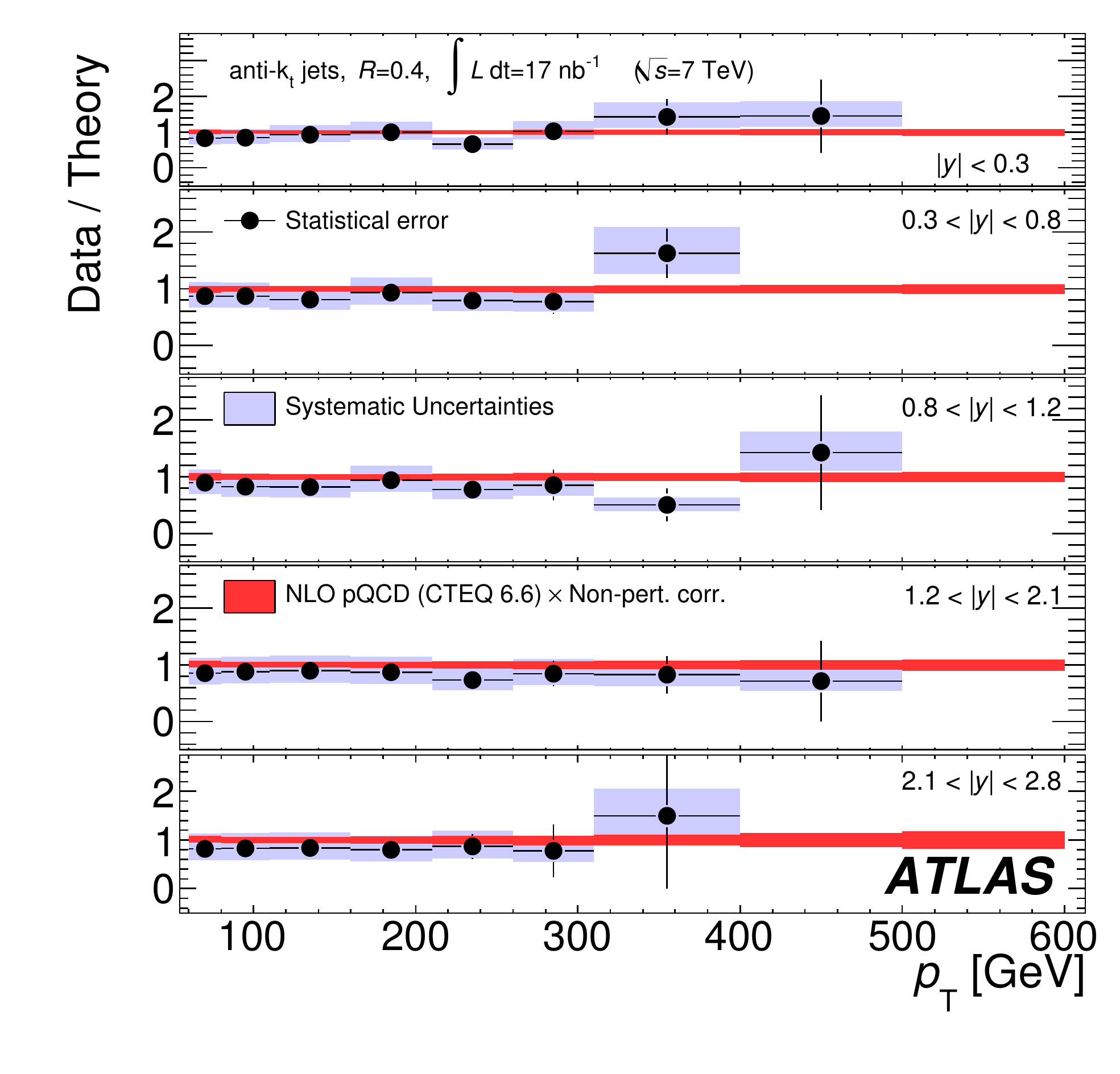}
\caption{Inclusive jet double-differential cross section as a function of jet $\pt$ in different regions of $|y|$ 
for jets identified using the \AKT algorithm with $R=0.4$.
The ratio of the data to the theoretical prediction is shown, indicating the total systematic uncertainty on the measurement.
The uncertainties on the data and theory are shown as described in Fig.~\protect\ref{fig:incjetptfullsummary04}.
}
\label{fig:incjetratio04data}
\end{center}
\end{figure*}

\begin{figure*}
\begin{center}
\includegraphics[width=0.9\textwidth]{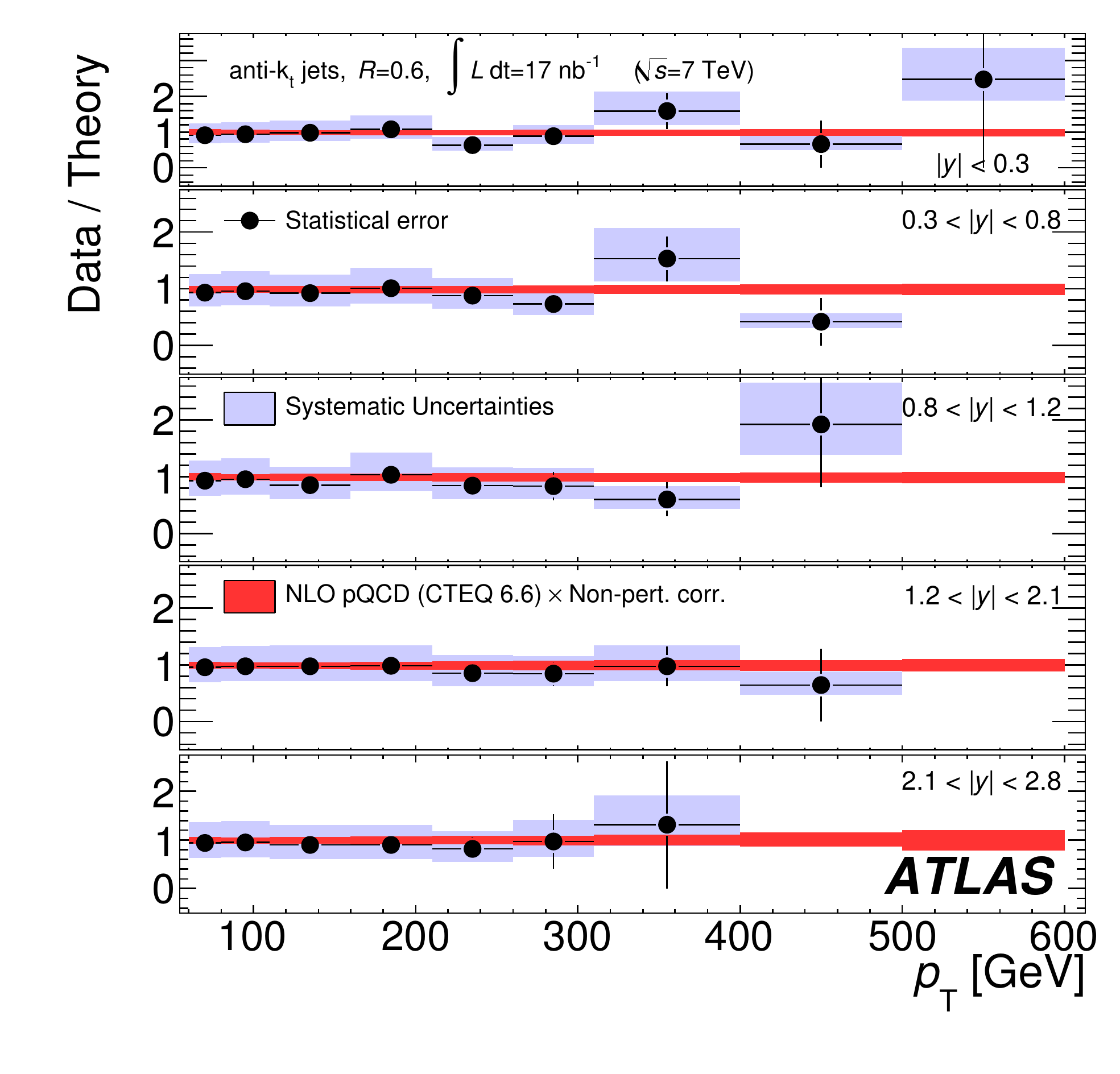}
\caption{Inclusive jet double-differential cross section as a function of jet $\pt$ in different regions of $|y|$ 
for jets identified using the \AKT algorithm with $R=0.6$.
The ratio of the data to the theoretical prediction is shown, indicating the total systematic uncertainty on the measurement.
The uncertainties on the data and theory are shown as described in Fig.~\protect\ref{fig:incjetptfullsummary04}.}
\label{fig:incjetratio06data}
\end{center}
\end{figure*}

Figures~\ref{fig:incjetptsummary04} and \ref{fig:incjetptsummary06} show the double-differential cross section
as a function of jet $\pt$ in several different regions of rapidity. 
Tables~\ref{tab:inclusiveresults1a}--\ref{tab:inclusiveresults1c} and \ref{tab:inclusiveresults2a}--\ref{tab:inclusiveresults2c} detail the same data. 
A selection of
the same cross sections expressed as a function of rapidity in different $\pt$ ranges is 
shown in Figs.~\ref{fig:incjetysummary04} and \ref{fig:incjetysummary06}. 
In Figs.~\ref{fig:incjetratio04data} and \ref{fig:incjetratio06data} the ratio of the measurement 
to the theoretical prediction is shown for the double-differential distribution in jet $\pt$ for $R=0.4$ and $R=0.6$ 
respectively.  
The data are again compared to NLO pQCD predictions to which soft corrections have been applied, where the predictions are also given in the tables.
In all regions, the theory is consistent with the data.

\begin{figure*}
\begin{center}
\includegraphics[width=0.9\textwidth]{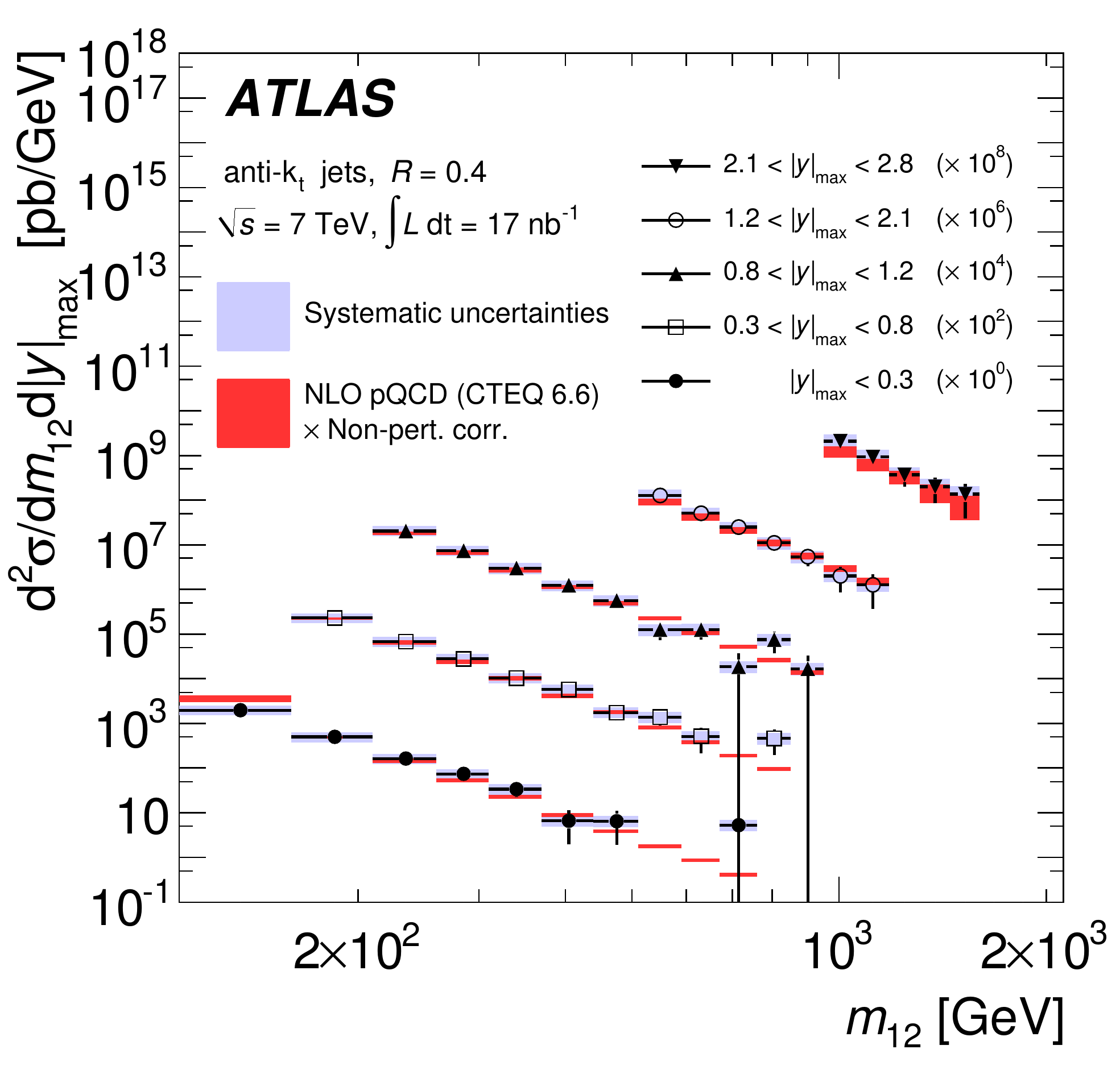}
\caption{Dijet double-differential cross section as a function of dijet mass, binned in the maximum rapidity of the two leading jets, $|y|_{\mathrm{max}}$.  The results are shown for jets identified using the \AKT algorithm with $R=0.4$.
The data are compared to NLO pQCD calculations to which soft QCD corrections have been applied. 
The uncertainties on the data and theory are shown as described in Fig.~\protect\ref{fig:incjetptfullsummary04}.}
\label{fig:dijetmasssummary04}
\end{center}
\end{figure*}

\begin{figure*}
\begin{center}
\includegraphics[width=0.9\textwidth]{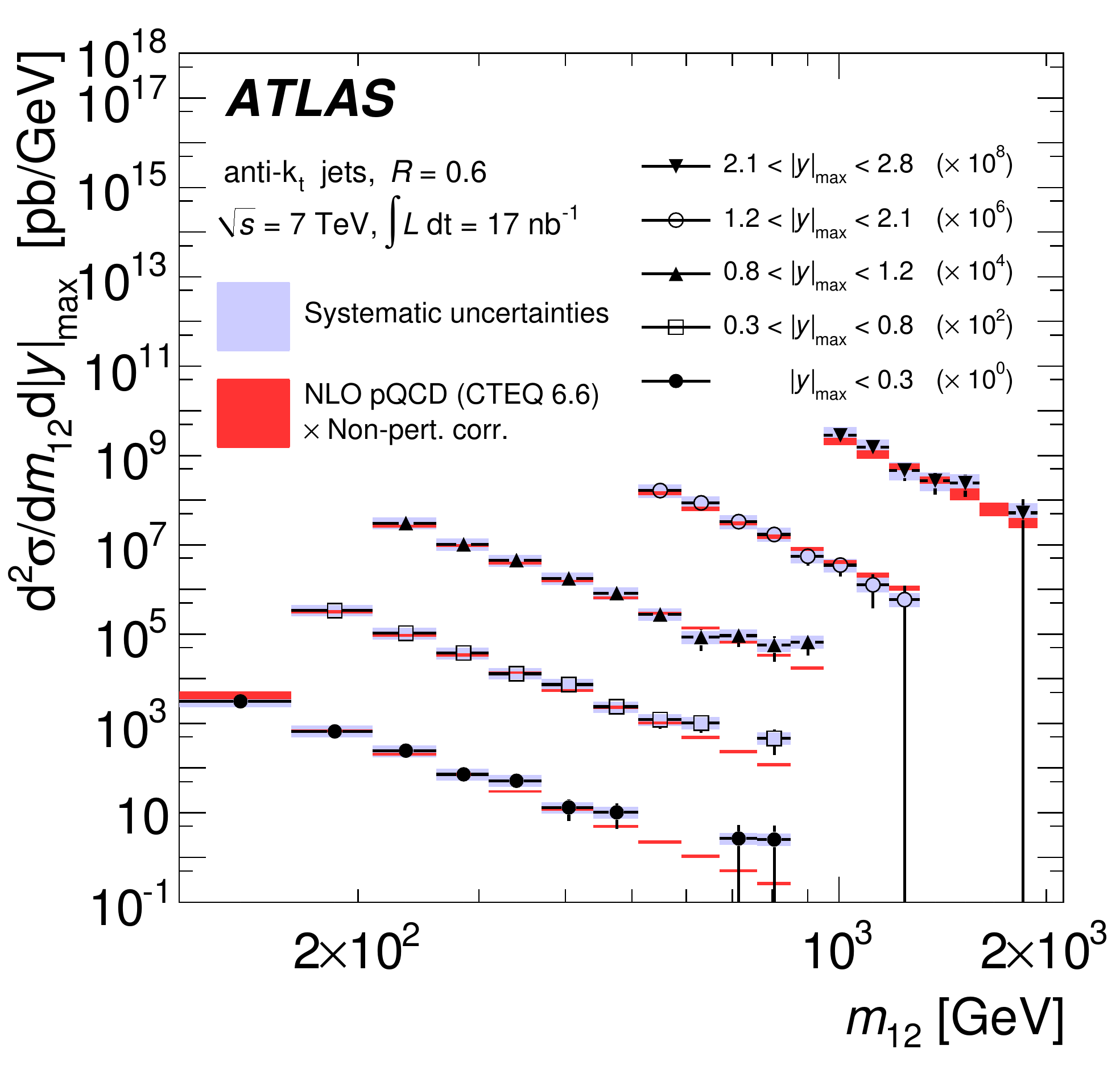}
\caption{Dijet double-differential cross section as a function of dijet mass, binned in the maximum rapidity of the two leading jets, $|y|_{\mathrm{max}}$.  The results are shown for jets identified using the \AKT algorithm with $R=0.6$.
The data are compared to NLO pQCD calculations to which soft QCD corrections have been applied.
The uncertainties on the data and theory are shown as described in Fig.~\protect\ref{fig:incjetptfullsummary04}.}
\label{fig:dijetmasssummary06}
\end{center}
\end{figure*}

\begin{figure*}
\begin{center}
\includegraphics[width=0.8\textwidth]{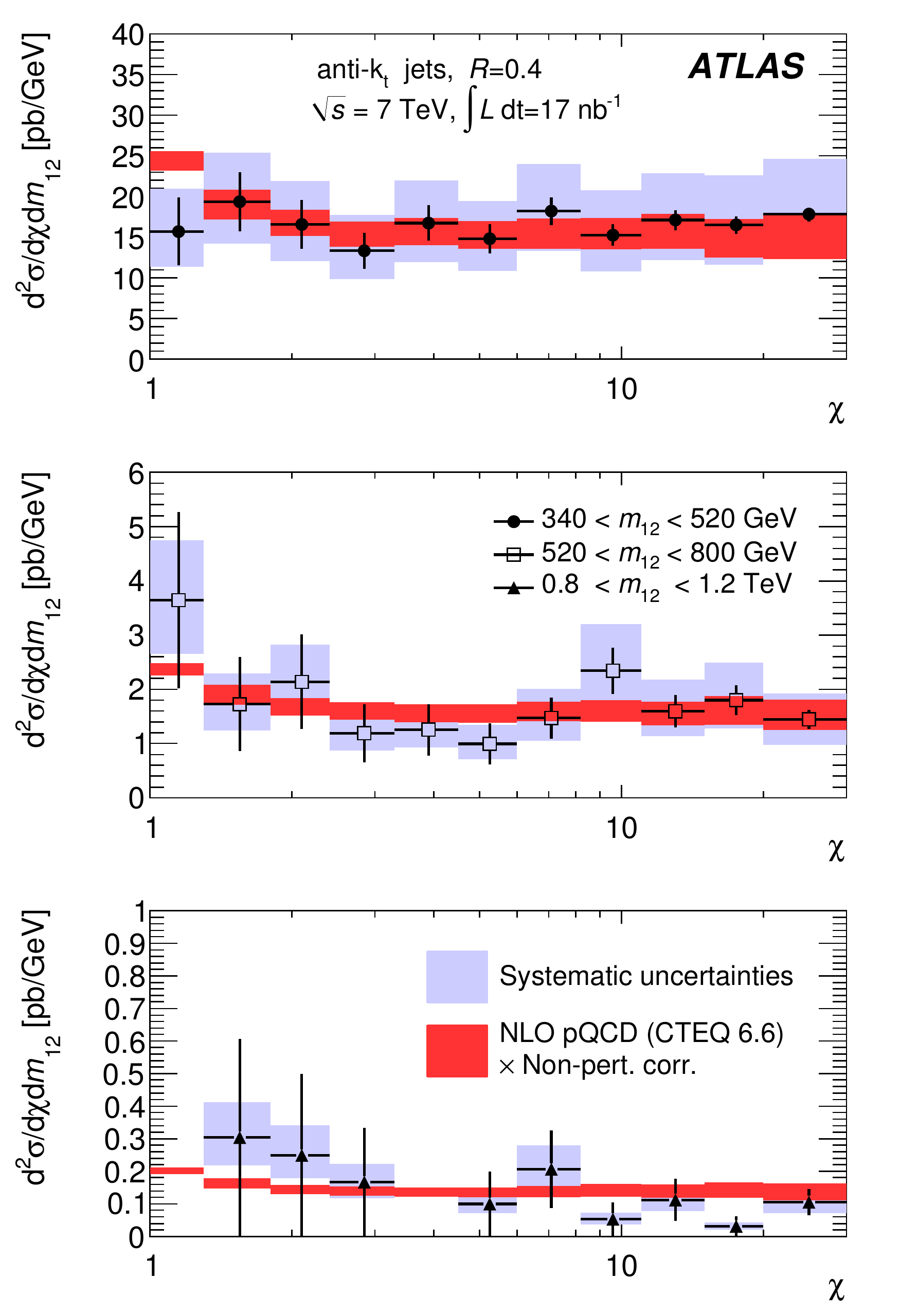}
\caption{Dijet double-differential cross section as a function of angular variable $\chi$ in different bins of dijet mass $\twomass{1}{2}$, for jets identified using the \AKT algorithm with $R=0.4$.
The data are compared to NLO pQCD calculations to which soft QCD corrections have been applied.
The uncertainties on the data and theory are shown as described in Fig.~\protect\ref{fig:incjetptfullsummary04}.}
\label{fig:dijetchisummary04}
\end{center}
\end{figure*}

\begin{figure*}
\begin{center}
\includegraphics[width=0.8\textwidth]{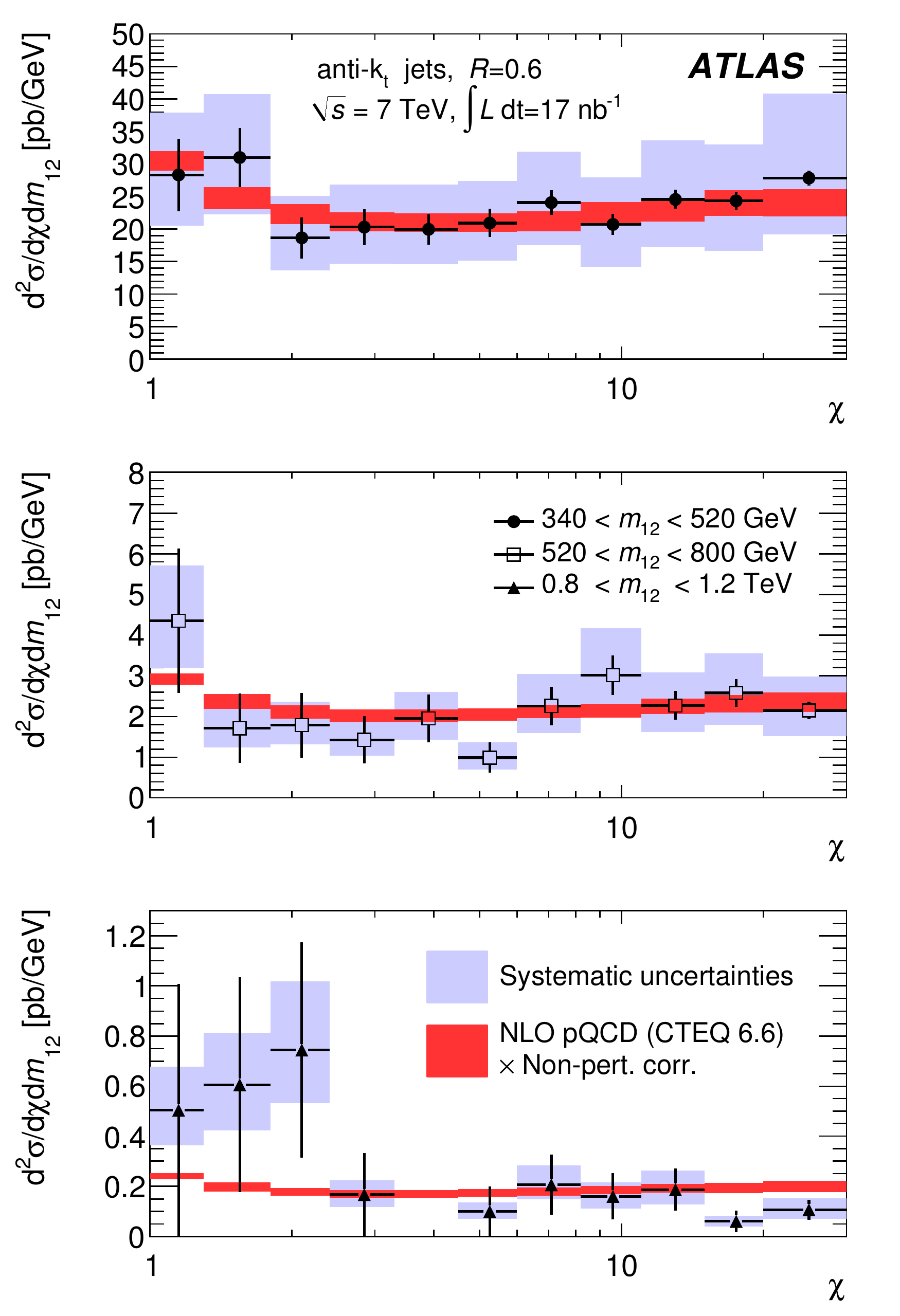}
\caption{Dijet double-differential cross section as a function of angular variable $\chi$ in different bins of dijet mass $\twomass{1}{2}$, for jets identified using the \AKT algorithm with $R=0.6$.
The data are compared to NLO pQCD calculations to which soft QCD corrections have been applied. 
The uncertainties on the data and theory are shown as described in Fig.~\protect\ref{fig:incjetptfullsummary04}.}
\label{fig:dijetchisummary06}
\end{center}
\end{figure*}

In Figs.~\ref{fig:dijetmasssummary04} and \ref{fig:dijetmasssummary06}, the double-differential dijet cross section 
is shown as a function of the dijet mass, for different bins in $|y|_{\mathrm{max}}$.
The cross section falls rapidly with mass, and extends up to masses of nearly 2~TeV. 
Figures~\ref{fig:dijetchisummary04} and \ref{fig:dijetchisummary06} show the cross section as a 
function of the dijet angular variable $\chi$ for different ranges of the dijet mass $\twomass{1}{2}$.  
The data are compared to NLO pQCD calculations 
corrected for non-perturbative effects. The theory is consistent with the data.
The dijet mass measurements and the theory predictions are also given in Tables~\ref{tab:DijetMassResults04_0}-\ref{tab:DijetMassResults04_4}
and \ref{tab:DijetMassResults06_0}-\ref{tab:DijetMassResults06_4} for $R=0.4$ and $R=0.6$ respectively.  Those for $\chi$ are given in Tables~\ref{tab:DijetChiResults04_3}-\ref{tab:DijetChiResults04_5}
and \ref{tab:DijetChiResults06_3}-\ref{tab:DijetChiResults06_5}. 

\begin{figure*}
\begin{center}
\includegraphics[width=0.49\textwidth]{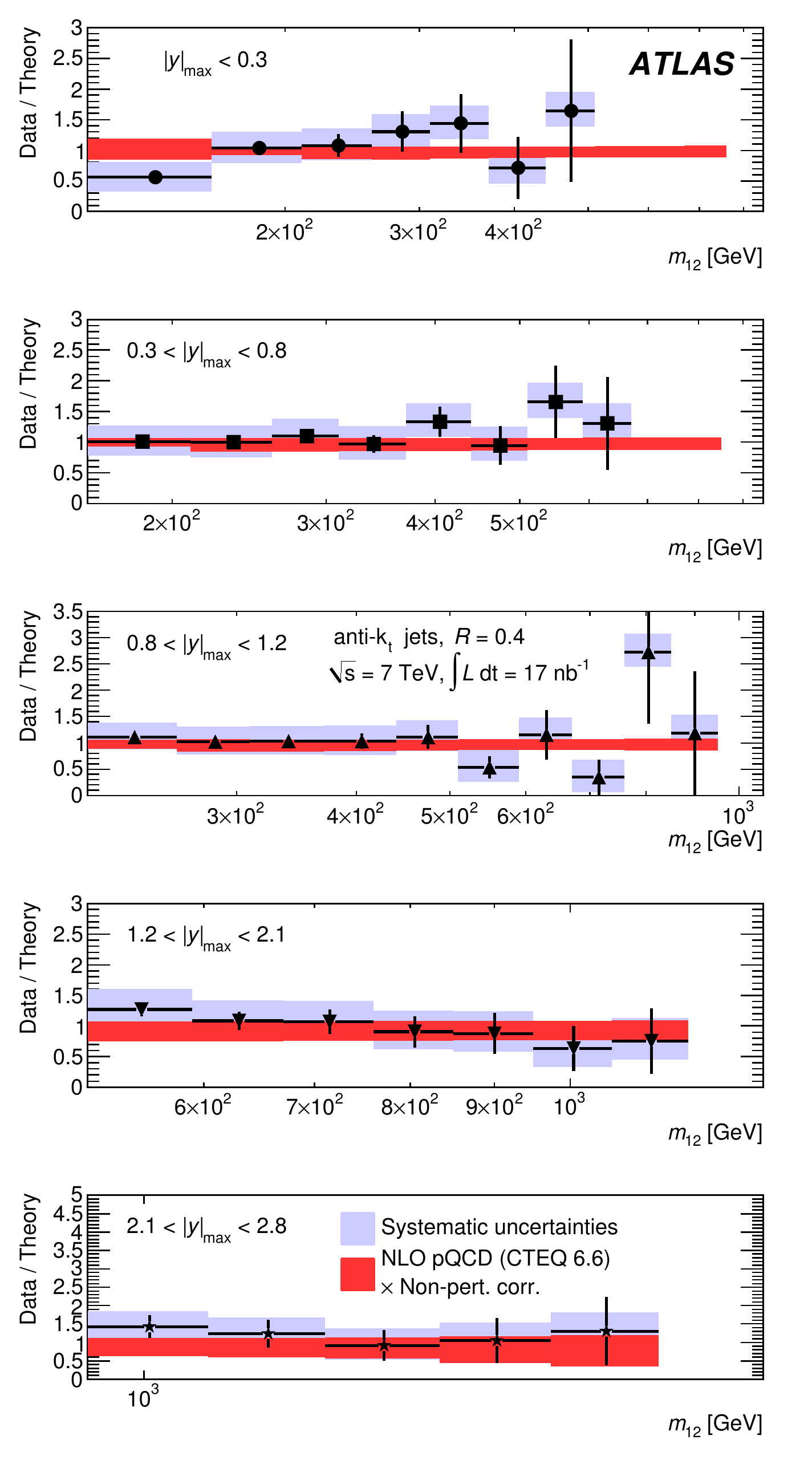}
\includegraphics[width=0.49\textwidth]{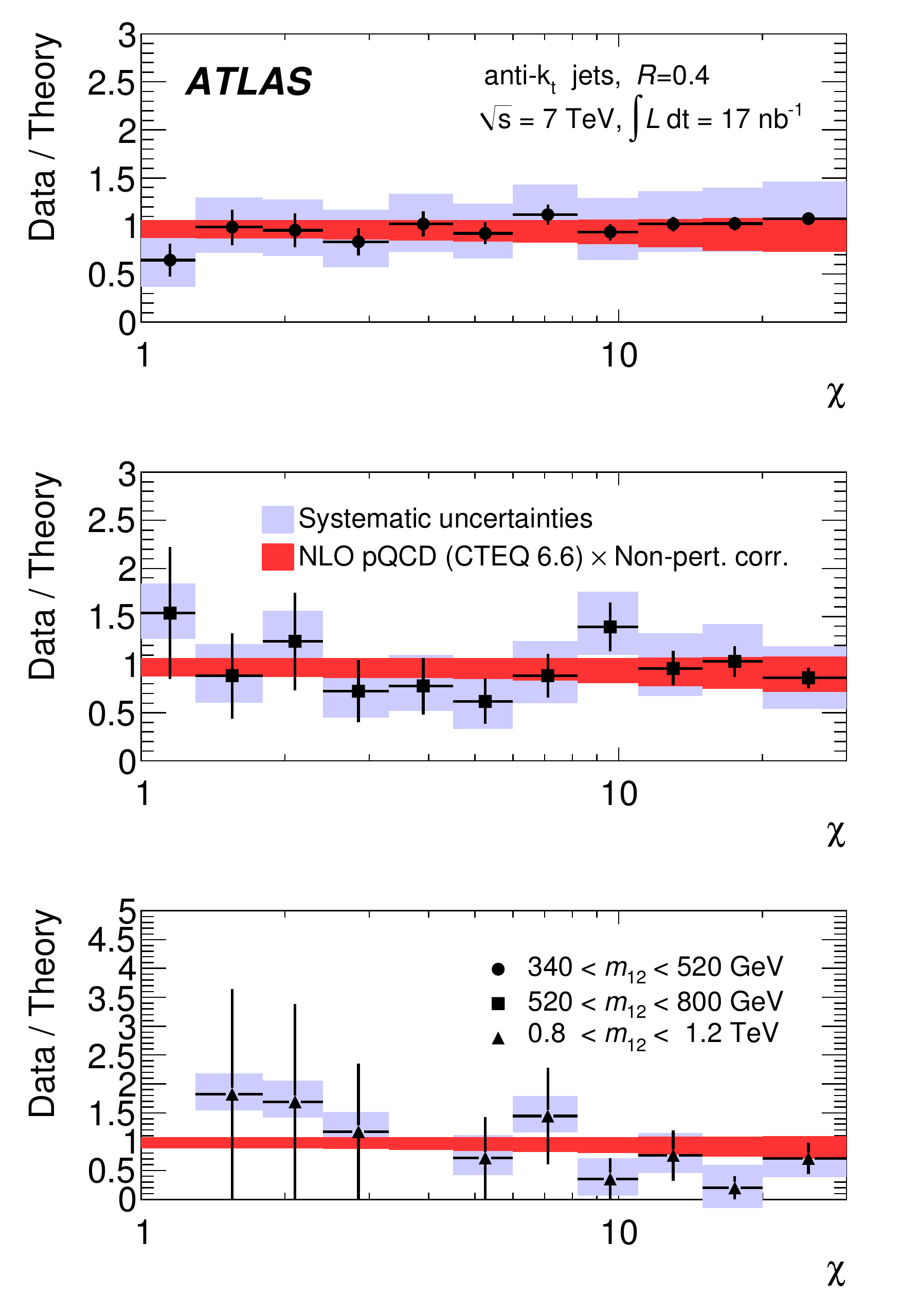}
\caption{Dijet double-differential cross sections as a function of dijet mass $\twomass{1}{2}$ and $\chi$ as shown in 
Fig.~\protect\ref{fig:dijetmasssummary04} and
Fig.~\protect\ref{fig:dijetchisummary04} respectively, expressed as a ratio to the theoretical prediction.  These are shown for jets identified using the \AKT algorithm with $R=0.4$.  The uncertainties on the data and theory are displayed as described in Fig.~\protect\ref{fig:incjetptfullsummary04}.  For each of the two lowest $|y|_{\mathrm{max}}$ bins of the dijet mass spectrum, a statistically insignificant data point at high mass lies outside the plotted range of the ratio.}
\label{fig:dijetratios04}
\end{center}
\end{figure*}

\begin{figure*}
\begin{center}
\includegraphics[width=0.49\textwidth]{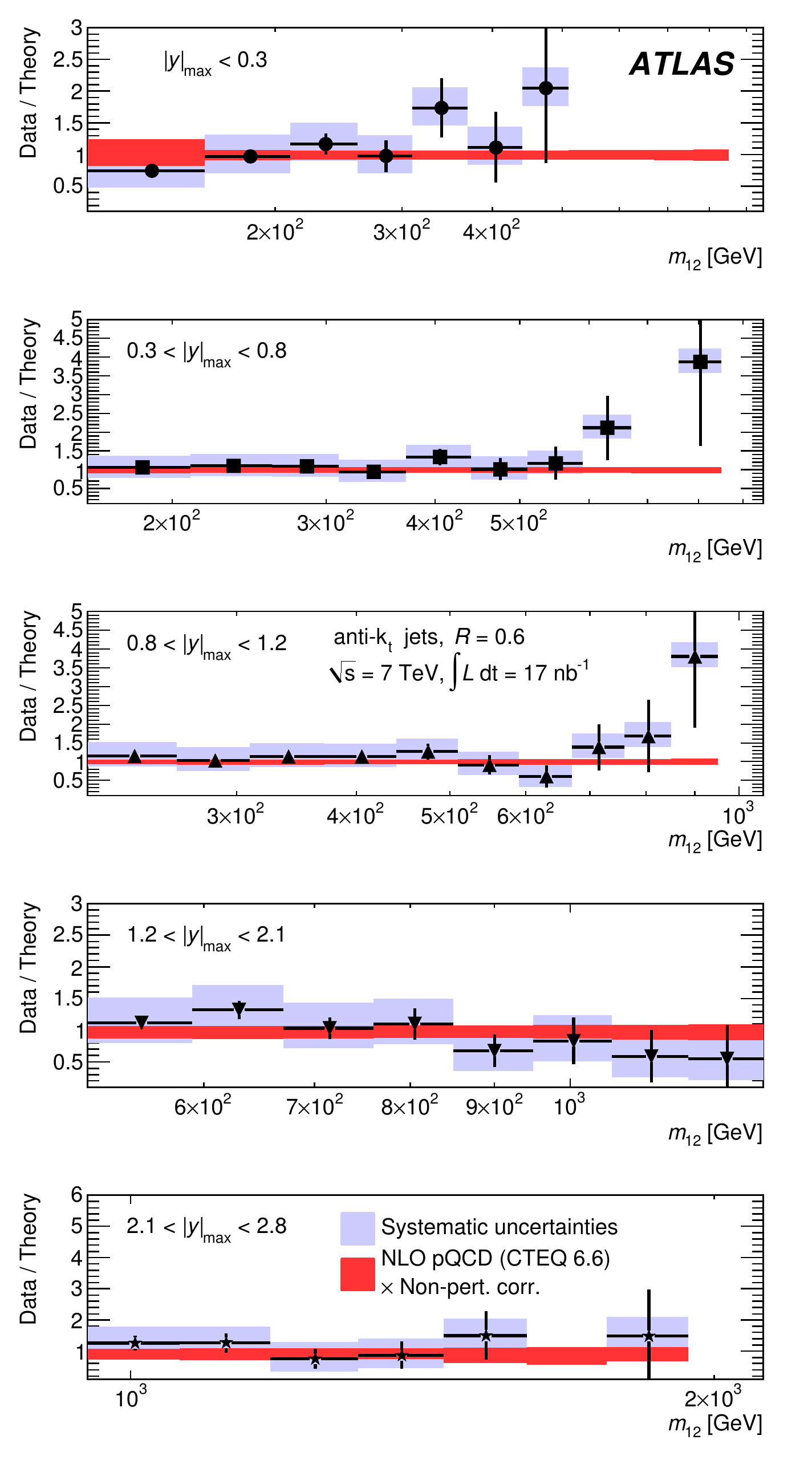}
\includegraphics[width=0.49\textwidth]{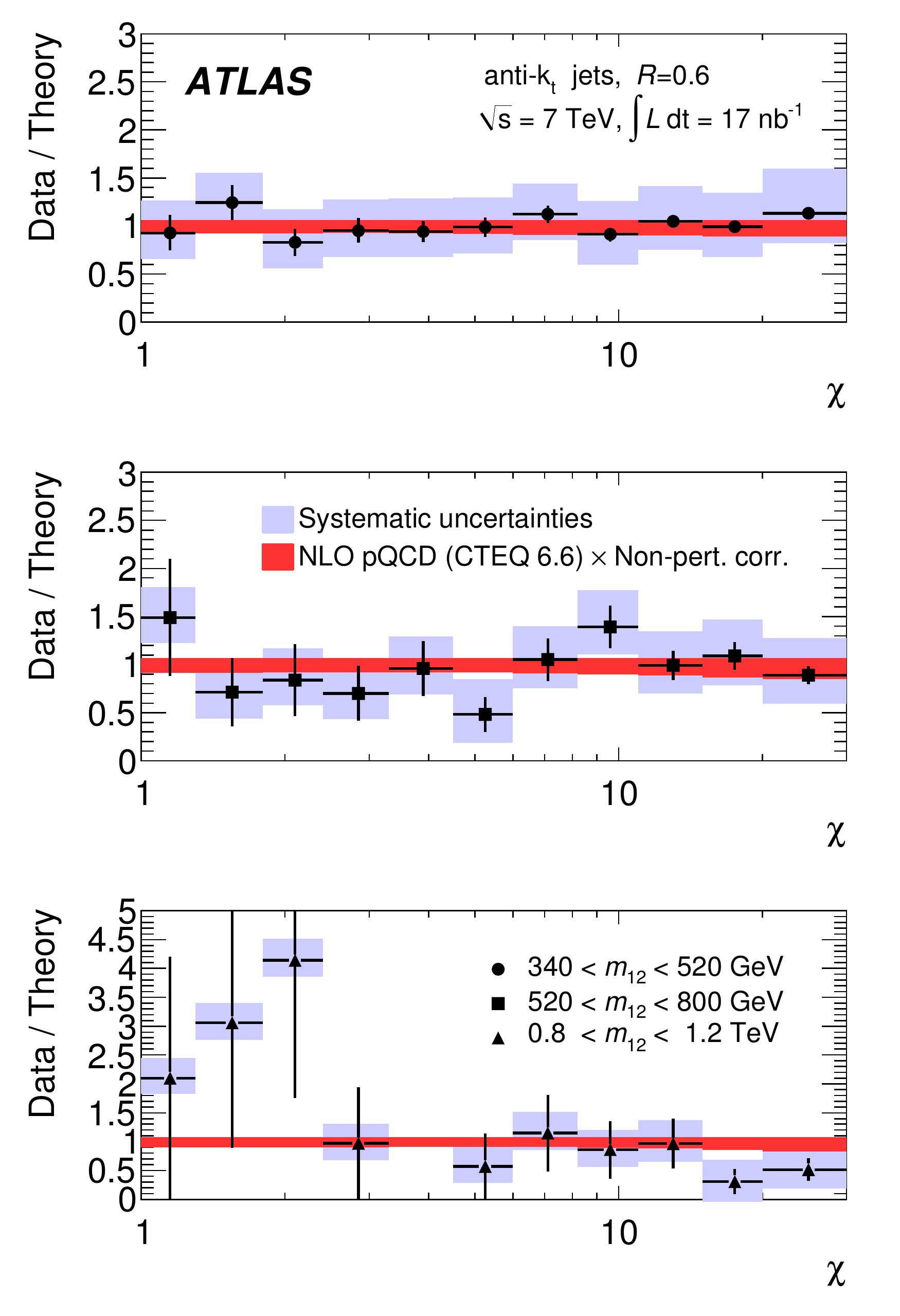}
\caption{Dijet double-differential cross sections as a function of dijet mass $\twomass{1}{2}$ and $\chi$ as shown in 
Fig.~\protect\ref{fig:dijetmasssummary06} and
Fig.~\protect\ref{fig:dijetchisummary06} respectively, expressed as a ratio to the theoretical prediction.  These are shown for jets identified using the \AKT algorithm with $R=0.6$.  The uncertainties on the data and theory are displayed as described in Fig.~\protect\ref{fig:incjetptfullsummary04}.  For the lowest $|y|_{\mathrm{max}}$ bin of the dijet mass spectrum, two statistically insignificant data points at high mass lie outside the plotted range of the ratio.}
\label{fig:dijetratios06}
\end{center}
\end{figure*}

In Figs.~\ref{fig:dijetratios04} and \ref{fig:dijetratios06} the ratio of the measurement 
to the theoretical prediction is shown for the double-differential dijet cross sections
for $R=0.4$ and $R=0.6$ 
respectively.  
The data are again compared to NLO pQCD predictions to which soft corrections have been applied, 
also included in the tables.
In all regions, the theory is consistent with the data.

\section{Conclusion}

Inclusive and dijet cross sections have been measured for the first time in proton-proton collisions 
with the ATLAS detector, at a centre-of-mass energy of 7~TeV, using an integrated luminosity of 17~nb$^{-1}$. 

The cross sections have been measured with the \AKT algorithm using two different $R$ parameters, 
with different sensitivity to soft QCD corrections. 
This is the first cross section measurement in hadron-hadron collisions using this jet algorithm.

The cross sections extend into previously unmeasured kinematic regimes.
For inclusive jets, the
double-differen\-tial   
 cross section has been measured for jets with $|y|<2.8$ and $\pt > 60\gev$.
The $\pt$ distribution extends up to 600~\gev.
For dijet events, containing a jet with $\pt > 30$~GeV in the same rapidity region, 
the cross section has been measured as a function of the dijet mass and of the angular variable
$\chi$. The dijet mass distribution extends up to nearly 2~TeV.

The dominant systematic uncertainty in these measurements comes from the jet energy response of the 
calo\-ri\-meter. 
This scale uncertainty has been determined to be below 10\% over the whole kinematic range of these measurements, 
and to be below 7\% for central jets with $\pt > 60$~GeV, leading to a systematic uncertainty 
in the cross sections of around 40\%.

The measurements use only 17~nb$^{-1}$ of integrated luminosity, but the statistical errors are not 
the dominant contribution to the uncertainty below around 300~GeV in transverse momentum. 
Data already recorded by ATLAS
will extend the reach of subsequent measurements and their precision at high transverse 
momenta.

Leading-logarithmic parton-shower MC generators provide a reasonable description of the energy 
flow around the jets, and of the shapes of the measured distributions.

The differential cross sections have been compared to NLO pQCD calculations 
corrected for non-perturbative effects. 
The inclusive jet measurements are sensitive to 
the combination of the QCD matrix element and parton densities within the 
proton, evolved from determinations made using measurements from previous experiments at 
lower energy scales. 
The dijet measurements have been made in a region where the sensitivity to the parton distributions is
reduced, and thus primarily test the structure of the QCD matrix element. 
For both inclusive and dijet measurements, the theory agrees well with the data, 
validating this perturbative QCD approach in a new kinematic regime.



\section*{Acknowledgements}

We deeply thank everybody at CERN involved in operating the LHC in such a superb way during this initial high-energy data-taking period. We acknowledge equally warmly all the technical and administrative staff in the collaborating institutions without whom ATLAS could not be operated so efficiently. 

We acknowledge the support of ANPCyT, Argentina; Yerevan Physics Institute, Armenia; ARC and DEST, Australia; Bundesministerium f\"ur Wissenschaft und Forschung, Austria; National Academy of Sciences of Azerbaijan; State Committee on Science \& Technologies of the Republic of Belarus; CNPq and FINEP, Brazil; NSERC, NRC, and CFI, Canada; CERN; CONICYT, Chile; NSFC, China; COLCIENCIAS, Colombia; Ministry of Education, Youth and Sports of the Czech Republic, Ministry of Industry and Trade of the Czech Republic, and Committee for Collaboration of the Czech Republic with CERN; DNRF, DNSRC and the Lundbeck Foundation, Denmark; European Commission, through the ARTEMIS Research Training Network; IN2P3-CNRS and CEA-DSM/IRFU, France; Georgian Academy of Sciences; BMBF, DFG, HGF and MPG, Germany; Ministry of Education and Religion, through the EPEAEK program PYTHAGORAS II and GSRT, Greece; ISF, MINERVA, GIF, DIP, and Benoziyo Center, Israel; INFN, Italy; MEXT, Japan; CNRST, Morocco; FOM and NWO, Netherlands; The Research Council of Norway; Ministry of Science and Higher Education, Poland; GRICES and FCT, Portugal; Ministry of Education and Research, Romania; Ministry of Education and Science of the Russian Federation and State Atomic Energy Corporation ROSATOM; JINR; Ministry of Science, Serbia; Department of International Science and Technology Cooperation, Ministry of Education of the Slovak Republic; Slovenian Research Agency, Ministry of Higher Education, Science and Technology, Slovenia; Ministerio de Educaci\'{o}n y Ciencia, Spain; The Swedish Research Council, The Knut and Alice Wallenberg Foundation, Sweden; State Secretariat for Education and Science, Swiss National Science Foundation, and Cantons of Bern and Geneva, Switzerland; National Science Council, Taiwan; TAEK, Turkey; The STFC, the Royal Society and The Leverhulme Trust, United Kingdom; DOE and NSF, United States of America. 


\bibliographystyle{atlasnote}
\bibliography{%
JetXSPaper}

\providecommand{\href}[2]{#2}\begingroup\raggedright\begin{thebibliography}{10}

\bibitem{Arnison:1983dk}
{UA1} Collaboration, G.~Arnison et al., {\em {Angular Distributions and
  Structure Functions from Two Jet Events at the CERN SPS $p\bar{p}$
  Collider}\/},
\href{http://dx.doi.org/10.1016/0370-2693(84)91164-X}{Phys. Lett. {\bf B136}
  (1984)  294}.

\bibitem{Alitti:1990aa}
{UA2} Collaboration, J.~Alitti et al., {\em {Inclusive jet cross-section and a
  search for quark compositeness at the CERN $p\bar{p}$ collider}\/},
\href{http://dx.doi.org/10.1016/0370-2693(91)90887-V}{Phys. Lett. {\bf B257}
  (1991)  232--240}.

\bibitem{Adeva:1990nu}
{L3} Collaboration, B.~Adeva et al., {\em {Determination of $\alpha_s$ from jet
  multiplicities measured on the $Z$ resonance}\/},
\href{http://dx.doi.org/10.1016/0370-2693(90)90323-X}{Phys. Lett. {\bf B248}
  (1990)  464--472}.

\bibitem{Chekanov:2001bw}
{ZEUS} Collaboration, S.~Chekanov et al., {\em {Dijet photoproduction at HERA
  and the structure of the photon}\/},
  \href{http://dx.doi.org/10.1007/s100520200936}{Eur. Phys. J. {\bf C23} (2002)
   615--631},
\href{http://arxiv.org/abs/hep-ex/0112029}{{\tt arXiv:hep-ex/0112029}}.

\bibitem{Heister:2002tq}
{ALEPH} Collaboration, A.~Heister et al., {\em {Measurements of the strong
  coupling constant and the QCD colour factors using four-jet observables from
  hadronic Z decays}\/},
\href{http://dx.doi.org/10.1140/epjc/s2002-01114-2}{Eur. Phys. J. {\bf C27}
  (2003)  1--17}.

\bibitem{Chekanov:2002be}
{ZEUS} Collaboration, S.~Chekanov et al., {\em {Inclusive jet cross sections in
  the Breit frame in neutral current deep inelastic scattering at HERA and
  determination of $\alpha_s$}\/},
  \href{http://dx.doi.org/10.1016/S0370-2693(02)02763-6}{Phys. Lett. {\bf B547}
  (2002)  164--180},
\href{http://arxiv.org/abs/hep-ex/0208037}{{\tt arXiv:hep-ex/0208037}}.

\bibitem{Abdallah:2004uq}
{DELPHI} Collaboration, J.~Abdallah et al., {\em {Measurement of the energy
  dependence of hadronic jet rates and the strong coupling $\alpha_s$ from the
  four-jet rate with the DELPHI detector at LEP}\/},
  \href{http://dx.doi.org/10.1140/epjc/s2004-02060-7}{Eur. Phys. J. {\bf C38}
  (2005)  413--426},
\href{http://arxiv.org/abs/hep-ex/0410071}{{\tt arXiv:hep-ex/0410071}}.

\bibitem{Abbiendi:2005vd}
{OPAL} Collaboration, G.~Abbiendi et al., {\em {Measurement of the strong
  coupling $\alpha_s$ from four-jet observables in $\ee$ annihilation}\/},
  \href{http://dx.doi.org/10.1140/epjc/s2006-02581-y}{Eur. Phys. J. {\bf C47}
  (2006)  295--307},
\href{http://arxiv.org/abs/hep-ex/0601048}{{\tt arXiv:hep-ex/0601048}}.

\bibitem{Chekanov:2005nn}
{ZEUS} Collaboration, S.~Chekanov et al., {\em {An NLO QCD analysis of
  inclusive cross-section and jet- production data from the ZEUS
  experiment}\/},  \href{http://dx.doi.org/10.1140/epjc/s2005-02293-x}{Eur.
  Phys. J. {\bf C42} (2005)  1--16},
\href{http://arxiv.org/abs/hep-ph/0503274}{{\tt arXiv:hep-ph/0503274}}.

\bibitem{Abulencia:2007ez}
{CDF} Collaboration, A.~Abulencia et al., {\em {Measurement of the inclusive
  jet cross section using the $k_{T}$ algorithm in $p \bar{p}$ collisions at
  $\sqrt{s}$ = 1.96~TeV with the CDF II detector}\/},
  \href{http://dx.doi.org/10.1103/PhysRevD.75.092006}{Phys. Rev. {\bf D75}
  (2007)  092006},
\href{http://arxiv.org/abs/hep-ex/0701051}{{\tt hep-ex/0701051}}.

\bibitem{:2007jx}
{OPAL} Collaboration, G.~Abbiendi et al., {\em {Inclusive Jet Production in
  Photon-Photon Collisions at $\sqrt{s_{ee}}$ from 189 to 209 GeV}\/},
  \href{http://dx.doi.org/10.1016/j.physletb.2007.08.096}{Phys. Lett. {\bf
  B658} (2008)  185--192},
\href{http://arxiv.org/abs/0706.4382}{{\tt arXiv:0706.4382 [hep-ex]}}.

\bibitem{:2008hua}
{D0} Collaboration, V.~M. Abazov et al., {\em {Measurement of the inclusive jet
  cross-section in $p \bar{p}$ collisions at $\sqrt{s}$ =1.96 TeV}\/},
  \href{http://dx.doi.org/10.1103/PhysRevLett.101.062001}{Phys. Rev. Lett. {\bf
  101} (2008)  062001},
\href{http://arxiv.org/abs/0802.2400}{{\tt arXiv:0802.2400 [hep-ex]}}.

\bibitem{Aaltonen:2008eq}
{CDF} Collaboration, T.~Aaltonen et al., {\em {Measurement of the Inclusive Jet
  Cross Section at the Fermilab Tevatron p-pbar Collider Using a Cone-Based Jet
  Algorithm}\/},  \href{http://dx.doi.org/10.1103/PhysRevD.78.052006}{Phys.
  Rev. {\bf D78} (2008)  052006},
\href{http://arxiv.org/abs/0807.2204}{{\tt arXiv:0807.2204 [hep-ex]}}.

\bibitem{Abazov:2009nc}
{D0} Collaboration, V.~M. Abazov et al., {\em {Determination of the strong
  coupling constant from the inclusive jet cross section in ppbar collisions at
  $\sqrt{s}=1.96$ TeV}\/},
  \href{http://dx.doi.org/10.1103/PhysRevD.80.111107}{Phys. Rev. {\bf D80}
  (2009)  111107},
\href{http://arxiv.org/abs/0911.2710}{{\tt arXiv:0911.2710 [hep-ex]}}.

\bibitem{:2009he}
{H1} Collaboration, F.~D. Aaron et al., {\em {Jet Production in ep Collisions
  at Low $Q^2$ and Determination of $\alpha_s$}\/},
  \href{http://dx.doi.org/10.1140/epjc/s10052-010-1282-x}{Eur. Phys. J. {\bf
  C67} (2010)  1--24},
\href{http://arxiv.org/abs/0911.5678}{{\tt arXiv:0911.5678 [hep-ex]}}.

\bibitem{Aaron:2009vs}
{H1} Collaboration, F.~D. Aaron et al., {\em {Jet production in $ep$ collisions
  at high $Q^2$ and determination of $\alpha_s$}\/},  Eur. Phys. J. {\bf C}
  (2009)  , \href{http://arxiv.org/abs/0904.3870}{{\tt arXiv:0904.3870
  [hep-ex]}}.
DESY-09-032.

\bibitem{:2009mh}
{D0} Collaboration, V.~M. Abazov et al., {\em {Measurement of dijet angular
  distributions at $\sqrt{s} = 1.96$~TeV and searches for quark compositeness
  and extra spatial dimensions}\/},
  \href{http://dx.doi.org/10.1103/PhysRevLett.103.191803}{Phys. Rev. Lett. {\bf
  103} (2009)  191803},
\href{http://arxiv.org/abs/0906.4819}{{\tt arXiv:0906.4819 [hep-ex]}}.

\bibitem{Abramowicz:2010ke}
{ZEUS} Collaboration, H.~Abramowicz et al., {\em {Inclusive-jet cross sections
  in NC DIS at HERA and a comparison of the $k_T$, anti-$k_T$ and SIScone jet
  algorithms}\/},
  \href{http://dx.doi.org/10.1016/j.physletb.2010.06.015}{Phys. Lett. {\bf
  B691} (2010)  127--137},
\href{http://arxiv.org/abs/1003.2923}{{\tt arXiv:1003.2923 [hep-ex]}}.

\bibitem{Abazov:2010fr}
{D0} Collaboration, V.~M. Abazov et al., {\em {Measurement of the dijet
  invariant mass cross section in $p\overline{p}$ collisions at $\sqrt{s} =$
  1.96 TeV}\/},  \href{http://arxiv.org/abs/1002.4594}{{\tt arXiv:1002.4594
  [hep-ex]}}.
Submitted to Phys. Lett. B.

\bibitem{Collaboration:2010bc}
{ATLAS} Collaboration, {\em {Search for New Particles in Two-Jet Final States
  in 7 TeV Proton-Proton Collisions with the ATLAS Detector at the LHC}\/},
  \href{http://arxiv.org/abs/1008.2461}{{\tt arXiv:1008.2461 [hep-ex]}}.
Accepted by Phys. Rev. Lett.

\bibitem{angles}
{ATLAS Collaboration}, {\em {Search for Quark Contact Interactions in Dijet
  Angular Distributions in $pp$ Collisions at $\sqrt{s} = 7$~TeV Measured with
  the ATLAS Detector}\/},  \href{http://arxiv.org/abs/1009.5069}{{\tt
  arXiv:1009.5069 [hep-ex]}}.
Submitted to Phys. Lett. B.

\bibitem{Aad:2008zzm}
{ATLAS} Collaboration, {\em {The ATLAS Experiment at the CERN Large Hadron
  Collider}\/},
\href{http://dx.doi.org/10.1088/1748-0221/3/08/S08003}{JINST {\bf 3} (2008)
  S08003}.

\bibitem{Aad:2010rd}
{ATLAS} Collaboration, {\em {Charged-particle multiplicities in pp interactions
  at $\sqrt{s} = 900$~GeV measured with the ATLAS detector at the LHC}\/},
  \href{http://dx.doi.org/10.1016/j.physletb.2010.03.064}{Phys. Lett. {\bf
  B688} (2010)  21--42},
\href{http://arxiv.org/abs/1003.3124}{{\tt arXiv:1003.3124 [hep-ex]}}.

\bibitem{Cacciari:2008gp}
M.~Cacciari, G.~Salam, and G.~Soyez, {\em The anti-$k_t$ jet clustering
  algorithm\/},  JHEP {\bf 0804} (2008)  063,
\href{http://arxiv.org/abs/0802.1189}{{\tt arXiv:0802.1189}}.

\bibitem{Cacciari:2005hq}
M.~Cacciari and G.~P. Salam, {\em Dispelling the $N^3$ myth for the $k_t$
  jet-finder\/},  Phys. Lett. {\bf B641} (2006)  57--61,
\href{http://arxiv.org/abs/hep-ph/0512210}{{\tt hep-ph/0512210}}.

\bibitem{fastjet}
G.~P. Salam, M.~Cacciari, and G.~Soyez.
\newblock \url{http://www.lpthe.jussieu.fr/~salam/fastjet/}.

\bibitem{Buttar:2008jx}
C.~Buttar et al., {\em {Standard Model Handles and Candles Working Group: Tools
  and Jets Summary Report}\/},
\href{http://arxiv.org/abs/arXiv:0803.0678}{{\tt arXiv:arXiv:0803.0678
  [hep-ph]}}.

\bibitem{Frixione:1997ks}
S.~Frixione and G.~Ridolfi, {\em {Jet photoproduction at HERA}\/},
  \href{http://dx.doi.org/10.1016/S0550-3213(97)00575-0}{Nucl. Phys. {\bf B507}
  (1997)  315--333},
\href{http://arxiv.org/abs/hep-ph/9707345}{{\tt arXiv:hep-ph/9707345}}.

\bibitem{pythia}
T.~Sj{\"o}strand, S.~Mrenna, and P.~Skands, {\em PYTHIA 6.4 physics and
  manual\/},  JHEP {\bf 05} (2006)  026,
\href{http://arxiv.org/abs/hep-ph/0603175}{{\tt hep-ph/0603175}}.

\bibitem{herwig2}
G.~Corcella et al., {\em HERWIG 6: An event generator for hadron emission
  reactions with interfering gluons (including supersymmetric processes)\/},
  JHEP {\bf 01} (2001)  010,
\href{http://arxiv.org/abs/hep-ph/0011363}{{\tt hep-ph/0011363}}.

\bibitem{Butterworth:1996zw}
J.~M. Butterworth, J.~R. Forshaw, and M.~H. Seymour, {\em {Multiparton
  interactions in photoproduction at HERA}\/},
  \href{http://dx.doi.org/10.1007/s002880050286}{Z. Phys. {\bf C72} (1996)
  637--646},
\href{http://arxiv.org/abs/hep-ph/9601371}{{\tt arXiv:hep-ph/9601371}}.

\bibitem{Bahr:2008pv}
M.~Bahr et al., {\em {Herwig++ Physics and Manual}\/},
  \href{http://dx.doi.org/10.1140/epjc/s10052-008-0798-9}{Eur. Phys. J. {\bf
  C58} (2008)  639--707},
\href{http://arxiv.org/abs/0803.0883}{{\tt arXiv:0803.0883 [hep-ph]}}.

\bibitem{Mangano:2002ea}
M.~L. Mangano, M.~Moretti, F.~Piccinini, R.~Pittau, and A.~D. Polosa, {\em
  ALPGEN, a generator for hard multiparton processes in hadronic collisions\/},
   JHEP {\bf 07} (2003)  001,
\href{http://arxiv.org/abs/hep-ph/0206293}{{\tt hep-ph/0206293}}.

\bibitem{Gleisberg:2008ta}
T.~Gleisberg et al., {\em {Event generation with SHERPA 1.1}\/},
  \href{http://dx.doi.org/10.1088/1126-6708/2009/02/007}{JHEP {\bf 02} (2009)
  007},
\href{http://arxiv.org/abs/0811.4622}{{\tt arXiv:0811.4622 [hep-ph]}}.

\bibitem{ATL-PHYS-PUB-2010-002}
{ATLAS Collaboration}, {\em ATLAS Monte Carlo tunes for MC09\/},
  {ATL-PHYS-PUB-2010-002}, 2010.

\bibitem{Martin:2009iq}
A.~D. Martin, W.~J. Stirling, R.~S. Thorne, and G.~Watt, {\em {Parton
  distributions for the LHC}\/},
  \href{http://dx.doi.org/10.1140/epjc/s10052-009-1072-5}{Eur. Phys. J. {\bf
  C63} (2009)  189--285},
\href{http://arxiv.org/abs/0901.0002}{{\tt arXiv:0901.0002 [hep-ph]}}.

\bibitem{Sherstnev:2007nd}
A.~Sherstnev and R.~S. Thorne, {\em {Parton Distributions for LO
  Generators}\/},  \href{http://dx.doi.org/10.1140/epjc/s10052-008-0610-x}{Eur.
  Phys. J. {\bf C55} (2008)  553--575},
\href{http://arxiv.org/abs/0711.2473}{{\tt arXiv:0711.2473 [hep-ph]}}.

\bibitem{Collaboration:2010wq}
{ATLAS} Collaboration, {\em {The ATLAS Simulation Infrastructure}\/},
  \href{http://arxiv.org/abs/1005.4568}{{\tt arXiv:1005.4568}}.
Accepted by Eur. Phys. J.

\bibitem{Agostinelli:2002hh}
{GEANT4} Collaboration, S.~Agostinelli et al., {\em {GEANT4: A simulation
  toolkit}\/},
\href{http://dx.doi.org/10.1016/S0168-9002(03)01368-8}{Nucl. Instrum. Meth.
  {\bf A506} (2003)  250--303}.

\bibitem{QGS}
G.~Folger and J.~P. Wellisch, {\em {String parton models in Geant4}\/},
\href{http://arxiv.org/abs/nucl-th/0306007}{{\tt arXiv:nucl-th/0306007}}.

\bibitem{Bertini}
H.~Bertini, {\em Intranuclear Cascade Calculation of the Secondary Nucleon
  Spectra from Nucleon Nucleus Interactions in the Energy Range 340 to 2900 MeV
  and Comparisons with Experiment\/},  Phys. Rev. {\bf 188} (1969)  1711--1730.

\bibitem{geanthadronic}
A.~Ribon et al., {\em Status of Geant4 hadronic physics for the simulation of
  LHC experiments at the LHC physics program\/},  Tech. Rep.
  CERN-LCGAPP-2010-02, CERN, Geneva, May, 2010.

\bibitem{Abat:1263861}
E.~Abat et al., {\em Response and Shower Topology of 2 to 180 GeV Pions
  Measured with the ATLAS Barrel Calorimeter at the CERN Test-beam and
  Comparison to Monte Carlo Simulations\/},  Tech. Rep. ATL-CAL-PUB-2010-001,
  CERN, Geneva, May, 2010.

\bibitem{Adragna:1214935}
P.~Adragna et al., {\em Measurement of Pion and Proton Response and
  Longitudinal Shower Profiles up to 20 Nuclear Interaction Lengths with the
  ATLAS Tile Calorimeter\/},  {CERN-PH-EP-2009-019. ATL-TILECAL-PUB-2009-009},
  2009.

\bibitem{TB1}
E.~Abat et al., {\em {Study of the response of the ATLAS central calorimeter to
  pions of energies from 3 to 9 GeV}\/},
\href{http://dx.doi.org/10.1016/j.nima.2009.05.158}{Nucl. Instrum. Meth. {\bf
  A607} (2009)  372--386}.

\bibitem{NIMA621134}
E.~Abat et al., {\em {Study of energy response and resolution of the ATLAS
  barrel calorimeter to hadrons of energies from 20 to 350 GeV}\/},  Nuclear
  Instruments and Methods in Physics Research A {\bf 621} (2010)  134--150.

\bibitem{TB2}
J.~Pinfold et al., {\em Performance of the ATLAS liquid argon endcap
  calorimeter in the pseudorapidity region 2.5 to 4.0 in beam tests\/},  Nucl.
  Instrum. Meth. {\bf A593} (2008)  324--342.

\bibitem{1748-0221-2-05-P05005}
D.~M. Gingrich et al., {\em Construction, assembly and testing of the ATLAS
  hadronic end-cap calorimeter\/},  J. Inst. {\bf 2} (2007) no.~05, P05005.

\bibitem{Nagy:2003tz}
Z.~Nagy, {\em {Next-to-leading order calculation of three jet observables in
  hadron hadron collision}\/},
  \href{http://dx.doi.org/10.1103/PhysRevD.68.094002}{Phys. Rev. {\bf D68}
  (2003)  094002},
\href{http://arxiv.org/abs/hep-ph/0307268}{{\tt arXiv:hep-ph/0307268}}.

\bibitem{Giele:1993dj}
W.~T. Giele, E.~W.~N. Glover, and D.~A. Kosower, {\em Higher order corrections
  to jet cross-sections in hadron colliders\/},  Nucl. Phys. {\bf B403} (1993)
  633--670,
\href{http://arXiv.org/abs/hep-ph/9302225}{{\tt hep-ph/9302225}}.

\bibitem{Nadolsky:2008zw}
P.~M. Nadolsky et al., {\em {Implications of CTEQ global analysis for collider
  observables}\/},  \href{http://dx.doi.org/10.1103/PhysRevD.78.013004}{Phys.
  Rev. {\bf D78} (2008)  013004},
\href{http://arxiv.org/abs/0802.0007}{{\tt arXiv:0802.0007 [hep-ph]}}.

\bibitem{Ball:2008by}
{NNPDF} Collaboration, R.~D. Ball et al., {\em {A determination of parton
  distributions with faithful uncertainty estimation}\/},
  \href{http://dx.doi.org/10.1016/j.nuclphysb.2008.09.037}{Nucl. Phys. {\bf
  B809} (2009)  1--63},
\href{http://arxiv.org/abs/0808.1231}{{\tt arXiv:0808.1231 [hep-ph]}}.

\bibitem{H1:2009wt}
{The H1 and ZEUS Collaborations}, {\em {Combined Measurement and QCD Analysis
  of the Inclusive $e^{\pm}p$ Scattering Cross Sections at HERA}\/},  JHEP
  (2010)  ,
\href{http://arxiv.org/abs/0911.0884}{{\tt arXiv:0911.0884 [hep-ex]}}.

\bibitem{Carli:2010rw}
T.~Carli et al., {\em {A posteriori inclusion of parton density functions in
  NLO QCD final-state calculations at hadron colliders: The APPLGRID
  Project}\/},  \href{http://dx.doi.org/10.1140/epjc/s10052-010-1255-0}{Eur.
  Phys. J. {\bf C66} (2010)  503--524},
\href{http://arxiv.org/abs/0911.2985}{{\tt arXiv:0911.2985 [hep-ph]}}.

\bibitem{Skands:2009zm}
P.~Z. Skands, {\em {The Perugia Tunes}\/},
\href{http://arxiv.org/abs/0905.3418}{{\tt arXiv:0905.3418 [hep-ph]}}.

\bibitem{Buckley:2009bj}
A.~Buckley, H.~Hoeth, H.~Lacker, H.~Schulz, and J.~E. von Seggern, {\em
  {Systematic event generator tuning for the LHC}\/},
  \href{http://dx.doi.org/10.1140/epjc/s10052-009-1196-7}{Eur. Phys. J. {\bf
  C65} (2010)  331--357},
\href{http://arxiv.org/abs/0907.2973}{{\tt arXiv:0907.2973 [hep-ph]}}.

\bibitem{Buckley:2010ar}
A.~Buckley et al., {\em {Rivet user manual}\/},
\href{http://arxiv.org/abs/1003.0694}{{\tt arXiv:1003.0694 [hep-ph]}}.

\bibitem{ATLAS-LARG-PUB-2008-002}
W.~Lampl et al., {\em Calorimeter Clustering algorithms: Description and
  Performance\/},  Tech. Rep. ATLAS-LARG-PUB-2008-002, CERN, Geneva, December,
  2008.

\bibitem{Ctbelectron}
M.~Aleksa et al., {\em ATLAS Combined Testbeam: Computation and Validation of
  the Electronic Calibration Constants for the Electromagnetic Calorimeter\/},
  {ATL-LARG-PUB-2006-003}, 2006.

\bibitem{LArReadiness}
{ATLAS} Collaboration, M.~Aharrouche et al., {\em Readiness of the ATLAS Liquid
  Argon Calorimeter for LHC Collisions\/},
  \href{http://arxiv.org/abs/0912.2642}{{\tt arXiv:0912.2642}}.

\bibitem{ATLAS-TCAL-2010-01-003}
M.~Aharrouche et al., {\em {Readiness of the ATLAS Tile Calorimeter for LHC
  collisions}\/},  \href{http://arxiv.org/abs/1007.5423}{{\tt arXiv:1007.5423
  [physics.ins-det]}}.
ATLAS-TCAL-2010-01-006,CERN-PH-EP-2010-024.

\bibitem{Aharrouche:2010zz}
M.~Aharrouche et al., {\em {Measurement of the response of the ATLAS liquid
  argon barrel calorimeter to electrons at the 2004 combined test- beam}\/},
\href{http://dx.doi.org/10.1016/j.nima.2009.12.055}{Nucl. Instrum. Meth. {\bf
  A614} (2010)  400--432}.

\bibitem{atlas:2010knc}
{ATLAS} Collaboration, {\em {Performance of the ATLAS Detector using First
  Collision Data}\/},
\href{http://arxiv.org/abs/1005.5254}{{\tt arXiv:1005.5254 [hep-ex]}}.

\bibitem{Aharrouche:2008zz}
M.~Aharrouche et al., {\em {Time resolution of the ATLAS barrel liquid argon
  electromagnetic calorimeter}\/},
\href{http://dx.doi.org/10.1016/j.nima.2008.08.142}{Nucl. Instrum. Meth. {\bf
  A597} (2008)  178--188}.

\bibitem{ATLAS-CONF-2010-055}
{ATLAS Collaboration}, {\em In-situ pseudo-rapidity inter-calibration to
  evaluate jet energy scale uncertainty and calorimeter performance in the
  forward region\/},  {ATLAS-CONF-2010-055}, 2010.

\bibitem{Fritiof}
B.~Andersson, G.~Gustafson, and B.~Nilsson-Almqvist, {\em A model for low-pT
  hadronic reactions with generalizations to hadron-nucleus and nucleus-nucleus
  collisions\/},  \href{http://dx.doi.org/DOI:
  10.1016/0550-3213(87)90257-4}{Nucl. Phys. B {\bf 281} (1987) no.~1-2, 289 --
  309}.

\bibitem{ATL-CAL-PUB-2010-001}
E.~Abat et al., {\em Response and Shower Topology of 2 to 180 GeV Pions
  Measured with the ATLAS Barrel Calorimeter at the CERN Test-beam and
  Comparison to Monte Carlo Simulations\/},  {ATL-CAL-PUB-2010-001}, 2010.

\bibitem{ATLAS-CONF-2010-052}
{ATLAS Collaboration}, {\em {ATLAS Calorimeter Response to Single Isolated
  Hadrons and Estimation of the Calorimeter Jet Scale Uncertainty}\/},
  {ATLAS-CONF-2010-052}, 2010.

\bibitem{ATLAS-CONF-2010-053}
{ATLAS Collaboration}, {\em Properties of Jets and Inputs to Jet Reconstruction
  and Calibration with the ATLAS Detector Using Proton-Proton Collisions at
  $\sqrt{s}=7$ TeV\/},  {ATLAS-CONF-2010-053}, 2010.

\bibitem{ATLAS-CONF-2010-038}
{ATLAS Collaboration}, {\em Data-Quality Requirements and Event Cleaning for
  Jets and Missing Transverse Energy Reconstruction with the ATLAS Detector in
  Proton-Proton Collisions at a Center-of-Mass Energy of $\sqrt{s}=7$~TeV\/},
  {ATLAS-CONF-2010-038}, 2010.

\bibitem{ATLAS-CONF-2010-054}
{ATLAS Collaboration}, {\em {Jet Energy Resolution and Selection Efficiency
  Relative to Track Jets from In-situ Techniques with the ATLAS Detector Using
  Proton-Proton Collisions at a Center of Mass Energy $\sqrt{s}=7$ TeV}\/},
  {ATLAS-CONF-2010-054}, 2010.

\bibitem{ATLAS-CONF-2010-060}
{ATLAS Collaboration}, {\em Luminosity Determination Using the ATLAS
  Detector\/},  {ATLAS-CONF-2010-060}, 2010.

\end{thebibliography}\endgroup

\clearpage

\clearpage

\clearpage

\clearpage

\clearpage

\clearpage


\onecolumn

\begin{sidewaystable*}
\begin{center}
\vspace{16cm}
\begin{tabular}{|c|c|c|c|c|c|c|c|c|c|}
\hline
\multicolumn{10}{|l|}{$0< |y| <0.3$ } \\ 
\hline
$\pt$ [GeV]   & 60-80 & 80-110 & 110-160 & 160-210 & 210-260 & 260-310 & 310-400 & 400-500 & 500-600 \\ 
\hline
Measured cross section [pb/GeV]  & 3.5e+04 & 7.9e+03 & 1.4e+03 & 2.7e+02 & 43 & 22 & 8.8 & 2.0 & -- \\ 
NLO pQCD (CTEQ 6.6) $\times$ non-pert. corr. [pb/GeV]  & 4.1e+04 & 9.3e+03 & 1.6e+03 & 2.7e+02 & 66 & 21 & 6.2 & 1.4 & 0.36 \\ 
Non-perturbative correction  & 0.92 & 0.93 & 0.94 & 0.95 & 0.95 & 0.96 & 0.96 & 0.97 & 0.97 \\ 
\hline
Statistical uncertainty   & 0.011 & 0.020 & 0.036 & 0.085 & 0.21 & 0.30 & 0.35 & 0.71 & -- \\ 
Absolute JES uncertainty   & $^{+0.25}_{-0.22}$ & $^{+0.29}_{-0.21}$ & $^{+0.27}_{-0.24}$ & $^{+0.30}_{-0.24}$ & $^{+0.31}_{-0.23}$ & $^{+0.29}_{-0.28}$ & $^{+0.35}_{-0.25}$ & $^{+0.32}_{-0.26}$ & -- \\ 
Unfolding uncertainty   & 0.04 & 0.03 & 0.02 & 0.02 & 0.02 & 0.02 & 0.02 & 0.02 & -- \\ 
Total systematic uncertainty   & $^{+0.3}_{-0.2}$ & $^{+0.3}_{-0.2}$ & $^{+0.3}_{-0.2}$ & $^{+0.3}_{-0.2}$ & $^{+0.3}_{-0.2}$ & 0.3 & $^{+0.3}_{-0.2}$ & 0.3 & -- \\ 
\hline
PDF uncertainty   & 0.02  & 0.02 & 0.03 & 0.03  & 0.04  & 0.05 & $^{+0.06}_{-0.05}$ & $^{+0.08}_{-0.06}$  & $^{+0.09}_{-0.07}$  \\ 
Scale uncertainty   & $^{+0.006}_{-0.04}$  & $^{+0.004}_{-0.05}$  & $^{+0.003}_{-0.04}$  & $^{+0.003}_{-0.05}$  & $^{+0.005}_{-0.06}$  & $^{+0.007}_{-0.06}$  & $^{+0.01}_{-0.06}$  & $^{+0.02}_{-0.07}$  & $^{+0.02}_{-0.08}$  \\ 
$\alpha_{s}$ uncertainty   & 0.03  & 0.03  & 0.04 & 0.04  & 0.04  & 0.04  & 0.04  & 0.04  & 0.04  \\
Non-perturbative correction uncertainty   & $^{+0.06}_{-0}$  & $^{+0.05}_{-0}$  & $^{+0.04}_{-0}$  & $^{+0.04}_{-0}$  & $^{+0.03}_{-0}$  & $^{+0.03}_{-0}$  & $^{+0.03}_{-0}$  & $^{+0.03}_{-0}$  & $^{+0.02}_{-0}$  \\ 
Total theory uncertainty   & $^{+0.07}_{-0.05}$  & 0.06  & 0.06  & $^{+0.06}_{-0.07}$  & $^{+0.07}_{-0.08}$  & $^{+0.08}_{-0.09}$  & $^{+0.08}_{-0.09}$  & $^{+0.09}_{-0.1}$  & 0.1  \\ 
\hline

\multicolumn{10}{|l|}{$0.3< |y| <0.8$ } \\ 
\hline
$\pt$ [GeV]   & 60-80 & 80-110 & 110-160 & 160-210 & 210-260 & 260-310 & 310-400 & 400-500 & 500-600 \\ 
\hline
Measured cross section [pb/GeV]  & 3.4e+04 & 7.9e+03 & 1.2e+03 & 2.4e+02 & 49 & 15 & 9.3 & -- & -- \\ 
NLO pQCD (CTEQ 6.6) $\times$ non-pert. corr. [pb/GeV]  & 3.9e+04 & 9.1e+03 & 1.5e+03 & 2.6e+02 & 63 & 20 & 5.7 & 1.3 & 0.32 \\ 
Non-perturbative correction  & 0.94 & 0.95 & 0.95 & 0.96 & 0.96 & 0.96 & 0.96 & 0.97 & 0.97 \\ 
\hline
Statistical uncertainty   & 0.0089 & 0.015 & 0.030 & 0.069 & 0.15 & 0.28 & 0.27 & -- & -- \\ 
Absolute JES uncertainty   & $^{+0.27}_{-0.22}$ & $^{+0.28}_{-0.22}$ & $^{+0.28}_{-0.22}$ & $^{+0.27}_{-0.24}$ & $^{+0.31}_{-0.23}$ & $^{+0.29}_{-0.27}$ & $^{+0.34}_{-0.25}$ & -- & -- \\ 
Unfolding uncertainty   & 0.04 & 0.03 & 0.02 & 0.02 & 0.02 & 0.02 & 0.02 & -- & -- \\ 
Total systematic uncertainty   & $^{+0.3}_{-0.2}$ & $^{+0.3}_{-0.2}$ & $^{+0.3}_{-0.2}$ & $^{+0.3}_{-0.2}$ & $^{+0.3}_{-0.2}$ & 0.3 & $^{+0.3}_{-0.2}$ & -- & -- \\ 
\hline
PDF uncertainty   & 0.02  & 0.02  & 0.03  & 0.03  & 0.04  & 0.05  & $^{+0.07}_{-0.05}$  & $^{+0.08}_{-0.06}$  & $^{+0.1}_{-0.07}$  \\ 
Scale uncertainty   & $^{+0.006}_{-0.04}$  & $^{+0.005}_{-0.05}$  & $^{+0.002}_{-0.04}$  & $^{+0.003}_{-0.05}$  & $^{+0.004}_{-0.05}$  & $^{+0.006}_{-0.06}$  & $^{+0.007}_{-0.06}$  & $^{+0.01}_{-0.07}$  & $^{+0.02}_{-0.08}$  \\ 
$\alpha_{s}$ uncertainty   & 0.03  & 0.03  & 0.04 & 0.04  & 0.04 & 0.04 & 0.04 & 0.04 & 0.04 \\ 
Non-perturbative correction uncertainty   & $^{+0.05}_{-0}$  & $^{+0.03}_{-0.0004}$  & $^{+0.02}_{-0.002}$  & $^{+0.02}_{-0.001}$  & $^{+0.02}_{-0.0003}$  & $^{+0.02}_{-0}$  & $^{+0.02}_{-0}$  & $^{+0.02}_{-0}$  & $^{+0.02}_{-0}$  \\ 
Total theory uncertainty   & $^{+0.06}_{-0.05}$  & $^{+0.05}_{-0.06}$  & $^{+0.05}_{-0.06}$  & $^{+0.06}_{-0.07}$  & $^{+0.06}_{-0.08}$  & $^{+0.07}_{-0.09}$  & $^{+0.08}_{-0.09}$  & $^{+0.09}_{-0.1}$  & 0.1 \\ 
\hline

\end{tabular}
\end{center}
\caption{Measured inclusive jet double-differential cross section per GeV and per unit rapidity as a function of $\pt$ for \AKT jets with $R=0.4$, compared to NLO pQCD calculations corrected for non-perturbative effects.  All uncertainties listed are fractional uncertainties.  There is an additional overall uncertainty of 11\% due to the measurement of the integrated luminosity which is not included in the systematic uncertainties given above.  The statistical uncertainty is calculated as $1/\sqrt{N}$, where $N$ is the number of jets in a given bin.}
  \label{tab:inclusiveresults1a}
\end{sidewaystable*}

\begin{sidewaystable*}[htbp]
\begin{center}
\vspace{16cm}
\begin{tabular}{|c|c|c|c|c|c|c|c|c|c|}
\hline
\multicolumn{10}{|l|}{$0.8< |y| <1.2$ } \\ 
\hline
$\pt$ [GeV]   & 60-80 & 80-110 & 110-160 & 160-210 & 210-260 & 260-310 & 310-400 & 400-500 & 500-600 \\ 
\hline
Measured cross section [pb/GeV]  & 3.2e+04 & 6.8e+03 & 1.1e+03 & 2.1e+02 & 43 & 15 & 2.4 & 1.5 & -- \\ 
NLO pQCD (CTEQ 6.6) $\times$ non-pert. corr. [pb/GeV]  & 3.5e+04 & 8.3e+03 & 1.4e+03 & 2.2e+02 & 56 & 17 & 4.8 & 1.0 & 0.25 \\ 
Non-perturbative correction  & 0.92 & 0.93 & 0.95 & 0.95 & 0.96 & 0.96 & 0.97 & 0.97 & 0.97 \\ 
\hline
Statistical uncertainty   & 0.010 & 0.018 & 0.035 & 0.083 & 0.18 & 0.32 & 0.58 & 0.71 & -- \\ 
Absolute JES uncertainty   & $^{+0.27}_{-0.21}$ & $^{+0.26}_{-0.22}$ & $^{+0.29}_{-0.25}$ & $^{+0.32}_{-0.24}$ & $^{+0.29}_{-0.28}$ & $^{+0.38}_{-0.26}$ & $^{+0.34}_{-0.25}$ & $^{+0.35}_{-0.27}$ & -- \\ 
Unfolding uncertainty   & 0.04 & 0.03 & 0.02 & 0.02 & 0.02 & 0.02 & 0.02 & 0.03 & -- \\ 
Total systematic uncertainty   & $^{+0.3}_{-0.2}$ & $^{+0.3}_{-0.2}$ & $^{+0.3}_{-0.2}$ & $^{+0.3}_{-0.2}$ & 0.3 & $^{+0.4}_{-0.3}$ & 0.3 & $^{+0.4}_{-0.3}$ & -- \\ 
\hline
PDF uncertainty   & $^{+0.01}_{-0.02}$  & 0.02  & 0.03  & $^{+0.04}_{-0.03}$  & $^{+0.05}_{-0.04}$  & $^{+0.06}_{-0.05}$  & $^{+0.07}_{-0.05}$  & $^{+0.08}_{-0.06}$  & $^{+0.1}_{-0.08}$  \\ 
Scale uncertainty   & $^{+0.005}_{-0.04}$  & $^{+0.004}_{-0.05}$  & $^{+0.005}_{-0.05}$  & $^{+0.004}_{-0.05}$  & $^{+0.007}_{-0.06}$  & $^{+0.01}_{-0.06}$  & $^{+0.01}_{-0.07}$  & $^{+0.01}_{-0.07}$  & $^{+0.02}_{-0.08}$  \\ 
$\alpha_{s}$ uncertainty   & 0.03 & 0.03 & 0.04 & 0.04 & 0.04 & 0.04 & 0.04 & 0.04 & 0.04 \\ 
Non-perturbative correction uncertainty   & $^{+0.06}_{-0}$  & $^{+0.04}_{-0}$  & $^{+0.03}_{-0}$  & $^{+0.03}_{-0}$  & $^{+0.03}_{-0.0004}$  & $^{+0.03}_{-0.003}$  & $^{+0.03}_{-0.005}$  & $^{+0.03}_{-0.007}$  & $^{+0.02}_{-0.009}$  \\ 
Total theory uncertainty   & $^{+0.07}_{-0.05}$  & 0.06  & $^{+0.06}_{-0.07}$  & $^{+0.06}_{-0.07}$  & $^{+0.07}_{-0.08}$  & $^{+0.08}_{-0.09}$  & $^{+0.09}_{-0.1}$  & 0.1  & 0.1  \\ 
\hline

\multicolumn{10}{|l|}{$1.2< |y| <2.1$ } \\ 
\hline
$\pt$ [GeV]   & 60-80 & 80-110 & 110-160 & 160-210 & 210-260 & 260-310 & 310-400 & 400-500 & 500-600 \\ 
\hline
Measured cross section [pb/GeV]  & 2.5e+04 & 5.6e+03 & 9.2e+02 & 1.4e+02 & 27 & 9.0 & 2.2 & 0.32 & -- \\ 
NLO pQCD (CTEQ 6.6) $\times$ non-pert. corr. [pb/GeV]  & 3.0e+04 & 6.4e+03 & 1.0e+03 & 1.6e+02 & 37 & 11 & 2.6 & 0.45 & 0.086 \\ 
Non-perturbative correction  & 0.92 & 0.93 & 0.94 & 0.95 & 0.96 & 0.96 & 0.96 & 0.97 & 0.97 \\ 
\hline
Statistical uncertainty   & 0.0078 & 0.014 & 0.026 & 0.069 & 0.15 & 0.27 & 0.41 & 1.0 & -- \\ 
Absolute JES uncertainty   & $^{+0.29}_{-0.25}$ & $^{+0.30}_{-0.24}$ & $^{+0.31}_{-0.23}$ & $^{+0.31}_{-0.26}$ & $^{+0.34}_{-0.26}$ & $^{+0.33}_{-0.29}$ & $^{+0.40}_{-0.28}$ & $^{+0.43}_{-0.32}$ & -- \\ 
Unfolding uncertainty   & 0.03 & 0.03 & 0.02 & 0.02 & 0.02 & 0.02 & 0.02 & 0.02 & -- \\ 
Total systematic uncertainty   & 0.3 & $^{+0.3}_{-0.2}$ & $^{+0.3}_{-0.2}$ & 0.3 & 0.3 & 0.3 & $^{+0.4}_{-0.3}$ & $^{+0.4}_{-0.3}$ & -- \\ 
\hline
PDF uncertainty   & $^{+0.01}_{-0.02}$  & $^{+0.03}_{-0.02}$  & 0.03  & 0.04  & $^{+0.05}_{-0.04}$  & $^{+0.07}_{-0.05}$  & $^{+0.08}_{-0.06}$  & $^{+0.1}_{-0.07}$  & $^{+0.1}_{-0.09}$  \\ 
Scale uncertainty   & $^{+0.006}_{-0.04}$  & $^{+0.003}_{-0.05}$  & $^{+0.005}_{-0.05}$  & $^{+0.006}_{-0.05}$  & $^{+0.006}_{-0.06}$  & $^{+0.007}_{-0.06}$  & $^{+0.009}_{-0.07}$  & $^{+0.009}_{-0.07}$  & $^{+0.003}_{-0.07}$  \\ 
$\alpha_{s}$ uncertainty   & 0.03 & 0.04  & 0.04 & 0.04  & 0.04  & 0.04  & 0.04 & 0.04 & 0.04 \\ 
Non-perturbative correction uncertainty   & $^{+0.06}_{-0}$  & $^{+0.04}_{-0}$  & $^{+0.03}_{-0}$  & $^{+0.03}_{-0}$  & $^{+0.03}_{-0}$  & $^{+0.02}_{-0}$  & $^{+0.02}_{-0}$  & $^{+0.02}_{-0}$  & $^{+0.02}_{-0}$  \\ 
Total theory uncertainty   & $^{+0.07}_{-0.06}$  & 0.06  & $^{+0.06}_{-0.07}$  & $^{+0.07}_{-0.08}$  & $^{+0.07}_{-0.08}$  & $^{+0.08}_{-0.09}$  & 0.1  & 0.1  & 0.1  \\ 
\hline

\end{tabular}
\end{center}
\caption{Measured inclusive jet double-differential cross section  per GeV and per unit rapidity as a function of $\pt$ for \AKT jets with $R=0.4$, compared to NLO pQCD calculations corrected for non-perturbative effects.    All uncertainties listed are fractional uncertainties.  There is an additional overall uncertainty of 11\% due to the measurement of the integrated luminosity which is not included in the systematic uncertainies given above.  The statistical uncertainty is calculated as $1/\sqrt{N}$, where $N$ is the number of jets in a given bin.}
  \label{tab:inclusiveresults1b}
\end{sidewaystable*}

\begin{sidewaystable*}[htbp]
\begin{center}
\vspace{16cm}
\begin{tabular}{|c|c|c|c|c|c|c|c|c|c|}
\hline
\multicolumn{10}{|l|}{$2.1< |y| <2.8$ } \\ 
\hline
$\pt$ [GeV]   & 60-80 & 80-110 & 110-160 & 160-210 & 210-260 & 260-310 & 310-400 & 400-500 & 500-600 \\ 
\hline
Measured cross section [pb/GeV]  & 1.6e+04 & 3.2e+03 & 4.6e+02 & 52 & 9.3 & 1.7 & 0.48 & -- & -- \\ 
NLO pQCD (CTEQ 6.6) $\times$ non-pert. corr. [pb/GeV]  & 1.9e+04 & 3.9e+03 & 5.5e+02 & 66 & 11 & 2.2 & 0.32 & 0.021 & 0.00097 \\ 
Non-perturbative correction  & 0.91 & 0.92 & 0.93 & 0.94 & 0.94 & 0.95 & 0.95 & 0.96 & 0.96 \\ 
\hline
Statistical uncertainty   & 0.011 & 0.021 & 0.043 & 0.13 & 0.30 & 0.71 & 1.0 & -- & -- \\ 
Absolute JES uncertainty   & $^{+0.37}_{-0.27}$ & $^{+0.38}_{-0.29}$ & $^{+0.39}_{-0.31}$ & $^{+0.45}_{-0.33}$ & $^{+0.56}_{-0.36}$ & $^{+0.58}_{-0.43}$ & $^{+0.70}_{-0.44}$ & -- & -- \\ 
Unfolding uncertainty   & 0.03 & 0.03 & 0.02 & 0.03 & 0.03 & 0.03 & 0.03 & -- & -- \\ 
Total systematic uncertainty   & $^{+0.4}_{-0.3}$ & $^{+0.4}_{-0.3}$ & $^{+0.4}_{-0.3}$ & $^{+0.4}_{-0.3}$ & $^{+0.6}_{-0.4}$ & $^{+0.6}_{-0.4}$ & $^{+0.7}_{-0.4}$ & -- & -- \\ 
\hline
PDF uncertainty   & $^{+0.03}_{-0.02}$  & $^{+0.04}_{-0.03}$  & $^{+0.06}_{-0.05}$  & $^{+0.08}_{-0.06}$  & $^{+0.09}_{-0.06}$  & $^{+0.1}_{-0.08}$  & $^{+0.2}_{-0.1}$  & $^{+0.3}_{-0.2}$  & $^{+1}_{-2}$  \\ 
Scale uncertainty   & $^{+0.004}_{-0.04}$  & $^{+0.004}_{-0.05}$  & $^{+0.0009}_{-0.07}$  & $^{+0}_{-0.09}$  & $^{+0}_{-0.09}$  & $^{+0}_{-0.1}$  & $^{+0}_{-0.1}$  & $^{+0.003}_{-0.2}$  & $^{+0.3}_{-0.7}$  \\ 
$\alpha_{s}$ uncertainty   & 0.04  & 0.04  & 0.04  & 0.04  & 0.04  & 0.04  & 0.04  & 0.03  & 0.1  \\ 
Non-perturbative correction uncertainty   & $^{+0.07}_{-0}$  & $^{+0.05}_{-0}$  & $^{+0.04}_{-0}$  & $^{+0.04}_{-0}$  & $^{+0.04}_{-0}$  & $^{+0.03}_{-0}$  & $^{+0.03}_{-0}$  & $^{+0.03}_{-0}$  & $^{+0.03}_{-0}$  \\ 
Total theory uncertainty   & $^{+0.08}_{-0.06}$  & $^{+0.08}_{-0.07}$  & $^{+0.08}_{-0.1}$  & $^{+0.09}_{-0.1}$  & 0.1  & $^{+0.1}_{-0.2}$  & 0.2  & 0.3  & $^{+1}_{-2}$  \\ 
\hline

\end{tabular}
\end{center}
\caption{Measured inclusive jet double-differential cross section  per GeV and per unit rapidity as a function of $\pt$ for \AKT jets with $R=0.4$, compared to NLO pQCD calculations corrected for non-perturbative effects.    All uncertainties listed are fractional uncertainties.  There is an additional overall uncertainty of 11\% due to the measurement of the integrated luminosity which is not included in the systematic uncertainies given above.  The statistical uncertainty is calculated as $1/\sqrt{N}$, where $N$ is the number of jets in a given bin.}
  \label{tab:inclusiveresults1c}
\end{sidewaystable*}


\begin{sidewaystable*}[htbp]
\begin{center}
\vspace{16cm}
\begin{tabular}{|c|c|c|c|c|c|c|c|c|c|}
\hline
\multicolumn{10}{|l|}{$0< |y| <0.3$ } \\ 
\hline
$\pt$ [GeV]   & 60-80 & 80-110 & 110-160 & 160-210 & 210-260 & 260-310 & 310-400 & 400-500 & 500-600 \\ 
\hline
Measured cross section [pb/GeV]  & 5.0e+04 & 1.1e+04 & 1.9e+03 & 3.4e+02 & 49 & 22 & 11 & 1.0 & 1.0 \\ 
NLO pQCD (CTEQ 6.6) $\times$ non-pert. corr. [pb/GeV]  & 5.4e+04 & 1.1e+04 & 1.9e+03 & 3.1e+02 & 78 & 25 & 7.0 & 1.5 & 0.41 \\ 
Non-perturbative correction  & 1.0 & 1.0 & 1.0 & 1.0 & 0.99 & 0.99 & 0.99 & 0.99 & 0.99 \\ 
\hline
Statistical uncertainty   & 0.0096 & 0.017 & 0.032 & 0.076 & 0.20 & 0.30 & 0.32 & 1.0 & 1.0 \\ 
Absolute JES uncertainty   & $^{+0.41}_{-0.28}$ & $^{+0.35}_{-0.24}$ & $^{+0.30}_{-0.26}$ & $^{+0.33}_{-0.27}$ & $^{+0.36}_{-0.24}$ & $^{+0.31}_{-0.29}$ & $^{+0.37}_{-0.26}$ & $^{+0.35}_{-0.26}$ & $^{+0.34}_{-0.3}$ \\ 
Unfolding uncertainty   & 0.04 & 0.03 & 0.03 & 0.03 & 0.02 & 0.02 & 0.02 & 0.02 & 0.03 \\ 
Total systematic uncertainty   & $^{+0.4}_{-0.3}$ & $^{+0.4}_{-0.2}$ & 0.3 & 0.3 & $^{+0.4}_{-0.2}$ & 0.3 & $^{+0.4}_{-0.3}$ & $^{+0.4}_{-0.3}$ & 0.3 \\ 
\hline
PDF uncertainty   & 0.02  & 0.02  & 0.03  & 0.03  & 0.04  & 0.05  & $^{+0.07}_{-0.05}$  & $^{+0.08}_{-0.06}$  & $^{+0.09}_{-0.07}$  \\ 
Scale uncertainty   & $^{+0.02}_{-0.06}$  & $^{+0.02}_{-0.07}$  & $^{+0.02}_{-0.07}$  & $^{+0.03}_{-0.07}$  & $^{+0.03}_{-0.08}$  & $^{+0.02}_{-0.07}$  & $^{+0.04}_{-0.09}$  & $^{+0.04}_{-0.09}$  & $^{+0.04}_{-0.09}$  \\ 
$\alpha_{s}$ uncertainty   & 0.03  & 0.03 & 0.04 & 0.04  & 0.04  & 0.04  & 0.05  & 0.05  & 0.05  \\ 
Non-perturbative correction uncertainty   & $^{+0.06}_{-0}$  & $^{+0.05}_{-0}$  & $^{+0.04}_{-0}$  & $^{+0.03}_{-0}$  & $^{+0.03}_{-0}$  & $^{+0.02}_{-0.001}$  & $^{+0.02}_{-0.003}$  & $^{+0.02}_{-0.005}$  & $^{+0.01}_{-0.006}$  \\ 
Total theory uncertainty   & 0.07  & $^{+0.07}_{-0.08}$  & $^{+0.07}_{-0.08}$  & $^{+0.07}_{-0.09}$  & $^{+0.07}_{-0.09}$  & $^{+0.08}_{-0.1}$  & $^{+0.09}_{-0.1}$  & 0.1  & 0.1 \\ 
\hline

\multicolumn{10}{|l|}{$0.3< |y| <0.8$ } \\ 
\hline
$\pt$ [GeV]   & 60-80 & 80-110 & 110-160 & 160-210 & 210-260 & 260-310 & 310-400 & 400-500 & 500-600 \\ 
\hline
Measured cross section [pb/GeV]  & 4.8e+04 & 1.1e+04 & 1.7e+03 & 3.0e+02 & 64 & 17 & 10 & 0.60 & -- \\ 
NLO pQCD (CTEQ 6.6) $\times$ non-pert. corr. [pb/GeV]  & 5.2e+04 & 1.1e+04 & 1.8e+03 & 3.0e+02 & 73 & 23 & 6.5 & 1.4 & 0.36 \\ 
Non-perturbative correction  & 1.1 & 1.0 & 1.0 & 1.0 & 1.0 & 0.99 & 0.99 & 0.99 & 0.99 \\ 
\hline
Statistical uncertainty   & 0.0074 & 0.013 & 0.026 & 0.062 & 0.13 & 0.27 & 0.26 & 1.0 & -- \\ 
Absolute JES uncertainty   & $^{+0.39}_{-0.27}$ & $^{+0.35}_{-0.26}$ & $^{+0.33}_{-0.23}$ & $^{+0.29}_{-0.27}$ & $^{+0.36}_{-0.24}$ & $^{+0.32}_{-0.29}$ & $^{+0.37}_{-0.27}$ & $^{+0.37}_{-0.27}$ & -- \\ 
Unfolding uncertainty   & 0.04 & 0.03 & 0.03 & 0.02 & 0.02 & 0.03 & 0.02 & 0.02 & -- \\ 
Total systematic uncertainty   & $^{+0.4}_{-0.3}$ & $^{+0.4}_{-0.3}$ & $^{+0.3}_{-0.2}$ & 0.3 & $^{+0.4}_{-0.2}$ & 0.3 & $^{+0.4}_{-0.3}$ & $^{+0.4}_{-0.3}$ & -- \\ 
\hline
PDF uncertainty   & 0.02  & 0.02  & 0.03  & 0.03  & 0.04  & $^{+0.06}_{-0.05}$  & $^{+0.07}_{-0.05}$  & $^{+0.08}_{-0.06}$  & $^{+0.1}_{-0.07}$  \\ 
Scale uncertainty   & $^{+0.02}_{-0.06}$  & $^{+0.02}_{-0.07}$  & $^{+0.02}_{-0.07}$  & $^{+0.02}_{-0.07}$  & $^{+0.03}_{-0.08}$  & $^{+0.03}_{-0.08}$  & $^{+0.03}_{-0.08}$  & $^{+0.03}_{-0.09}$  & $^{+0.04}_{-0.1}$  \\ 
$\alpha_{s}$ uncertainty   & 0.03  & 0.03  & 0.04  & 0.04  & 0.04  & 0.04  & 0.04  & 0.05  & 0.05  \\ 
Non-perturbative correction uncertainty   & $^{+0.04}_{-0.02}$  & $^{+0.02}_{-0.03}$  & 0.02  & $^{+0.02}_{-0.01}$  & $^{+0.02}_{-0.006}$  & $^{+0.02}_{-0.003}$  & $^{+0.02}_{-0}$  & $^{+0.02}_{-0}$  & $^{+0.02}_{-0}$  \\ 
Total theory uncertainty   & $^{+0.06}_{-0.08}$  & $^{+0.05}_{-0.08}$  & $^{+0.06}_{-0.09}$  & $^{+0.06}_{-0.09}$  & $^{+0.07}_{-0.1}$  & $^{+0.08}_{-0.1}$  & $^{+0.09}_{-0.1}$  & 0.1 & 0.1  \\ 
\hline

\end{tabular}
\end{center}
\caption{Measured inclusive jet double-differential cross section  per GeV and per unit rapidity as a function of $\pt$ for \AKT jets with $R=0.6$, compared to NLO pQCD calculations corrected for non-perturbative effects. All uncertainties listed are fractional uncertainties.  There is an additional overall uncertainty of 11\% due to the measurement of the integrated luminosity which is not included in the systematic uncertainies given above.  The statistical uncertainty is calculated as $1/\sqrt{N}$, where $N$ is the number of jets in a given bin.}
\label{tab:inclusiveresults2a}
\end{sidewaystable*}

\begin{sidewaystable*}[htbp]
\begin{center}
\vspace{16cm}
\begin{tabular}{|c|c|c|c|c|c|c|c|c|c|}
\hline
\multicolumn{10}{|l|}{$0.8< |y| <1.2$ } \\ 
\hline
$\pt$ [GeV]   & 60-80 & 80-110 & 110-160 & 160-210 & 210-260 & 260-310 & 310-400 & 400-500 & 500-600 \\ 
\hline
Measured cross section [pb/GeV]  & 4.5e+04 & 9.5e+03 & 1.4e+03 & 2.7e+02 & 53 & 16 & 3.2 & 2.2 & -- \\ 
NLO pQCD (CTEQ 6.6) $\times$ non-pert. corr. [pb/GeV]  & 4.8e+04 & 1.0e+04 & 1.7e+03 & 2.6e+02 & 63 & 19 & 5.4 & 1.1 & 0.27 \\ 
Non-perturbative correction  & 1.1 & 1.0 & 1.0 & 1.0 & 0.99 & 0.99 & 0.99 & 0.99 & 0.99 \\ 
\hline
Statistical uncertainty   & 0.0086 & 0.015 & 0.031 & 0.073 & 0.16 & 0.30 & 0.50 & 0.58 & -- \\ 
Absolute JES uncertainty   & $^{+0.37}_{-0.27}$ & $^{+0.38}_{-0.28}$ & $^{+0.37}_{-0.27}$ & $^{+0.36}_{-0.25}$ & $^{+0.33}_{-0.28}$ & $^{+0.38}_{-0.27}$ & $^{+0.35}_{-0.27}$ & $^{+0.38}_{-0.28}$ & -- \\ 
Unfolding uncertainty   & 0.04 & 0.04 & 0.03 & 0.02 & 0.02 & 0.03 & 0.02 & 0.02 & -- \\ 
Total systematic uncertainty   & $^{+0.4}_{-0.3}$ & $^{+0.4}_{-0.3}$ & $^{+0.4}_{-0.3}$ & $^{+0.4}_{-0.3}$ & 0.3 & $^{+0.4}_{-0.3}$ & $^{+0.4}_{-0.3}$ & $^{+0.4}_{-0.3}$ & -- \\ 
\hline
PDF uncertainty   & $^{+0.01}_{-0.02}$  & 0.02  & 0.03  & $^{+0.04}_{-0.03}$  & $^{+0.05}_{-0.04}$  & $^{+0.06}_{-0.05}$  & $^{+0.07}_{-0.06}$  & $^{+0.08}_{-0.06}$  & $^{+0.1}_{-0.08}$  \\ 
Scale uncertainty   & $^{+0.03}_{-0.07}$  & $^{+0.02}_{-0.07}$  & $^{+0.03}_{-0.07}$  & $^{+0.02}_{-0.07}$  & $^{+0.03}_{-0.08}$  & $^{+0.03}_{-0.08}$  & $^{+0.03}_{-0.08}$  & $^{+0.03}_{-0.09}$  & $^{+0.03}_{-0.1}$  \\ 
$\alpha_{s}$ uncertainty   & 0.03  & 0.04  & 0.04  & 0.04  & 0.04  & 0.04  & 0.04  & 0.04 & 0.04  \\ 
Non-perturbative correction uncertainty   & $^{+0.05}_{-0}$  & $^{+0.03}_{-0.009}$  & $^{+0.03}_{-0.01}$  & $^{+0.03}_{-0.01}$  & $^{+0.02}_{-0.01}$  & $^{+0.02}_{-0.01}$  & $^{+0.02}_{-0.01}$  & $^{+0.02}_{-0.009}$  & $^{+0.02}_{-0.009}$  \\ 
Total theory uncertainty   & $^{+0.07}_{-0.08}$  & $^{+0.06}_{-0.08}$  & $^{+0.06}_{-0.09}$  & $^{+0.06}_{-0.09}$  & $^{+0.07}_{-0.1}$  & $^{+0.08}_{-0.1}$  & $^{+0.09}_{-0.1}$  & 0.1 & 0.1  \\ 
\hline

\multicolumn{10}{|l|}{$1.2< |y| <2.1$ } \\ 
\hline
$\pt$ [GeV]   & 60-80 & 80-110 & 110-160 & 160-210 & 210-260 & 260-310 & 310-400 & 400-500 & 500-600 \\ 
\hline
Measured cross section [pb/GeV]  & 3.6e+04 & 7.6e+03 & 1.2e+03 & 1.8e+02 & 36 & 10 & 2.9 & 0.32 & -- \\ 
NLO pQCD (CTEQ 6.6) $\times$ non-pert. corr. [pb/GeV]  & 3.8e+04 & 7.8e+03 & 1.2e+03 & 1.8e+02 & 42 & 12 & 3.0 & 0.50 & 0.095 \\ 
Non-perturbative correction  & 1.1 & 1.0 & 1.0 & 1.0 & 0.99 & 0.99 & 0.99 & 0.99 & 0.99 \\ 
\hline
Statistical uncertainty   & 0.0065 & 0.012 & 0.023 & 0.060 & 0.13 & 0.25 & 0.35 & 1.0 & -- \\ 
Absolute JES uncertainty   & $^{+0.43}_{-0.29}$ & $^{+0.37}_{-0.27}$ & $^{+0.36}_{-0.25}$ & $^{+0.33}_{-0.29}$ & $^{+0.40}_{-0.27}$ & $^{+0.35}_{-0.32}$ & $^{+0.44}_{-0.30}$ & $^{+0.47}_{-0.33}$ & -- \\ 
Unfolding uncertainty   & 0.04 & 0.03 & 0.03 & 0.02 & 0.02 & 0.02 & 0.02 & 0.02 & -- \\ 
Total systematic uncertainty   & $^{+0.4}_{-0.3}$ & $^{+0.4}_{-0.3}$ & $^{+0.4}_{-0.3}$ & 0.3 & $^{+0.4}_{-0.3}$ & 0.3 & $^{+0.4}_{-0.3}$ & $^{+0.5}_{-0.3}$ & -- \\ 
\hline
PDF uncertainty   & $^{+0.01}_{-0.02}$  & $^{+0.03}_{-0.02}$  & $^{+0.04}_{-0.03}$  & 0.04  & $^{+0.06}_{-0.04}$  & $^{+0.07}_{-0.05}$  & $^{+0.09}_{-0.06}$  & $^{+0.1}_{-0.07}$  & $^{+0.1}_{-0.09}$  \\ 
Scale uncertainty   & $^{+0.02}_{-0.06}$  & $^{+0.02}_{-0.07}$  & $^{+0.02}_{-0.07}$  & $^{+0.03}_{-0.08}$  & $^{+0.02}_{-0.08}$  & $^{+0.02}_{-0.08}$  & $^{+0.03}_{-0.09}$  & $^{+0.02}_{-0.09}$  & $^{+0.02}_{-0.1}$  \\ 
$\alpha_{s}$ uncertainty   & 0.04  & 0.04 & 0.04  & 0.04  & 0.04 & 0.04 & 0.04 & 0.05 & 0.04 \\ 
Non-perturbative correction uncertainty   & $^{+0.04}_{-0.008}$  & $^{+0.03}_{-0.01}$  & $^{+0.02}_{-0.007}$  & $^{+0.02}_{-0.004}$  & $^{+0.02}_{-0.002}$  & $^{+0.02}_{-0.0002}$  & $^{+0.02}_{-0}$  & $^{+0.02}_{-0}$  & $^{+0.02}_{-0}$  \\ 
Total theory uncertainty   & $^{+0.06}_{-0.07}$  & $^{+0.06}_{-0.08}$  & $^{+0.06}_{-0.09}$  & $^{+0.07}_{-0.1}$  & $^{+0.08}_{-0.1}$  & $^{+0.09}_{-0.1}$  & 0.1 & 0.1  & 0.1  \\ 
\hline

\end{tabular}
\end{center}
\caption{Measured inclusive jet double-differential cross section per GeV and per unit rapidity as a function of $\pt$ for \AKT jets with $R=0.6$, compared to NLO pQCD calculations corrected for non-perturbative effects.    All uncertainties listed are fractional uncertainties.  There is an additional overall uncertainty of 11\% due to the measurement of the integrated luminosity which is not included in the systematic uncertainies given above.  The statistical uncertainty is calculated as $1/\sqrt{N}$, where $N$ is the number of jets in a given bin.}
  \label{tab:inclusiveresults2b}
\end{sidewaystable*}

\begin{sidewaystable*}[htbp]
\begin{center}
\vspace{16cm}
\begin{tabular}{|c|c|c|c|c|c|c|c|c|c|}
\hline

\multicolumn{10}{|l|}{$2.1< |y| <2.8$ } \\ 
\hline
$\pt$ [GeV]   & 60-80 & 80-110 & 110-160 & 160-210 & 210-260 & 260-310 & 310-400 & 400-500 & 500-600 \\ 
\hline
Measured cross section [pb/GeV]  & 2.3e+04 & 4.3e+03 & 5.7e+02 & 68 & 10 & 2.5 & 0.49 & -- & -- \\ 
NLO pQCD (CTEQ 6.6) $\times$ non-pert. corr. [pb/GeV]  & 2.4e+04 & 4.6e+03 & 6.3e+02 & 76 & 12 & 2.6 & 0.37 & 0.023 & 0.00096 \\ 
Non-perturbative correction  & 1.0 & 1.0 & 1.0 & 0.99 & 0.99 & 0.99 & 0.99 & 0.99 & 0.99 \\ 
\hline
Statistical uncertainty   & 0.0096 & 0.018 & 0.038 & 0.11 & 0.29 & 0.58 & 1.0 & -- & -- \\ 
Absolute JES uncertainty   & $^{+0.49}_{-0.32}$ & $^{+0.46}_{-0.33}$ & $^{+0.45}_{-0.32}$ & $^{+0.47}_{-0.36}$ & $^{+0.60}_{-0.38}$ & $^{+0.62}_{-0.45}$ & $^{+0.78}_{-0.46}$ & -- & -- \\ 
Unfolding uncertainty   & 0.03 & 0.03 & 0.03 & 0.03 & 0.03 & 0.04 & 0.03 & -- & -- \\ 
Total systematic uncertainty   & $^{+0.5}_{-0.3}$ & $^{+0.5}_{-0.3}$ & $^{+0.5}_{-0.3}$ & $^{+0.5}_{-0.4}$ & $^{+0.6}_{-0.4}$ & $^{+0.6}_{-0.5}$ & $^{+0.8}_{-0.5}$ & -- & -- \\ 
\hline
PDF uncertainty   & $^{+0.03}_{-0.02}$  & $^{+0.05}_{-0.04}$  & $^{+0.06}_{-0.05}$  & $^{+0.08}_{-0.06}$  & $^{+0.1}_{-0.07}$  & $^{+0.1}_{-0.09}$  & $^{+0.2}_{-0.1}$  & $^{+0.3}_{-0.2}$  & 0.8  \\ 
Scale uncertainty   & $^{+0.01}_{-0.06}$  & $^{+0.02}_{-0.07}$  & $^{+0.01}_{-0.07}$  & $^{+0.02}_{-0.08}$  & $^{+0.02}_{-0.08}$  & $^{+0.01}_{-0.08}$  & $^{+0}_{-0.2}$  & $^{+0.02}_{-0.3}$  & $^{+0.1}_{-0.4}$  \\ 
$\alpha_{s}$ uncertainty   & 0.04  & 0.04  & 0.04  & 0.04  & 0.05  & 0.05  & 0.05  & 0.04  & 0.1  \\ 
Non-perturbative correction uncertainty   & $^{+0.04}_{-0}$  & $^{+0.03}_{-0}$  & $^{+0.03}_{-0}$  & $^{+0.03}_{-0}$  & $^{+0.02}_{-0.001}$  & $^{+0.02}_{-0.01}$  & 0.02  & $^{+0.02}_{-0.03}$  & $^{+0.02}_{-0.04}$  \\ 
Total theory uncertainty   & $^{+0.07}_{-0.08}$  & $^{+0.07}_{-0.09}$  & $^{+0.08}_{-0.09}$  & $^{+0.09}_{-0.1}$  & 0.1 & 0.1 & 0.2  & 0.3  & $^{+0.8}_{-1}$  \\ 
\hline

\end{tabular}
\end{center}
\caption{Measured inclusive jet double-differential cross section  per GeV and per unit rapidity as a function of $\pt$ for \AKT jets with $R=0.6$, compared to NLO pQCD calculations corrected for non-perturbative effects.    All uncertainties listed are fractional uncertainties.  There is an additional overall uncertainty of 11\% due to the measurement of the integrated luminosity which is not included in the systematic uncertainies given above.  The statistical uncertainty is calculated as $1/\sqrt{N}$, where $N$ is the number of jets in a given bin.}
  \label{tab:inclusiveresults2c}
\end{sidewaystable*}

\clearpage


\begin{sidewaystable*}
\begin{center}
\vspace{16cm}
\begin{tabular}{|c|c|c|c|c|c|c|c|c|c|c|c|}
\hline
\multicolumn{12}{|l|}{$     |y|_{\mathrm{max}} <0.3$ } \\
\hline
$\twomass{1}{2}$ [GeV] & 110-160 & 160-210 & 210-260 & 260-310 & 310-370 & 370-440 & 440-510 & 510-590 & 590-670 & 670-760 & 760-850 \\
\hline
Measured cross section [pb/GeV] & 2000 & 500 & 170 & 73 & 34 & 6.6 & 6.5 & -- & -- & 5.2 & -- \\
NLO pQCD (CTEQ 6.6) $\times$ non-pert. corr. [pb/GeV] & 3500 & 480 & 150 & 56 & 24 & 9.2 & 4.0 & 1.8 & 0.87 & 0.42 & 0.21 \\
Non-perturbative correction & 0.94 & 0.98 & 0.99 & 0.97 & 0.98 & 0.96 & 0.97 & 0.99 & 0.98 & 0.98 & 0.96 \\
\hline
Statistical uncertainty & 0.047 & 0.094 & 0.17 & 0.25 & 0.33 & 0.71 & 0.71 & -- & -- & 0.71 & -- \\
Absolute JES uncertainty & $^{+0.23}_{-0.21}$ & $^{+0.25}_{-0.22}$ & $^{+0.26}_{-0.23}$ & $^{+0.27}_{-0.23}$ & $^{+0.28}_{-0.24}$ & $^{+0.29}_{-0.24}$ & $^{+0.29}_{-0.24}$ & -- & -- & $^{+0.31}_{-0.25}$ & -- \\
Relative JES uncertainty & $^{+0.0031}_{-0}$ & $^{+0.0046}_{-0.0016}$ & $^{+0.0056}_{-0.0028}$ & $^{+0.0063}_{-0.0036}$ & $^{+0.0069}_{-0.0043}$ & $^{+0.0073}_{-0.0050}$ & $^{+0.0075}_{-0.0054}$ & -- & -- & $^{+0.0073}_{-0.0062}$ & -- \\
Unfolding uncertainty & 0.1 & 0.1 & 0.1 & 0.1 & 0.1 & 0.1 & 0.1 & -- & -- & 0.1 & -- \\
Total systematic uncertainty & $^{+0.3}_{-0.2}$ & $^{+0.3}_{-0.2}$ & $^{+0.3}_{-0.2}$ & 0.3 & 0.3 & 0.3 & 0.3 & -- & -- & 0.3 & -- \\
\hline
PDF uncertainty & 0.03 & $^{+0.03}_{-0.04}$ & $^{+0.03}_{-0.04}$ & 0.03 & 0.03 & $^{+0.02}_{-0.03}$ & $^{+0.02}_{-0.03}$ & 0.03 & 0.03 & 0.03 & 0.04 \\
Scale uncertainty & $^{+0.4}_{-0.3}$ & $^{+0.2}_{-0.1}$ & $^{+0.01}_{-0.04}$ & $^{+0.02}_{-0.1}$ & $^{+0.02}_{-0.1}$ & $^{+0.01}_{-0.1}$ & $^{+0.01}_{-0.1}$ & $^{+0.01}_{-0.1}$ & $^{+0.01}_{-0.1}$ & $^{+0.01}_{-0.09}$ & $^{+0.009}_{-0.09}$ \\
$\alpha_{s}$ uncertainty & 0.04 & 0.03 & 0.02 & 0.02 & 0.02 & 0.03 & 0.03 & 0.03 & 0.03 & 0.04 & 0.04 \\
Non-perturbative correction uncertainty & 0.05 & 0.05 & 0.05 & 0.05 & 0.05 & 0.05 & 0.05 & 0.05 & 0.05 & 0.05 & 0.05 \\
Total theory uncertainty & $^{+0.4}_{-0.3}$ & 0.2 & $^{+0.06}_{-0.08}$ & $^{+0.07}_{-0.1}$ & $^{+0.06}_{-0.1}$ & $^{+0.06}_{-0.1}$ & $^{+0.06}_{-0.1}$ & $^{+0.07}_{-0.1}$ & $^{+0.07}_{-0.1}$ & $^{+0.07}_{-0.1}$ & $^{+0.07}_{-0.1}$ \\
\hline
\end{tabular}
\end{center}
\caption{Measured dijet double-differential cross section per GeV and per unit absolute rapidity as a function of dijet mass $\twomass{1}{2}$ for \AKT jets with $R=0.4$, compared to NLO pQCD calculations corrected for non-perturbative effects.  All uncertainties listed are fractional uncertainties.  There is an additional overall uncertainty of 11\% due to the measurement of the integrated luminosity which is not included in the systematic uncertainties given above.  The statistical uncertainty is calculated as $1/\sqrt{N}$, where $N$ is the number of events in a given bin.}
\label{tab:DijetMassResults04_0}
\end{sidewaystable*}

\begin{sidewaystable*}
\begin{center}
\vspace{16cm}
\begin{tabular}{|c|c|c|c|c|c|c|c|c|c|c|}
\hline
\multicolumn{11}{|l|}{$0.3< |y|_{\mathrm{max}} <0.8$ } \\
\hline
$\twomass{1}{2}$ [GeV] & 160-210 & 210-260 & 260-310 & 310-370 & 370-440 & 440-510 & 510-590 & 590-670 & 670-760 & 760-850 \\
\hline
Measured cross section [pb/GeV] & 2300 & 680 & 280 & 100 & 57 & 18 & 14 & 5.1 & -- & 4.6 \\
NLO pQCD (CTEQ 6.6) $\times$ non-pert. corr. [pb/GeV] & 2200 & 680 & 250 & 110 & 43 & 18 & 8.3 & 3.9 & 2.0 & 0.99 \\
Non-perturbative correction & 0.98 & 0.97 & 0.96 & 0.97 & 0.97 & 0.98 & 0.98 & 0.97 & 0.98 & 0.98 \\
\hline
Statistical uncertainty & 0.034 & 0.063 & 0.098 & 0.15 & 0.19 & 0.33 & 0.35 & 0.58 & -- & 0.58 \\
Absolute JES uncertainty & $^{+0.25}_{-0.22}$ & $^{+0.26}_{-0.22}$ & $^{+0.26}_{-0.22}$ & $^{+0.27}_{-0.23}$ & $^{+0.28}_{-0.23}$ & $^{+0.29}_{-0.23}$ & $^{+0.30}_{-0.24}$ & $^{+0.30}_{-0.24}$ & -- & $^{+0.32}_{-0.25}$ \\
Relative JES uncertainty & $^{+0.018}_{-0.017}$ & 0.019 & $^{+0.019}_{-0.020}$ & $^{+0.020}_{-0.021}$ & $^{+0.020}_{-0.022}$ & $^{+0.021}_{-0.022}$ & $^{+0.022}_{-0.023}$ & $^{+0.023}_{-0.024}$ & -- & 0.025 \\
Unfolding uncertainty & 0.1 & 0.1 & 0.1 & 0.1 & 0.1 & 0.1 & 0.1 & 0.1 & -- & 0.1 \\
Total systematic uncertainty & $^{+0.3}_{-0.2}$ & $^{+0.3}_{-0.2}$ & $^{+0.3}_{-0.2}$ & $^{+0.3}_{-0.2}$ & 0.3 & 0.3 & 0.3 & 0.3 & -- & 0.3 \\
\hline
PDF uncertainty & $^{+0.03}_{-0.04}$ & $^{+0.03}_{-0.04}$ & 0.03 & $^{+0.02}_{-0.03}$ & $^{+0.02}_{-0.03}$ & 0.02 & $^{+0.02}_{-0.03}$ & 0.03 & 0.03 & 0.04 \\
Scale uncertainty & 0.2 & $^{+0.01}_{-0.04}$ & $^{+0.03}_{-0.1}$ & $^{+0.02}_{-0.1}$ & $^{+0.02}_{-0.1}$ & $^{+0.02}_{-0.1}$ & $^{+0.02}_{-0.1}$ & $^{+0.01}_{-0.1}$ & $^{+0.01}_{-0.1}$ & $^{+0.01}_{-0.1}$ \\
$\alpha_{s}$ uncertainty & 0.03 & 0.02 & 0.02 & 0.02 & 0.03 & 0.03 & 0.03 & 0.03 & 0.04 & 0.04 \\
Non-perturbative correction uncertainty & 0.05 & 0.05 & 0.05 & 0.05 & 0.05 & 0.05 & 0.05 & 0.05 & 0.05 & 0.05 \\
Total theory uncertainty & 0.2  & $^{+0.06}_{-0.07}$ & $^{+0.07}_{-0.1}$ & $^{+0.06}_{-0.1}$ & $^{+0.06}_{-0.1}$ & $^{+0.06}_{-0.1}$ & $^{+0.07}_{-0.1}$ & $^{+0.07}_{-0.1}$ & $^{+0.07}_{-0.1}$ & $^{+0.07}_{-0.1}$ \\
\hline
\end{tabular}
\end{center}
\caption{Measured dijet double-differential cross section per GeV and per unit absolute rapidity as a function of dijet mass $\twomass{1}{2}$ for \AKT jets with $R=0.4$, compared to NLO pQCD calculations corrected for non-perturbative effects.  All uncertainties listed are fractional uncertainties.  There is an additional overall uncertainty of 11\% due to the measurement of the integrated luminosity which is not included in the systematic uncertainties given above.  The statistical uncertainty is calculated as $1/\sqrt{N}$, where $N$ is the number of events in a given bin.}
\label{tab:DijetMassResults04_1}
\end{sidewaystable*}

\begin{sidewaystable*}
\begin{center}
\vspace{16cm}
\begin{tabular}{|c|c|c|c|c|c|c|c|c|c|c|}
\hline
\multicolumn{11}{|l|}{$0.8< |y|_{\mathrm{max}} <1.2$ } \\
\hline
$\twomass{1}{2}$ [GeV] & 210-260 & 260-310 & 310-370 & 370-440 & 440-510 & 510-590 & 590-670 & 670-760 & 760-850 & 850-950 \\
\hline
Measured cross section [pb/GeV] & 2100 & 730 & 300 & 120 & 57 & 12 & 13 & 1.8 & 7.5 & 1.7 \\
NLO pQCD (CTEQ 6.6) $\times$ non-pert. corr. [pb/GeV] & 1900 & 710 & 290 & 120 & 51 & 23 & 11 & 5.4 & 2.7 & 1.4 \\
Non-perturbative correction & 0.96 & 0.98 & 0.97 & 0.97 & 0.98 & 0.97 & 0.99 & 0.98 & 0.99 & 0.98 \\
\hline
Statistical uncertainty & 0.040 & 0.067 & 0.096 & 0.14 & 0.20 & 0.41 & 0.41 & 1.0 & 0.50 & 1.0 \\
Absolute JES uncertainty & $^{+0.25}_{-0.21}$ & $^{+0.26}_{-0.22}$ & $^{+0.28}_{-0.23}$ & $^{+0.29}_{-0.24}$ & $^{+0.30}_{-0.24}$ & $^{+0.31}_{-0.25}$ & $^{+0.32}_{-0.25}$ & $^{+0.32}_{-0.26}$ & $^{+0.33}_{-0.26}$ & $^{+0.34}_{-0.26}$ \\
Relative JES uncertainty & $^{+0.038}_{-0.036}$ & $^{+0.037}_{-0.036}$ & $^{+0.037}_{-0.036}$ & 0.037 & 0.038 & 0.039 & $^{+0.041}_{-0.040}$ & $^{+0.044}_{-0.041}$ & $^{+0.046}_{-0.043}$ & $^{+0.049}_{-0.044}$ \\
Unfolding uncertainty & 0.1 & 0.1 & 0.1 & 0.1 & 0.1 & 0.1 & 0.1 & 0.1 & 0.1 & 0.1 \\
Total systematic uncertainty & $^{+0.3}_{-0.2}$ & $^{+0.3}_{-0.2}$ & 0.3 & 0.3 & 0.3 & 0.3 & 0.3 & 0.3 & 0.3 & $^{+0.4}_{-0.3}$ \\
\hline
PDF uncertainty & 0.03 & 0.03 & $^{+0.02}_{-0.03}$ & 0.02 & 0.02 & 0.02 & 0.03 & 0.03 & $^{+0.04}_{-0.03}$ & 0.04 \\
Scale uncertainty & $^{+0.06}_{-0.08}$ & $^{+0.01}_{-0.1}$ & $^{+0.03}_{-0.2}$ & $^{+0.02}_{-0.2}$ & $^{+0.02}_{-0.1}$ & $^{+0.02}_{-0.1}$ & $^{+0.01}_{-0.1}$ & $^{+0.01}_{-0.1}$ & $^{+0.01}_{-0.1}$ & $^{+0.01}_{-0.1}$ \\
$\alpha_{s}$ uncertainty & 0.02 & 0.02 & 0.02 & 0.03 & 0.03 & 0.03 & 0.03 & 0.04 & 0.04 & 0.04 \\
Non-perturbative correction uncertainty & 0.05 & 0.05 & 0.05 & 0.05 & 0.05 & 0.05 & 0.05 & 0.05 & 0.05 & 0.05 \\
Total theory uncertainty &  $^{+0.09}_{-0.1}$  & $^{+0.06}_{-0.1}$ & $^{+0.07}_{-0.2}$ & $^{+0.06}_{-0.2}$ & $^{+0.06}_{-0.2}$ & $^{+0.07}_{-0.1}$ & $^{+0.07}_{-0.1}$ & $^{+0.07}_{-0.1}$ & $^{+0.07}_{-0.1}$ & $^{+0.08}_{-0.1}$ \\
\hline
\end{tabular}
\end{center}
\caption{Measured dijet double-differential cross section per GeV and per unit absolute rapidity as a function of dijet mass $\twomass{1}{2}$ for \AKT jets with $R=0.4$, compared to NLO pQCD calculations corrected for non-perturbative effects.  All uncertainties listed are fractional uncertainties.  There is an additional overall uncertainty of 11\% due to the measurement of the integrated luminosity which is not included in the systematic uncertainties given above.  The statistical uncertainty is calculated as $1/\sqrt{N}$, where $N$ is the number of events in a given bin.}
\label{tab:DijetMassResults04_2}
\end{sidewaystable*}

\begin{sidewaystable*}
\begin{center}
\vspace{16cm}
\begin{tabular}{|c|c|c|c|c|c|c|c|c|}
\hline
\multicolumn{9}{|l|}{$1.2< |y|_{\mathrm{max}} <2.1$ } \\
\hline
$\twomass{1}{2}$ [GeV] & 510-590 & 590-670 & 670-760 & 760-850 & 850-950 & 950-1060 & 1060-1180 & 1180-1310 \\
\hline
Measured cross section [pb/GeV] & 130 & 51 & 25 & 11 & 5.4 & 2.0 & 1.3 & -- \\
NLO pQCD (CTEQ 6.6) $\times$ non-pert. corr. [pb/GeV] & 100 & 47 & 23 & 12 & 6.1 & 3.2 & 1.7 & 0.86 \\
Non-perturbative correction & 0.96 & 0.96 & 0.96 & 1.0 & 0.96 & 0.98 & 0.97 & 0.97 \\
\hline
Statistical uncertainty & 0.086 & 0.14 & 0.19 & 0.28 & 0.38 & 0.58 & 0.71 & -- \\
Absolute JES uncertainty & $^{+0.30}_{-0.26}$ & $^{+0.31}_{-0.26}$ & $^{+0.32}_{-0.26}$ & $^{+0.32}_{-0.26}$ & $^{+0.33}_{-0.26}$ & $^{+0.34}_{-0.27}$ & $^{+0.35}_{-0.27}$ & -- \\
Relative JES uncertainty & $^{+0.081}_{-0.083}$ & $^{+0.082}_{-0.081}$ & $^{+0.084}_{-0.080}$ & $^{+0.086}_{-0.080}$ & $^{+0.088}_{-0.080}$ & $^{+0.091}_{-0.081}$ & $^{+0.095}_{-0.082}$ & -- \\
Unfolding uncertainty & 0.1 & 0.1 & 0.1 & 0.1 & 0.1 & 0.1 & 0.1 & -- \\
Total systematic uncertainty & 0.3 & 0.3 & 0.3 & $^{+0.4}_{-0.3}$ & $^{+0.4}_{-0.3}$ & $^{+0.4}_{-0.3}$ & $^{+0.4}_{-0.3}$ & -- \\
\hline
PDF uncertainty & 0.02 & 0.03 & 0.03 & 0.03 & 0.04 & $^{+0.05}_{-0.04}$ & $^{+0.05}_{-0.04}$ & $^{+0.06}_{-0.05}$ \\
Scale uncertainty & $^{+0.02}_{-0.2}$ & $^{+0.04}_{-0.2}$ & $^{+0.05}_{-0.2}$ & $^{+0.04}_{-0.2}$ & $^{+0.03}_{-0.2}$ & $^{+0.03}_{-0.2}$ & $^{+0.03}_{-0.2}$ & $^{+0.03}_{-0.2}$ \\
$\alpha_{s}$ uncertainty & 0.03 & 0.03 & 0.03 & 0.04 & 0.04 & 0.04 & 0.04 & 0.04 \\
Non-perturbative correction uncertainty & 0.05 & 0.05 & 0.05 & 0.05 & 0.05 & 0.05 & 0.05 & 0.05 \\
Total theory uncertainty &  $^{+0.06}_{-0.2}$  & $^{+0.08}_{-0.3}$ & $^{+0.08}_{-0.3}$ & $^{+0.08}_{-0.2}$ & $^{+0.08}_{-0.2}$ & $^{+0.08}_{-0.2}$ & $^{+0.09}_{-0.2}$ & $^{+0.09}_{-0.2}$ \\
\hline
\end{tabular}
\end{center}
\caption{Measured dijet double-differential cross section per GeV and per unit absolute rapidity as a function of dijet mass $\twomass{1}{2}$ for \AKT jets with $R=0.4$, compared to NLO pQCD calculations corrected for non-perturbative effects.  All uncertainties listed are fractional uncertainties.  There is an additional overall uncertainty of 11\% due to the measurement of the integrated luminosity which is not included in the systematic uncertainties given above.  The statistical uncertainty is calculated as $1/\sqrt{N}$, where $N$ is the number of events in a given bin.}
\label{tab:DijetMassResults04_3}
\end{sidewaystable*}

\begin{sidewaystable*}
\begin{center}
\vspace{16cm}
\begin{tabular}{|c|c|c|c|c|c|c|c|}
\hline
\multicolumn{8}{|l|}{$2.1< |y|_{\mathrm{max}} <2.8$ } \\
\hline
$\twomass{1}{2}$ [GeV] & 950-1060 & 1060-1180 & 1180-1310 & 1310-1450 & 1450-1600 & 1600-1760 & 1760-1940 \\
\hline
Measured cross section [pb/GeV] & 21 & 9.3 & 3.7 & 2.0 & 1.4 & -- & -- \\
NLO pQCD (CTEQ 6.6) $\times$ non-pert. corr. [pb/GeV] & 15 & 7.5 & 4.0 & 1.9 & 1.0 & 0.52 & 0.24 \\
Non-perturbative correction & 0.96 & 0.91 & 0.93 & 0.97 & 0.98 & 0.98 & 0.94 \\
\hline
Statistical uncertainty & 0.20 & 0.29 & 0.45 & 0.58 & 0.71 & -- & -- \\
Absolute JES uncertainty & $^{+0.39}_{-0.30}$ & $^{+0.41}_{-0.32}$ & $^{+0.43}_{-0.33}$ & $^{+0.45}_{-0.35}$ & $^{+0.47}_{-0.37}$ & -- & -- \\
Relative JES uncertainty & $^{+0.13}_{-0.14}$ & $^{+0.14}_{-0.15}$ & $^{+0.15}_{-0.16}$ & $^{+0.15}_{-0.16}$ & 0.16 & -- & -- \\
Unfolding uncertainty & 0.1 & 0.1 & 0.1 & 0.1 & 0.1 & -- & -- \\
Total systematic uncertainty & $^{+0.4}_{-0.3}$ & 0.4 & $^{+0.5}_{-0.4}$ & $^{+0.5}_{-0.4}$ & $^{+0.5}_{-0.4}$ & -- & -- \\
\hline
PDF uncertainty & 0.04 & $^{+0.05}_{-0.04}$ & 0.05 & $^{+0.06}_{-0.05}$ & $^{+0.07}_{-0.05}$ & $^{+0.08}_{-0.06}$ & $^{+0.09}_{-0.07}$ \\
Scale uncertainty & $^{+0.04}_{-0.3}$ & $^{+0.07}_{-0.4}$ & $^{+0.08}_{-0.4}$ & $^{+0.09}_{-0.4}$ & $^{+0.1}_{-0.5}$ & $^{+0.2}_{-0.7}$ & $^{+0.1}_{-0.6}$ \\
$\alpha_{s}$ uncertainty & 0.04 & 0.04 & 0.04 & 0.04 & 0.04 & 0.03 & 0.04 \\
Non-perturbative correction uncertainty & 0.05 & 0.05 & 0.05 & 0.05 & 0.05 & 0.05 & 0.05 \\
Total theory uncertainty &  $^{+0.09}_{-0.3}$  & $^{+0.1}_{-0.4}$ & $^{+0.1}_{-0.4}$ & $^{+0.1}_{-0.4}$ & $^{+0.2}_{-0.6}$ & $^{+0.2}_{-0.7}$ & $^{+0.2}_{-0.6}$ \\
\hline
\end{tabular}
\end{center}
\caption{Measured dijet double-differential cross section per GeV and per unit absolute rapidity as a function of dijet mass $\twomass{1}{2}$ for \AKT jets with $R=0.4$, compared to NLO pQCD calculations corrected for non-perturbative effects.  All uncertainties listed are fractional uncertainties.  There is an additional overall uncertainty of 11\% due to the measurement of the integrated luminosity which is not included in the systematic uncertainties given above.  The statistical uncertainty is calculated as $1/\sqrt{N}$, where $N$ is the number of events in a given bin.}
\label{tab:DijetMassResults04_4}
\end{sidewaystable*}


\begin{sidewaystable*}
\begin{flushleft}
\vspace{16cm}
\begin{tabular}{|c|c|c|c|c|c|c|c|c|c|c|c|}
\hline
\multicolumn{12}{|l|}{$     |y|_{\mathrm{max}} <0.3$ } \\
\hline
$\twomass{1}{2}$ [GeV] & 110-160 & 160-210 & 210-260 & 260-310 & 310-370 & 370-440 & 440-510 & 510-590 & 590-670 & 670-760 & 760-850 \\
\hline
Measured cross section [pb/GeV] & 3100 & 670 & 240 & 72 & 53 & 13 & 10 & -- & -- & 2.6 & 2.5 \\
NLO pQCD (CTEQ 6.6) & & & & & & & & & & & \\ 
$\times$ non-pert. corr. [pb/GeV] & 4200 & 690 & 210 & 74 & 30 & 12 & 5.0 & 2.2 & 1.1 & 0.51 & 0.26 \\
Non-perturbative correction & 1.1 & 1.1 & 1.1 & 1.0 & 1.0 & 1.0 & 1.0 & 1.0 & 1.0 & 1.0 & 1.0 \\
\hline
Statistical uncertainty & 0.038 & 0.081 & 0.14 & 0.26 & 0.27 & 0.50 & 0.58 & -- & -- & 1.0 & 1.0 \\
Absolute JES uncertainty & $^{+0.33}_{-0.25}$ & $^{+0.32}_{-0.25}$ & $^{+0.32}_{-0.25}$ & $^{+0.31}_{-0.26}$ & $^{+0.31}_{-0.26}$ & $^{+0.31}_{-0.26}$ & $^{+0.31}_{-0.26}$ & -- & -- & $^{+0.33}_{-0.27}$ & $^{+0.33}_{-0.27}$ \\
Relative JES uncertainty & $^{+5.3e-05}_{-8.0e-05}$ & $^{+0.0012}_{-0.0023}$ & $^{+0.0023}_{-0.0039}$ & $^{+0.0032}_{-0.0051}$ & $^{+0.0040}_{-0.0060}$ & $^{+0.0049}_{-0.0067}$ & $^{+0.0058}_{-0.0072}$ & -- & -- & $^{+0.0083}_{-0.0075}$ & $^{+0.0091}_{-0.0072}$ \\
Unfolding uncertainty & 0.1 & 0.1 & 0.1 & 0.1 & 0.1 & 0.1 & 0.1 & -- & -- & 0.1 & 0.1 \\
Total systematic uncertainty & 0.3 & 0.3 & 0.3 & 0.3 & 0.3 & 0.3 & 0.3 & -- & -- & 0.3 & 0.3 \\
\hline
PDF uncertainty & 0.03 & $^{+0.03}_{-0.04}$ & $^{+0.03}_{-0.04}$ & 0.03 & 0.03 & $^{+0.02}_{-0.03}$ & $^{+0.02}_{-0.03}$ & 0.03 & 0.03 & 0.03 & 0.04 \\
Scale uncertainty & $^{+0.5}_{-0.3}$ & 0.2 & $^{+0.03}_{-0.06}$ & $^{+0.01}_{-0.06}$ & $^{+0.008}_{-0.05}$ & $^{+0.006}_{-0.04}$ & $^{+0.003}_{-0.05}$ & $^{+0.004}_{-0.05}$ & $^{+0.003}_{-0.05}$ & $^{+0.003}_{-0.05}$ & $^{+0.005}_{-0.06}$ \\
$\alpha_{s}$ uncertainty & 0.04 & 0.03 & 0.02 & 0.02 & 0.02 & 0.03 & 0.03 & 0.03 & 0.04 & 0.04 & 0.04 \\
Non-perturbative correction uncertainty & 0.05 & 0.05 & 0.05 & 0.05 & 0.05 & 0.05 & 0.05 & 0.05 & 0.05 & 0.05 & 0.05 \\
Total theory uncertainty &  $^{+0.5}_{-0.3}$  & 0.2 & $^{+0.07}_{-0.09}$ & $^{+0.06}_{-0.09}$ & $^{+0.06}_{-0.08}$ & $^{+0.06}_{-0.07}$ & $^{+0.06}_{-0.08}$ & $^{+0.07}_{-0.08}$ & $^{+0.07}_{-0.08}$ & $^{+0.07}_{-0.09}$ & $^{+0.08}_{-0.09}$ \\
\hline
\end{tabular}
\end{flushleft}
\caption{Measured dijet double-differential cross section per GeV and per unit absolute rapidity as a function of dijet mass $\twomass{1}{2}$ for \AKT jets with $R=0.6$, compared to NLO pQCD calculations corrected for non-perturbative effects.  All uncertainties listed are fractional uncertainties.  There is an additional overall uncertainty of 11\% due to the measurement of the integrated luminosity which is not included in the systematic uncertainties given above.  The statistical uncertainty is calculated as $1/\sqrt{N}$, where $N$ is the number of events in a given bin.}
\label{tab:DijetMassResults06_0}
\end{sidewaystable*}

\begin{sidewaystable*}
\begin{center}
\vspace{16cm}
\begin{tabular}{|c|c|c|c|c|c|c|c|c|c|c|}
\hline
\multicolumn{11}{|l|}{$0.3< |y|_{\mathrm{max}} <0.8$ } \\
\hline
$\twomass{1}{2}$ [GeV] & 160-210 & 210-260 & 260-310 & 310-370 & 370-440 & 440-510 & 510-590 & 590-670 & 670-760 & 760-850 \\
\hline
Measured cross section [pb/GeV] & 3400 & 1000 & 380 & 130 & 73 & 24 & 12 & 10 & -- & 4.6 \\
NLO pQCD (CTEQ 6.6) $\times$ non-pert. corr. [pb/GeV] & 3200 & 940 & 350 & 140 & 55 & 23 & 10 & 4.9 & 2.4 & 1.2 \\
Non-perturbative correction & 1.1 & 1.1 & 1.0 & 1.0 & 1.0 & 1.0 & 1.0 & 1.0 & 1.0 & 1.0 \\
\hline
Statistical uncertainty & 0.028 & 0.051 & 0.085 & 0.13 & 0.16 & 0.29 & 0.38 & 0.41 & -- & 0.58 \\
Absolute JES uncertainty & $^{+0.30}_{-0.25}$ & $^{+0.30}_{-0.25}$ & $^{+0.31}_{-0.25}$ & $^{+0.31}_{-0.25}$ & $^{+0.32}_{-0.26}$ & $^{+0.32}_{-0.26}$ & $^{+0.33}_{-0.26}$ & $^{+0.33}_{-0.26}$ & -- & $^{+0.34}_{-0.27}$ \\
Relative JES uncertainty & $^{+0.019}_{-0.020}$ & $^{+0.020}_{-0.021}$ & 0.021 & $^{+0.021}_{-0.022}$ & 0.022 & 0.023 & $^{+0.024}_{-0.023}$ & 0.024 & -- & $^{+0.026}_{-0.024}$ \\
Unfolding uncertainty & 0.1 & 0.1 & 0.1 & 0.1 & 0.1 & 0.1 & 0.1 & 0.1 & -- & 0.1 \\
Total systematic uncertainty & 0.3 & 0.3 & 0.3 & 0.3 & 0.3 & 0.3 & 0.3 & 0.3 & -- & $^{+0.4}_{-0.3}$ \\
\hline
PDF uncertainty & $^{+0.03}_{-0.04}$ & $^{+0.03}_{-0.04}$ & 0.03 & 0.03 & $^{+0.02}_{-0.03}$ & 0.02 & $^{+0.02}_{-0.03}$ & 0.03 & 0.03 & 0.04 \\
Scale uncertainty & 0.2 & $^{+0.03}_{-0.06}$ & $^{+0.01}_{-0.05}$ & $^{+0.007}_{-0.04}$ & $^{+0.005}_{-0.04}$ & $^{+0.004}_{-0.04}$ & $^{+0.003}_{-0.04}$ & $^{+0.002}_{-0.04}$ & $^{+0.003}_{-0.05}$ & $^{+0.005}_{-0.05}$ \\
$\alpha_{s}$ uncertainty & 0.03 & 0.02 & 0.02 & 0.02 & 0.03 & 0.03 & 0.03 & 0.04 & 0.04 & 0.04 \\
Non-perturbative correction uncertainty & 0.05 & 0.05 & 0.05 & 0.05 & 0.05 & 0.05 & 0.05 & 0.05 & 0.05 & 0.05 \\
Total theory uncertainty & 0.2  & $^{+0.07}_{-0.09}$ & $^{+0.06}_{-0.08}$ & $^{+0.06}_{-0.08}$ & $^{+0.06}_{-0.08}$ & $^{+0.06}_{-0.08}$ & $^{+0.06}_{-0.08}$ & $^{+0.07}_{-0.08}$ & $^{+0.07}_{-0.09}$ & $^{+0.07}_{-0.09}$ \\
\hline
\end{tabular}
\end{center}
\caption{Measured dijet double-differential cross section per GeV and per unit absolute rapidity as a function of dijet mass $\twomass{1}{2}$ for \AKT jets with $R=0.6$, compared to NLO pQCD calculations corrected for non-perturbative effects.  All uncertainties listed are fractional uncertainties.  There is an additional overall uncertainty of 11\% due to the measurement of the integrated luminosity which is not included in the systematic uncertainties given above.  The statistical uncertainty is calculated as $1/\sqrt{N}$, where $N$ is the number of events in a given bin.}
\label{tab:DijetMassResults06_1}
\end{sidewaystable*}

\begin{sidewaystable*}
\begin{center}
\vspace{16cm}
\begin{tabular}{|c|c|c|c|c|c|c|c|c|c|c|}
\hline
\multicolumn{11}{|l|}{$0.8< |y|_{\mathrm{max}} <1.2$ } \\
\hline
$\twomass{1}{2}$ [GeV] & 210-260 & 260-310 & 310-370 & 370-440 & 440-510 & 510-590 & 590-670 & 670-760 & 760-850 & 850-950 \\
\hline
Measured cross section [pb/GeV] & 3100 & 1000 & 450 & 180 & 83 & 27 & 8.5 & 9.3 & 5.6 & 6.7 \\
NLO pQCD (CTEQ 6.6) $\times$ non-pert. corr. [pb/GeV] & 2700 & 980 & 390 & 150 & 65 & 30 & 14 & 6.7 & 3.4 & 1.8 \\
Non-perturbative correction & 1.1 & 1.1 & 1.1 & 1.0 & 1.0 & 1.0 & 1.0 & 1.0 & 1.0 & 1.0 \\
\hline
Statistical uncertainty & 0.033 & 0.057 & 0.079 & 0.12 & 0.17 & 0.28 & 0.50 & 0.45 & 0.58 & 0.50 \\
Absolute JES uncertainty & $^{+0.35}_{-0.26}$ & $^{+0.34}_{-0.26}$ & $^{+0.33}_{-0.26}$ & $^{+0.33}_{-0.26}$ & $^{+0.33}_{-0.26}$ & $^{+0.33}_{-0.26}$ & $^{+0.34}_{-0.27}$ & $^{+0.34}_{-0.27}$ & $^{+0.35}_{-0.27}$ & $^{+0.36}_{-0.27}$ \\
Relative JES uncertainty & $^{+0.037}_{-0.035}$ & $^{+0.038}_{-0.036}$ & $^{+0.040}_{-0.037}$ & $^{+0.041}_{-0.038}$ & $^{+0.042}_{-0.039}$ & $^{+0.043}_{-0.040}$ & $^{+0.044}_{-0.041}$ & $^{+0.045}_{-0.042}$ & $^{+0.046}_{-0.043}$ & $^{+0.047}_{-0.044}$ \\
Unfolding uncertainty & 0.1 & 0.1 & 0.1 & 0.1 & 0.1 & 0.1 & 0.1 & 0.1 & 0.1 & 0.1 \\
Total systematic uncertainty & $^{+0.4}_{-0.3}$ & $^{+0.4}_{-0.3}$ & $^{+0.4}_{-0.3}$ & 0.3 & 0.3 & $^{+0.4}_{-0.3}$ & $^{+0.4}_{-0.3}$ & $^{+0.4}_{-0.3}$ & $^{+0.4}_{-0.3}$ & $^{+0.4}_{-0.3}$ \\
\hline
PDF uncertainty & 0.03 & 0.03 & $^{+0.02}_{-0.03}$ & 0.02 & 0.02 & 0.02 & 0.03 & 0.03 & $^{+0.04}_{-0.03}$ & 0.04 \\
Scale uncertainty & 0.1 & $^{+0.009}_{-0.04}$ & $^{+0.009}_{-0.06}$ & $^{+0.008}_{-0.06}$ & $^{+0.005}_{-0.04}$ & $^{+0.003}_{-0.04}$ & $^{+0.003}_{-0.04}$ & $^{+0.003}_{-0.05}$ & $^{+0.004}_{-0.05}$ & $^{+0.004}_{-0.05}$ \\
$\alpha_{s}$ uncertainty & 0.03 & 0.02 & 0.03 & 0.03 & 0.03 & 0.03 & 0.04 & 0.04 & 0.04 & 0.04 \\
Non-perturbative correction uncertainty & 0.05 & 0.05 & 0.05 & 0.05 & 0.05 & 0.05 & 0.05 & 0.05 & 0.05 & 0.05 \\
Total theory uncertainty & 0.1  & $^{+0.06}_{-0.08}$ & $^{+0.06}_{-0.09}$ & $^{+0.06}_{-0.08}$ & $^{+0.06}_{-0.08}$ & $^{+0.06}_{-0.08}$ & $^{+0.07}_{-0.08}$ & $^{+0.07}_{-0.08}$ & $^{+0.07}_{-0.09}$ & $^{+0.08}_{-0.09}$ \\
\hline
\end{tabular}
\end{center}
\caption{Measured dijet double-differential cross section per GeV and per unit absolute rapidity as a function of dijet mass $\twomass{1}{2}$ for \AKT jets with $R=0.6$, compared to NLO pQCD calculations corrected for non-perturbative effects.  All uncertainties listed are fractional uncertainties.  There is an additional overall uncertainty of 11\% due to the measurement of the integrated luminosity which is not included in the systematic uncertainties given above.  The statistical uncertainty is calculated as $1/\sqrt{N}$, where $N$ is the number of events in a given bin.}
\label{tab:DijetMassResults06_2}
\end{sidewaystable*}

\begin{sidewaystable*}
\begin{center}
\vspace{16cm}
\begin{tabular}{|c|c|c|c|c|c|c|c|c|}
\hline
\multicolumn{9}{|l|}{$1.2< |y|_{\mathrm{max}} <2.1$ } \\
\hline
$\twomass{1}{2}$ [GeV] & 510-590 & 590-670 & 670-760 & 760-850 & 850-950 & 950-1060 & 1060-1180 & 1180-1310 \\
\hline
Measured cross section [pb/GeV] & 160 & 86 & 33 & 17 & 5.6 & 3.5 & 1.3 & 0.6 \\
NLO pQCD (CTEQ 6.6) $\times$ non-pert. corr. [pb/GeV] & 150 & 66 & 32 & 16 & 8.3 & 4.2 & 2.2 & 1.1 \\
Non-perturbative correction & 1.1 & 1.1 & 1.0 & 1.0 & 1.0 & 1.0 & 1.1 & 1.0 \\
\hline
Statistical uncertainty & 0.076 & 0.11 & 0.16 & 0.22 & 0.38 & 0.45 & 0.71 & 1.0 \\
Absolute JES uncertainty & $^{+0.38}_{-0.29}$ & $^{+0.38}_{-0.29}$ & $^{+0.38}_{-0.29}$ & $^{+0.38}_{-0.29}$ & $^{+0.38}_{-0.29}$ & $^{+0.39}_{-0.29}$ & $^{+0.39}_{-0.30}$ & $^{+0.40}_{-0.30}$ \\
Relative JES uncertainty & 0.080 & $^{+0.082}_{-0.080}$ & $^{+0.084}_{-0.081}$ & $^{+0.086}_{-0.081}$ & $^{+0.089}_{-0.083}$ & $^{+0.091}_{-0.084}$ & $^{+0.094}_{-0.086}$ & $^{+0.098}_{-0.089}$ \\
Unfolding uncertainty & 0.1 & 0.1 & 0.1 & 0.1 & 0.1 & 0.1 & 0.1 & 0.1 \\
Total systematic uncertainty & $^{+0.4}_{-0.3}$ & $^{+0.4}_{-0.3}$ & $^{+0.4}_{-0.3}$ & $^{+0.4}_{-0.3}$ & $^{+0.4}_{-0.3}$ & $^{+0.4}_{-0.3}$ & $^{+0.4}_{-0.3}$ & $^{+0.4}_{-0.3}$ \\
\hline
PDF uncertainty & 0.02 & $^{+0.02}_{-0.03}$ & 0.03 & 0.03 & 0.04 & 0.04 & 0.05 & 0.05 \\
Scale uncertainty & $^{+0.004}_{-0.06}$ & $^{+0.01}_{-0.1}$ & $^{+0.01}_{-0.1}$ & $^{+0.009}_{-0.1}$ & $^{+0.006}_{-0.1}$ & $^{+0.004}_{-0.09}$ & $^{+0.005}_{-0.1}$ & $^{+0.008}_{-0.1}$ \\
$\alpha_{s}$ uncertainty & 0.03 & 0.04 & 0.04 & 0.04 & 0.04 & 0.04 & 0.05 & 0.04 \\
Non-perturbative correction uncertainty & 0.05 & 0.05 & 0.05 & 0.05 & 0.05 & 0.05 & 0.05 & 0.05 \\
Total theory uncertainty & $^{+0.06}_{-0.09}$  & $^{+0.07}_{-0.1}$ & $^{+0.07}_{-0.1}$ & $^{+0.07}_{-0.1}$ & $^{+0.08}_{-0.1}$ & $^{+0.08}_{-0.1}$ & $^{+0.09}_{-0.1}$ & $^{+0.09}_{-0.1}$ \\
\hline
\end{tabular}
\end{center}
\caption{Measured dijet double-differential cross section per GeV and per unit absolute rapidity as a function of dijet mass $\twomass{1}{2}$ for \AKT jets with $R=0.6$, compared to NLO pQCD calculations corrected for non-perturbative effects.  All uncertainties listed are fractional uncertainties.  There is an additional overall uncertainty of 11\% due to the measurement of the integrated luminosity which is not included in the systematic uncertainties given above.  The statistical uncertainty is calculated as $1/\sqrt{N}$, where $N$ is the number of events in a given bin.}
\label{tab:DijetMassResults06_3}
\end{sidewaystable*}

\begin{sidewaystable*}
\begin{center}
\vspace{16cm}
\begin{tabular}{|c|c|c|c|c|c|c|c|}
\hline
\multicolumn{8}{|l|}{$2.1< |y|_{\mathrm{max}} <2.8$ } \\
\hline
$\twomass{1}{2}$ [GeV] & 950-1060 & 1060-1180 & 1180-1310 & 1310-1450 & 1450-1600 & 1600-1760 & 1760-1940 \\
\hline
Measured cross section [pb/GeV] & 29 & 15 & 4.6 & 2.7 & 2.4 & -- & 0.52 \\
NLO pQCD (CTEQ 6.6) $\times$ non-pert. corr. [pb/GeV] & 23 & 12 & 6.1 & 3.1 & 1.6 & 0.79 & 0.35 \\
Non-perturbative correction & 1.1 & 1.0 & 1.1 & 1.1 & 1.0 & 1.2 & 0.93 \\
\hline
Statistical uncertainty & 0.17 & 0.22 & 0.41 & 0.50 & 0.50 & -- & 1.0 \\
Absolute JES uncertainty & $^{+0.50}_{-0.35}$ & $^{+0.49}_{-0.35}$ & $^{+0.50}_{-0.36}$ & $^{+0.50}_{-0.37}$ & $^{+0.52}_{-0.37}$ & -- & $^{+0.57}_{-0.40}$ \\
Relative JES uncertainty & $^{+0.14}_{-0.13}$ & $^{+0.15}_{-0.14}$ & $^{+0.15}_{-0.14}$ & $^{+0.16}_{-0.15}$ & $^{+0.17}_{-0.15}$ & -- & $^{+0.19}_{-0.15}$ \\
Unfolding uncertainty & 0.1 & 0.1 & 0.1 & 0.1 & 0.1 & -- & 0.1 \\
Total systematic uncertainty & $^{+0.5}_{-0.4}$ & $^{+0.5}_{-0.4}$ & $^{+0.5}_{-0.4}$ & $^{+0.5}_{-0.4}$ & $^{+0.6}_{-0.4}$ & -- & $^{+0.6}_{-0.4}$ \\
\hline
PDF uncertainty & 0.04 & $^{+0.05}_{-0.04}$ & $^{+0.06}_{-0.05}$ & $^{+0.06}_{-0.05}$ & $^{+0.07}_{-0.05}$ & $^{+0.08}_{-0.06}$ & $^{+0.08}_{-0.06}$ \\
Scale uncertainty & $^{+0.005}_{-0.1}$ & $^{+0.03}_{-0.3}$ & $^{+0.04}_{-0.3}$ & $^{+0.02}_{-0.2}$ & $^{+0.02}_{-0.2}$ & $^{+0.06}_{-0.4}$ & $^{+0.08}_{-0.4}$ \\
$\alpha_{s}$ uncertainty & 0.05 & 0.05 & 0.04 & 0.05 & 0.05 & 0.05 & 0.03 \\
Non-perturbative correction uncertainty & 0.05 & 0.05 & 0.05 & 0.05 & 0.05 & 0.05 & 0.05 \\
Total theory uncertainty &  $^{+0.08}_{-0.1}$  & $^{+0.09}_{-0.3}$ & $^{+0.1}_{-0.3}$ & $^{+0.1}_{-0.2}$ & $^{+0.1}_{-0.2}$ & $^{+0.1}_{-0.4}$ & $^{+0.1}_{-0.4}$ \\
\hline
\end{tabular}
\end{center}
\caption{Measured dijet double-differential cross section per GeV and per unit absolute rapidity as a function of dijet mass $\twomass{1}{2}$ for \AKT jets with $R=0.6$, compared to NLO pQCD calculations corrected for non-perturbative effects.  All uncertainties listed are fractional uncertainties.  There is an additional overall uncertainty of 11\% due to the measurement of the integrated luminosity which is not included in the systematic uncertainties given above.  The statistical uncertainty is calculated as $1/\sqrt{N}$, where $N$ is the number of events in a given bin.}
\label{tab:DijetMassResults06_4}
\end{sidewaystable*}


\begin{sidewaystable*}
\begin{center}
\vspace{16cm}
\begin{tabular}{|c|c|c|c|c|c|c|c|c|c|c|c|}
\hline
\multicolumn{12}{|l|}{$340< \twomass{1}{2} <520 $ GeV } \\
\hline
$\chi$ & 1-1.3 & 1.3-1.8 & 1.8-2.4 & 2.4-3.3 & 3.3-4.5 & 4.5-6 & 6-8.2 & 8.2-11 & 11-15 & 15-20 & 20-30 \\
\hline
Measured cross section [pb/GeV] & 16 & 19 & 17 & 13 & 17 & 15 & 18 & 15 & 17 & 16 & 18 \\
NLO pQCD (CTEQ 6.6) $\times$ non-pert. corr. [pb/GeV] & 24 & 20 & 17 & 16 & 16 & 16 & 16 & 16 & 17 & 16 & 17 \\
Non-perturbative correction & 0.98 & 0.98 & 0.97 & 0.97 & 0.97 & 0.97 & 0.98 & 0.98 & 0.97 & 0.95 & 0.97 \\
\hline
Statistical uncertainty & 0.27 & 0.19 & 0.18 & 0.17 & 0.13 & 0.12 & 0.091 & 0.088 & 0.070 & 0.063 & 0.044 \\
Absolute JES uncertainty & $^{+0.32}_{-0.26}$ & $^{+0.29}_{-0.25}$ & $^{+0.30}_{-0.25}$ & $^{+0.32}_{-0.24}$ & $^{+0.30}_{-0.27}$ & $^{+0.29}_{-0.24}$ & $^{+0.29}_{-0.25}$ & $^{+0.34}_{-0.27}$ & $^{+0.32}_{-0.26}$ & $^{+0.34}_{-0.27}$ & $^{+0.36}_{-0.28}$ \\
Relative JES uncertainty & $^{+0.027}_{-0.025}$ & $^{+0.028}_{-0.026}$ & $^{+0.026}_{-0.034}$ & $^{+0.030}_{-0.028}$ & $^{+0.035}_{-0.040}$ & $^{+0.043}_{-0.040}$ & $^{+0.049}_{-0.040}$ & $^{+0.051}_{-0.052}$ & $^{+0.065}_{-0.067}$ & $^{+0.092}_{-0.058}$ & $^{+0.079}_{-0.074}$ \\
Unfolding uncertainty & 0.1 & 0.1 & 0.1 & 0.1 & 0.1 & 0.1 & 0.1 & 0.1 & 0.1 & 0.1 & 0.1 \\
Total systematic uncertainty & 0.3 & 0.3 & 0.3 & 0.3 & 0.3 & 0.3 & 0.3 & $^{+0.4}_{-0.3}$ & 0.3 & $^{+0.4}_{-0.3}$ & $^{+0.4}_{-0.3}$ \\
\hline
PDF uncertainty & 0.02 & 0.02 & 0.02 & 0.02 & 0.02 & 0.02 & 0.02 & 0.02 & 0.02 & 0.02 & 0.02 \\
Scale uncertainty & $^{+0.01}_{-0.1}$ & $^{+0.01}_{-0.1}$ & $^{+0.01}_{-0.1}$ & $^{+0.02}_{-0.1}$ & $^{+0.02}_{-0.1}$ & $^{+0.02}_{-0.2}$ & $^{+0.02}_{-0.2}$ & $^{+0.03}_{-0.2}$ & $^{+0.04}_{-0.2}$ & $^{+0.06}_{-0.2}$ & $^{+0.06}_{-0.3}$ \\
$\alpha_{s}$ uncertainty & 0.03 & 0.03 & 0.03 & 0.03 & 0.03 & 0.03 & 0.03 & 0.03 & 0.03 & 0.03 & 0.03 \\
Non-perturbative correction uncertainty & 0.05 & 0.05 & 0.05 & 0.05 & 0.05 & 0.05 & 0.05 & 0.05 & 0.05 & 0.05 & 0.05 \\
Total theory uncertainty & $^{+0.06}_{-0.1}$ & $^{+0.06}_{-0.1}$ & $^{+0.06}_{-0.1}$ & $^{+0.06}_{-0.1}$ & $^{+0.06}_{-0.1}$ & $^{+0.06}_{-0.2}$ & $^{+0.06}_{-0.2}$ & $^{+0.06}_{-0.2}$ & $^{+0.07}_{-0.2}$ & $^{+0.07}_{-0.2}$ & $^{+0.08}_{-0.3}$ \\
\hline
\end{tabular}
\end{center}
\caption{Measured dijet double-differential cross section per GeV and per unit $\chi$ as a function of the dijet angular variable $\chi$ for \AKT jets with $R=0.4$, compared to NLO pQCD calculations corrected for non-perturbative effects.  All uncertainties listed are fractional uncertainties.  There is an additional overall uncertainty of 11\% due to the measurement of the integrated luminosity which is not included in the systematic uncertainties given above.  The statistical uncertainty is calculated as $1/\sqrt{N}$, where $N$ is the number of events in a given bin.}
\label{tab:DijetChiResults04_3}
\end{sidewaystable*}

\begin{sidewaystable*}
\begin{center}
\vspace{16cm}
\begin{tabular}{|c|c|c|c|c|c|c|c|c|c|c|c|}
\hline
\multicolumn{12}{|l|}{$520< \twomass{1}{2} <800$ GeV } \\
\hline
$\chi$ & 1-1.3 & 1.3-1.8 & 1.8-2.4 & 2.4-3.3 & 3.3-4.5 & 4.5-6 & 6-8.2 & 8.2-11 & 11-15 & 15-20 & 20-30 \\
\hline
Measured cross section [pb/GeV] & 3.6 & 1.7 & 2.1 & 1.2 & 1.3 & 1.0 & 1.5 & 2.3 & 1.6 & 1.8 & 1.4 \\
NLO pQCD (CTEQ 6.6) $\times$ non-pert. corr. [pb/GeV] & 2.4 & 2.0 & 1.7 & 1.7 & 1.6 & 1.6 & 1.7 & 1.7 & 1.7 & 1.7 & 1.7 \\
Non-perturbative correction & 0.99 & 0.99 & 0.98 & 0.98 & 0.97 & 0.98 & 0.98 & 0.98 & 0.98 & 0.98 & 0.95 \\
\hline
Statistical uncertainty & 0.45 & 0.50 & 0.41 & 0.45 & 0.38 & 0.38 & 0.26 & 0.18 & 0.19 & 0.15 & 0.12 \\
Absolute JES uncertainty & $^{+0.29}_{-0.25}$ & $^{+0.31}_{-0.26}$ & $^{+0.30}_{-0.25}$ & $^{+0.29}_{-0.25}$ & $^{+0.30}_{-0.23}$ & $^{+0.33}_{-0.27}$ & $^{+0.34}_{-0.26}$ & $^{+0.35}_{-0.27}$ & $^{+0.35}_{-0.27}$ & $^{+0.37}_{-0.26}$ & $^{+0.31}_{-0.30}$ \\
Relative JES uncertainty & 0.028 & $^{+0.026}_{-0.029}$ & $^{+0.030}_{-0.020}$ & $^{+0.032}_{-0.031}$ & $^{+0.031}_{-0.035}$ & $^{+0.047}_{-0.041}$ & $^{+0.049}_{-0.054}$ & $^{+0.063}_{-0.058}$ & $^{+0.072}_{-0.056}$ & $^{+0.060}_{-0.082}$ & $^{+0.073}_{-0.080}$ \\
Unfolding uncertainty & 0.1 & 0.1 & 0.1 & 0.1 & 0.1 & 0.1 & 0.1 & 0.1 & 0.1 & 0.1 & 0.1 \\
Total systematic uncertainty & 0.3 & 0.3 & 0.3 & 0.3 & 0.3 & $^{+0.4}_{-0.3}$ & $^{+0.4}_{-0.3}$ & $^{+0.4}_{-0.3}$ & $^{+0.4}_{-0.3}$ & $^{+0.4}_{-0.3}$ & 0.3 \\
\hline
PDF uncertainty & 0.03 & 0.03 & 0.03 & 0.03 & 0.03 & 0.03 & 0.03 & 0.03 & 0.03 & 0.03 & 0.03 \\
Scale uncertainty & $^{+0.01}_{-0.1}$ & $^{+0.01}_{-0.1}$ & $^{+0.01}_{-0.1}$ & $^{+0.01}_{-0.1}$ & $^{+0.01}_{-0.1}$ & $^{+0.02}_{-0.1}$ & $^{+0.02}_{-0.2}$ & $^{+0.02}_{-0.2}$ & $^{+0.04}_{-0.2}$ & $^{+0.05}_{-0.2}$ & $^{+0.06}_{-0.3}$ \\
$\alpha_{s}$ uncertainty & 0.03 & 0.03 & 0.03 & 0.03 & 0.03 & 0.03 & 0.03 & 0.03 & 0.03 & 0.03 & 0.03 \\
Non-perturbative correction uncertainty & 0.05 & 0.05 & 0.05 & 0.05 & 0.05 & 0.05 & 0.05 & 0.05 & 0.05 & 0.05 & 0.05 \\
Total theory uncertainty & $^{+0.07}_{-0.1}$ & $^{+0.07}_{-0.1}$ & $^{+0.07}_{-0.1}$ & $^{+0.07}_{-0.1}$ & $^{+0.07}_{-0.1}$ & $^{+0.07}_{-0.1}$ & $^{+0.07}_{-0.2}$ & $^{+0.07}_{-0.2}$ & $^{+0.07}_{-0.2}$ & $^{+0.08}_{-0.2}$ & $^{+0.08}_{-0.3}$ \\
\hline
\end{tabular}
\end{center}
\caption{Measured dijet double-differential cross section per GeV and per unit $\chi$ as a function of the dijet angular variable $\chi$ for \AKT jets with $R=0.4$, compared to NLO pQCD calculations corrected for non-perturbative effects.  All uncertainties listed are fractional uncertainties.  There is an additional overall uncertainty of 11\% due to the measurement of the integrated luminosity which is not included in the systematic uncertainties given above.  The statistical uncertainty is calculated as $1/\sqrt{N}$, where $N$ is the number of events in a given bin.}
\label{tab:DijetChiResults04_4}
\end{sidewaystable*}

\begin{sidewaystable*}
\begin{center}
\vspace{16cm}
\begin{tabular}{|c|c|c|c|c|c|c|c|c|c|c|c|}
\hline
\multicolumn{12}{|l|}{$800< \twomass{1}{2} <1200$ GeV } \\
\hline
$\chi$ & 1-1.3 & 1.3-1.8 & 1.8-2.4 & 2.4-3.3 & 3.3-4.5 & 4.5-6 & 6-8.2 & 8.2-11 & 11-15 & 15-20 & 20-30 \\
\hline
Measured cross section [pb/GeV] & -- & 0.30 & 0.25 & 0.17 & -- & 0.10 & 0.21 & 0.053 & 0.11 & 0.031 & 0.11 \\
NLO pQCD (CTEQ 6.6) $\times$ non-pert. corr. [pb/GeV] & 0.20 & 0.17 & 0.15 & 0.14 & 0.14 & 0.14 & 0.14 & 0.15 & 0.15 & 0.15 & 0.15 \\
Non-perturbative correction & 0.98 & 0.98 & 0.98 & 0.99 & 0.98 & 0.97 & 0.98 & 0.99 & 0.99 & 0.99 & 0.96 \\
\hline
Statistical uncertainty & -- & 1.0 & 1.0 & 1.0 & -- & 1.0 & 0.58 & 1.0 & 0.58 & 1.0 & 0.38 \\
Absolute JES uncertainty & -- & $^{+0.34}_{-0.26}$ & $^{+0.35}_{-0.26}$ & $^{+0.32}_{-0.27}$ & -- & $^{+0.38}_{-0.27}$ & $^{+0.33}_{-0.27}$ & $^{+0.34}_{-0.27}$ & $^{+0.37}_{-0.28}$ & $^{+0.37}_{-0.32}$ & $^{+0.36}_{-0.29}$ \\
Relative JES uncertainty & -- & $^{+0.032}_{-0.030}$ & $^{+0.035}_{-0.032}$ & $^{+0.038}_{-0.031}$ & -- & $^{+0.048}_{-0.054}$ & $^{+0.066}_{-0.043}$ & $^{+0.064}_{-0.063}$ & $^{+0.078}_{-0.064}$ & $^{+0.11}_{-0.098}$ & $^{+0.11}_{-0.10}$ \\
Unfolding uncertainty & -- & 0.1 & 0.1 & 0.1 & -- & 0.1 & 0.1 & 0.1 & 0.1 & 0.1 & 0.1 \\
Total systematic uncertainty & -- & $^{+0.4}_{-0.3}$ & $^{+0.4}_{-0.3}$ & 0.3 & -- & $^{+0.4}_{-0.3}$ & $^{+0.4}_{-0.3}$ & $^{+0.4}_{-0.3}$ & $^{+0.4}_{-0.3}$ & $^{+0.4}_{-0.3}$ & $^{+0.4}_{-0.3}$ \\
\hline
PDF uncertainty & $^{+0.05}_{-0.04}$ & $^{+0.05}_{-0.04}$ & $^{+0.05}_{-0.04}$ & $^{+0.05}_{-0.04}$ & $^{+0.05}_{-0.04}$ & $^{+0.05}_{-0.04}$ & $^{+0.05}_{-0.04}$ & $^{+0.05}_{-0.04}$ & $^{+0.05}_{-0.04}$ & $^{+0.05}_{-0.04}$ & $^{+0.05}_{-0.04}$ \\
Scale uncertainty & $^{+0.006}_{-0.08}$ & $^{+0.007}_{-0.09}$ & $^{+0.007}_{-0.1}$ & $^{+0.009}_{-0.1}$ & $^{+0.01}_{-0.1}$ & $^{+0.02}_{-0.1}$ & $^{+0.02}_{-0.2}$ & $^{+0.02}_{-0.2}$ & $^{+0.03}_{-0.2}$ & $^{+0.04}_{-0.2}$ & $^{+0.06}_{-0.3}$ \\
$\alpha_{s}$ uncertainty & 0.04 & 0.04 & 0.04 & 0.04 & 0.04 & 0.04 & 0.04 & 0.04 & 0.04 & 0.04 & 0.04 \\
Non-perturbative correction uncertainty & 0.05 & 0.05 & 0.05 & 0.05 & 0.05 & 0.05 & 0.05 & 0.05 & 0.05 & 0.05 & 0.05 \\
Total theory uncertainty & $^{+0.08}_{-0.1}$ & $^{+0.08}_{-0.1}$ & $^{+0.08}_{-0.1}$ & $^{+0.08}_{-0.1}$ & $^{+0.08}_{-0.1}$ & $^{+0.08}_{-0.1}$ & $^{+0.08}_{-0.2}$ & $^{+0.08}_{-0.2}$ & $^{+0.08}_{-0.2}$ & $^{+0.08}_{-0.2}$ & $^{+0.09}_{-0.3}$ \\
\hline
\end{tabular}
\end{center}
\caption{Measured dijet double-differential cross section per GeV and per unit $\chi$ as a function of the dijet angular variable $\chi$ for \AKT jets with $R=0.4$, compared to NLO pQCD calculations corrected for non-perturbative effects.  All uncertainties listed are fractional uncertainties.  There is an additional overall uncertainty of 11\% due to the measurement of the integrated luminosity which is not included in the systematic uncertainties given above.  The statistical uncertainty is calculated as $1/\sqrt{N}$, where $N$ is the number of events in a given bin.}
\label{tab:DijetChiResults04_5}
\end{sidewaystable*}


\begin{sidewaystable*}
\begin{center}
\vspace{16cm}
\begin{tabular}{|c|c|c|c|c|c|c|c|c|c|c|c|}
\hline
\multicolumn{12}{|l|}{$340< \twomass{1}{2} <520 $ GeV } \\
\hline
$\chi$ & 1-1.3 & 1.3-1.8 & 1.8-2.4 & 2.4-3.3 & 3.3-4.5 & 4.5-6 & 6-8.2 & 8.2-11 & 11-15 & 15-20 & 20-30 \\
\hline
Measured cross section [pb/GeV] & 28 & 31 & 19 & 20 & 20 & 21 & 24 & 21 & 25 & 24 & 28 \\
NLO pQCD (CTEQ 6.6) $\times$ non-pert. corr. [pb/GeV] & 30 & 25 & 22 & 21 & 21 & 21 & 21 & 23 & 23 & 24 & 25 \\
Non-perturbative correction & 1.0 & 1.0 & 1.0 & 1.0 & 1.0 & 1.0 & 1.0 & 1.1 & 1.1 & 1.1 & 1.1 \\
\hline
Statistical uncertainty & 0.20 & 0.15 & 0.17 & 0.13 & 0.12 & 0.10 & 0.079 & 0.075 & 0.058 & 0.052 & 0.035 \\
Absolute JES uncertainty & $^{+0.32}_{-0.26}$ & $^{+0.30}_{-0.26}$ & $^{+0.33}_{-0.25}$ & $^{+0.31}_{-0.26}$ & $^{+0.33}_{-0.24}$ & $^{+0.29}_{-0.25}$ & $^{+0.30}_{-0.25}$ & $^{+0.32}_{-0.29}$ & $^{+0.35}_{-0.27}$ & $^{+0.33}_{-0.29}$ & $^{+0.44}_{-0.28}$ \\
Relative JES uncertainty & 0.027 & $^{+0.026}_{-0.024}$ & $^{+0.031}_{-0.023}$ & $^{+0.032}_{-0.034}$ & $^{+0.036}_{-0.034}$ & $^{+0.037}_{-0.038}$ & $^{+0.046}_{-0.048}$ & $^{+0.056}_{-0.062}$ & $^{+0.065}_{-0.072}$ & $^{+0.064}_{-0.069}$ & $^{+0.097}_{-0.076}$ \\
Unfolding uncertainty & 0.1 & 0.1 & 0.1 & 0.1 & 0.1 & 0.1 & 0.1 & 0.1 & 0.1 & 0.1 & 0.1 \\
Total systematic uncertainty & 0.3 & 0.3 & 0.3 & 0.3 & 0.3 & 0.3 & 0.3 & 0.3 & $^{+0.4}_{-0.3}$ & $^{+0.4}_{-0.3}$ & $^{+0.5}_{-0.3}$ \\
\hline
PDF uncertainty & 0.02 & 0.02 & 0.02 & 0.02 & 0.02 & 0.02 & 0.02 & 0.02 & 0.02 & 0.02 & 0.02 \\
Scale uncertainty & $^{+0.004}_{-0.04}$ & $^{+0.005}_{-0.04}$ & $^{+0.005}_{-0.04}$ & $^{+0.005}_{-0.04}$ & $^{+0.005}_{-0.04}$ & $^{+0.006}_{-0.05}$ & $^{+0.007}_{-0.06}$ & $^{+0.008}_{-0.07}$ & $^{+0.009}_{-0.08}$ & $^{+0.009}_{-0.09}$ & $^{+0.009}_{-0.09}$ \\
$\alpha_{s}$ uncertainty & 0.03 & 0.03 & 0.03 & 0.03 & 0.03 & 0.03 & 0.03 & 0.03 & 0.03 & 0.03 & 0.03 \\
Non-perturbative correction uncertainty & 0.05 & 0.05 & 0.05 & 0.05 & 0.05 & 0.05 & 0.05 & 0.05 & 0.05 & 0.05 & 0.05 \\
Total theory uncertainty & $^{+0.06}_{-0.07}$ & $^{+0.06}_{-0.08}$ & $^{+0.06}_{-0.07}$ & $^{+0.06}_{-0.07}$ & $^{+0.06}_{-0.07}$ & $^{+0.06}_{-0.08}$ & $^{+0.06}_{-0.08}$ & $^{+0.06}_{-0.09}$ & $^{+0.06}_{-0.09}$ & $^{+0.06}_{-0.1}$ & $^{+0.06}_{-0.1}$ \\
\hline
\end{tabular}
\end{center}
\caption{Measured dijet double-differential cross section per GeV and per unit $\chi$ as a function of the dijet angular variable $\chi$ for \AKT jets with $R=0.6$, compared to NLO pQCD calculations corrected for non-perturbative effects.  All uncertainties listed are fractional uncertainties.  There is an additional overall uncertainty of 11\% due to the measurement of the integrated luminosity which is not included in the systematic uncertainties given above.  The statistical uncertainty is calculated as $1/\sqrt{N}$, where $N$ is the number of events in a given bin.}
\label{tab:DijetChiResults06_3}
\end{sidewaystable*}

\begin{sidewaystable*}
\begin{center}
\vspace{16cm}
\begin{tabular}{|c|c|c|c|c|c|c|c|c|c|c|c|}
\hline
\multicolumn{12}{|l|}{$520< \twomass{1}{2} <800$ GeV } \\
\hline
$\chi$ & 1-1.3 & 1.3-1.8 & 1.8-2.4 & 2.4-3.3 & 3.3-4.5 & 4.5-6 & 6-8.2 & 8.2-11 & 11-15 & 15-20 & 20-30 \\
\hline
Measured cross section [pb/GeV] & 4.3 & 1.7 & 1.8 & 1.4 & 2.0 & 0.99 & 2.3 & 3.0 & 2.3 & 2.6 & 2.2 \\
NLO pQCD (CTEQ 6.6) $\times$ non-pert. corr. [pb/GeV] & 2.9 & 2.4 & 2.1 & 2.0 & 2.0 & 2.1 & 2.1 & 2.2 & 2.3 & 2.4 & 2.4 \\
Non-perturbative correction & 1.0 & 1.0 & 1.0 & 1.0 & 1.0 & 1.0 & 1.0 & 1.0 & 1.1 & 1.0 & 1.1 \\
\hline
Statistical uncertainty & 0.41 & 0.50 & 0.45 & 0.41 & 0.30 & 0.38 & 0.21 & 0.16 & 0.15 & 0.13 & 0.10 \\
Absolute JES uncertainty & $^{+0.30}_{-0.25}$ & $^{+0.33}_{-0.25}$ & $^{+0.31}_{-0.24}$ & $^{+0.33}_{-0.25}$ & $^{+0.31}_{-0.25}$ & $^{+0.35}_{-0.27}$ & $^{+0.33}_{-0.27}$ & $^{+0.36}_{-0.26}$ & $^{+0.33}_{-0.27}$ & $^{+0.36}_{-0.27}$ & $^{+0.36}_{-0.27}$ \\
Relative JES uncertainty & $^{+0.016}_{-0.021}$ & $^{+0.025}_{-0.031}$ & $^{+0.025}_{-0.026}$ & $^{+0.029}_{-0.036}$ & $^{+0.034}_{-0.040}$ & $^{+0.050}_{-0.040}$ & $^{+0.057}_{-0.051}$ & $^{+0.070}_{-0.055}$ & $^{+0.070}_{-0.058}$ & $^{+0.074}_{-0.085}$ & $^{+0.088}_{-0.072}$ \\
Unfolding uncertainty & 0.1 & 0.1 & 0.1 & 0.1 & 0.1 & 0.1 & 0.1 & 0.1 & 0.1 & 0.1 & 0.1 \\
Total systematic uncertainty & 0.3 & 0.3 & 0.3 & 0.3 & 0.3 & $^{+0.4}_{-0.3}$ & $^{+0.4}_{-0.3}$ & $^{+0.4}_{-0.3}$ & $^{+0.4}_{-0.3}$ & $^{+0.4}_{-0.3}$ & $^{+0.4}_{-0.3}$ \\
\hline
PDF uncertainty & 0.03 & 0.03 & 0.03 & 0.03 & 0.03 & 0.03 & 0.03 & 0.03 & 0.03 & 0.03 & 0.03 \\
Scale uncertainty & $^{+0.004}_{-0.05}$ & $^{+0.004}_{-0.05}$ & $^{+0.004}_{-0.05}$ & $^{+0.003}_{-0.05}$ & $^{+0.002}_{-0.04}$ & $^{+0.002}_{-0.05}$ & $^{+0.003}_{-0.06}$ & $^{+0.005}_{-0.07}$ & $^{+0.007}_{-0.09}$ & $^{+0.01}_{-0.1}$ & $^{+0.01}_{-0.1}$ \\
$\alpha_{s}$ uncertainty & 0.04 & 0.04 & 0.04 & 0.04 & 0.04 & 0.04 & 0.04 & 0.04 & 0.04 & 0.04 & 0.04 \\
Non-perturbative correction uncertainty & 0.05 & 0.05 & 0.05 & 0.05 & 0.05 & 0.05 & 0.05 & 0.05 & 0.05 & 0.05 & 0.05 \\
Total theory uncertainty & $^{+0.07}_{-0.09}$ & $^{+0.07}_{-0.08}$ & $^{+0.07}_{-0.08}$ & $^{+0.07}_{-0.08}$ & $^{+0.07}_{-0.08}$ & $^{+0.07}_{-0.08}$ & $^{+0.07}_{-0.08}$ & $^{+0.07}_{-0.09}$ & $^{+0.07}_{-0.1}$ & $^{+0.07}_{-0.1}$ & $^{+0.07}_{-0.1}$ \\
\hline
\end{tabular}
\end{center}
\caption{Measured dijet double-differential cross section per GeV and per unit $\chi$ as a function of the dijet angular variable $\chi$ for \AKT jets with $R=0.6$, compared to NLO pQCD calculations corrected for non-perturbative effects.  All uncertainties listed are fractional uncertainties.  There is an additional overall uncertainty of 11\% due to the measurement of the integrated luminosity which is not included in the systematic uncertainties given above.  The statistical uncertainty is calculated as $1/\sqrt{N}$, where $N$ is the number of events in a given bin.}
\label{tab:DijetChiResults06_4}
\end{sidewaystable*}

\begin{sidewaystable*}
\begin{center}
\vspace{16cm}
\begin{tabular}{|c|c|c|c|c|c|c|c|c|c|c|c|}
\hline
\multicolumn{12}{|l|}{$800< \twomass{1}{2} <1200$ GeV } \\
\hline
$\chi$ & 1-1.3 & 1.3-1.8 & 1.8-2.4 & 2.4-3.3 & 3.3-4.5 & 4.5-6 & 6-8.2 & 8.2-11 & 11-15 & 15-20 & 20-30 \\
\hline
Measured cross section [pb/GeV] & 0.50 & 0.61 & 0.74 & 0.17 & -- & 0.099 & 0.21 & 0.16 & 0.19 & 0.061 & 0.11 \\
NLO pQCD (CTEQ 6.6) $\times$ non-pert. corr. [pb/GeV] & 0.24 & 0.20 & 0.18 & 0.17 & 0.17 & 0.17 & 0.18 & 0.19 & 0.19 & 0.20 & 0.21 \\
Non-perturbative correction & 1.0 & 1.0 & 1.0 & 1.0 & 1.0 & 1.0 & 1.0 & 1.0 & 1.0 & 1.0 & 1.0 \\
\hline
Statistical uncertainty & 1.0 & 0.71 & 0.58 & 1.0 & -- & 1.0 & 0.58 & 0.58 & 0.45 & 0.71 & 0.38 \\
Absolute JES uncertainty & $^{+0.33}_{-0.26}$ & $^{+0.33}_{-0.28}$ & $^{+0.35}_{-0.26}$ & $^{+0.33}_{-0.28}$ & -- & $^{+0.35}_{-0.27}$ & $^{+0.34}_{-0.27}$ & $^{+0.32}_{-0.27}$ & $^{+0.39}_{-0.29}$ & $^{+0.35}_{-0.32}$ & $^{+0.41}_{-0.30}$ \\
Relative JES uncertainty & $^{+0.028}_{-0.032}$ & $^{+0.029}_{-0.033}$ & $^{+0.041}_{-0.035}$ & $^{+0.032}_{-0.033}$ & -- & $^{+0.046}_{-0.049}$ & $^{+0.064}_{-0.052}$ & $^{+0.062}_{-0.070}$ & $^{+0.081}_{-0.068}$ & $^{+0.083}_{-0.087}$ & $^{+0.096}_{-0.083}$ \\
Unfolding uncertainty & 0.1 & 0.1 & 0.1 & 0.1 & -- & 0.1 & 0.1 & 0.1 & 0.1 & 0.1 & 0.1 \\
Total systematic uncertainty & 0.3 & 0.3 & $^{+0.4}_{-0.3}$ & 0.3 & -- & $^{+0.4}_{-0.3}$ & $^{+0.4}_{-0.3}$ & 0.3 & $^{+0.4}_{-0.3}$ & $^{+0.4}_{-0.3}$ & $^{+0.4}_{-0.3}$ \\
\hline
PDF uncertainty & $^{+0.05}_{-0.04}$ & $^{+0.05}_{-0.04}$ & $^{+0.05}_{-0.04}$ & $^{+0.05}_{-0.04}$ & $^{+0.05}_{-0.04}$ & $^{+0.05}_{-0.04}$ & $^{+0.05}_{-0.04}$ & $^{+0.05}_{-0.04}$ & $^{+0.05}_{-0.04}$ & $^{+0.05}_{-0.04}$ & $^{+0.05}_{-0.04}$ \\
Scale uncertainty & $^{+0.008}_{-0.06}$ & $^{+0.009}_{-0.06}$ & $^{+0.007}_{-0.06}$ & $^{+0.006}_{-0.06}$ & $^{+0.005}_{-0.05}$ & $^{+0.003}_{-0.05}$ & $^{+0.002}_{-0.06}$ & $^{+0.0005}_{-0.07}$ & $^{+0.004}_{-0.09}$ & $^{+0.009}_{-0.1}$ & $^{+0.01}_{-0.2}$ \\
$\alpha_{s}$ uncertainty & 0.04 & 0.04 & 0.04 & 0.04 & 0.04 & 0.04 & 0.04 & 0.04 & 0.04 & 0.04 & 0.04 \\
Non-perturbative correction uncertainty & 0.05 & 0.05 & 0.05 & 0.05 & 0.05 & 0.05 & 0.05 & 0.05 & 0.05 & 0.05 & 0.05 \\
Total theory uncertainty & $^{+0.08}_{-0.1}$ & $^{+0.08}_{-0.1}$ & $^{+0.08}_{-0.1}$ & $^{+0.08}_{-0.1}$ & $^{+0.08}_{-0.1}$ & $^{+0.08}_{-0.09}$ & $^{+0.08}_{-0.09}$ & $^{+0.08}_{-0.09}$ & $^{+0.08}_{-0.1}$ & $^{+0.08}_{-0.1}$ & $^{+0.08}_{-0.1}$ \\
\hline
\end{tabular}
\end{center}
\caption{Measured dijet double-differential cross section per GeV and per unit $\chi$ as a function of the dijet angular variable $\chi$ for \AKT jets with $R=0.6$, compared to NLO pQCD calculations corrected for non-perturbative effects.  All uncertainties listed are fractional uncertainties.  There is an additional overall uncertainty of 11\% due to the measurement of the integrated luminosity which is not included in the systematic uncertainties given above.  The statistical uncertainty is calculated as $1/\sqrt{N}$, where $N$ is the number of events in a given bin.}
\label{tab:DijetChiResults06_5}
\end{sidewaystable*}

\clearpage

\begin{flushleft}
{\Large The ATLAS Collaboration}

\medskip

G.~Aad$^{\rm 48}$,
B.~Abbott$^{\rm 111}$,
J.~Abdallah$^{\rm 11}$,
A.A.~Abdelalim$^{\rm 49}$,
A.~Abdesselam$^{\rm 118}$,
O.~Abdinov$^{\rm 10}$,
B.~Abi$^{\rm 112}$,
M.~Abolins$^{\rm 88}$,
H.~Abramowicz$^{\rm 153}$,
H.~Abreu$^{\rm 115}$
% ...
\bigskip

% ...
$^{10}$ Institute of Physics, Azerbaijan Academy of Sciences, H. Javid Avenue 33, AZ 143 Baku, Azerbaijan\\
$^{11}$ Institut de F\'isica d'Altes Energies, IFAE, Edifici Cn, Universitat Aut\`onoma  de Barcelona,  ES - 08193 Bellaterra (Barcelona), Spain\\
% ...
$^{48}$ Albert-Ludwigs-Universit\"{a}t, Fakult\"{a}t f\"{u}r Mathematik und Physik, Hermann-Herder Str. 3, D - 79104 Freiburg i.Br., Germany\\
$^{49}$ Universit\'e de Gen\`eve, Section de Physique, 24 rue Ernest Ansermet, CH - 1211 Geneve 4, Switzerland\\
% ...
$^{88}$ Michigan State University, Department of Physics and Astronomy, High Energy Physics Group, East Lansing, MI 48824-2320, United States of America\\
% ...
$^{111}$ University of Oklahoma, Homer L. Dodge Department of Physics and Astronomy, 440 West Brooks, Room 100, Norman, OK 73019-0225, United States of America\\
$^{112}$ Oklahoma State University, Department of Physics, 145 Physical Sciences Building, Stillwater, OK 74078-3072, United States of America\\
% ...
$^{115}$ LAL, Univ. Paris-Sud, IN2P3/CNRS, Orsay, France\\
% ...
$^{118}$ Oxford University, Department of Physics, Denys Wilkinson Building, Keble Road, Oxford OX1 3RH, United Kingdom\\
% ...
$^{153}$ Tel Aviv University, Raymond and Beverly Sackler School of Physics and Astronomy, Ramat Aviv, IL - Tel Aviv 69978, Israel\\
% ...
$^{*}$ Deceased\end{flushleft}

\end{document}